\documentclass[final,3p,times]{elsarticle}
\pdfoutput=1

\usepackage{tabularx}
\usepackage{caption}
\usepackage{pbox}

\usepackage{amssymb}

\usepackage{graphicx}
\usepackage{epsfig}
\usepackage{dcolumn}
\usepackage{bm}
\usepackage{epstopdf}
\usepackage{url}
\usepackage{lineno}

\usepackage{tabu}
\usepackage{longtable}
\usepackage{lscape}

\newcommand{\photperelec}{\ensuremath{\gamma/e^-}}
\newcommand{\gevcc}{GeV\,c$^{-2}$}
\newcommand{\kevr}{keV$_{\rm r}$}
\newcommand{\kevee}{keV$_{\rm ee}$}
\newcommand{\cff}{CF$_{4}$}
\newcommand{\chf}{CH$_{4}$}
\newcommand{\cstwo}{CS$_{2}$}
\newcommand{\sfsix}{SF$_{6}$}
\newcommand{\ie}{{\em i.e.}}
\newcommand{\eg}{{\em e.g.}}
\newcommand{\etc}{{\em etc.}}
\newcommand{\etal}{{\em et al.}}
\newcommand{\xray}{X-ray}
\newcommand{\xrays}{\xray s}
\newcommand{\xy}{$x$-$y$}
\newcommand{\x}{$x$}
\newcommand{\y}{$y$}
\newcommand{\z}{$z$}
\newcommand{\threed}{3D}
\newcommand{\twod}{2D}
\newcommand{\isotope}[2]{$^{#2}${#1}}  
\newcommand{\degr}{\rm{o}}

\newcommand{\nocontentsline}[3]{}
\newcommand{\tocless}[2]{\bgroup\let\addcontentsline=\nocontentsline#1{#2}\egroup}

\newcommand{\Wellesley}{1}
\newcommand{\UNIZAR}{2}
\newcommand{\INFNNapoli}{3}
\newcommand{\METU}{4}
\newcommand{\Nagoya}{5}
\newcommand{\INFNLNF}{6}
\newcommand{\LPSC}{7}
\newcommand{\IPNL}{8}
\newcommand{\Napoli}{9}
\newcommand{\MITLNS}{10}
\newcommand{\CALTECH}{11}
\newcommand{\CEA}{12}
\newcommand{\INFNLNGS}{13}
\newcommand{\UCHICAGONOW}{14}
\newcommand{\RHUL}{15}
\newcommand{\sagaUniversity}{16}
\newcommand{\INFNGranSasso}{17}
\newcommand{\LBNL}{18}
\newcommand{\CSU}{19}
\newcommand{\KobeUniversity}{20}
\newcommand{\HAWAII}{21}
\newcommand{\UFL}{22}
\newcommand{\kyotouniversity}{23}
\newcommand{\Chiba}{24}
\newcommand{\UNM}{25}
\newcommand{\UCOLORADO}{26}
\newcommand{\INFNRoma}{27}
\newcommand{\kekTsukuba}{28}
\newcommand{\NSTC}{29}
\newcommand{\PARIS}{30}
\newcommand{\Roma}{31}
\newcommand{\kamiokaObservatory}{32}
\newcommand{\OXY}{33}
\newcommand{\Sheffield}{34}
\newcommand{\SILVERSIDE}{35}

\journal{Physics Reports}

\begin{document}

\begin{frontmatter}

\title{Readout technologies for directional WIMP Dark Matter detection}

\author[\Wellesley]{J.~B.~R.~Battat\corref{corresponding}}
\ead{jbattat@wellesley.edu}
\cortext[corresponding]{Corresponding author.}
\author[\UNIZAR]{I.~G.~Irastorza}
\author[\INFNNapoli]{A.~Aleksandrov}
\author[\METU]{M.~Ali Guler}
\author[\Nagoya]{T.~Asada} 
\author[\INFNLNF]{E.~Baracchini}
\author[\LPSC,\IPNL]{J.~Billard}
\author[\LPSC]{G.~Bosson}
\author[\LPSC]{O.~Bourrion}
\author[\LPSC]{J.~Bouvier}
\author[\INFNNapoli,\Napoli]{A.~Buonaura}
\author[\MITLNS,\CALTECH]{K.~Burdge}
\author[\UNIZAR]{S.~Cebri\'{a}n}
\author[\CEA]{P.~Colas}
\author[\INFNLNGS]{L.~Consiglio} 
\author[\UNIZAR]{T.~Dafni}
\author[\INFNLNGS]{N.~D'Ambrosio}
\author[\MITLNS,\UCHICAGONOW]{C.~Deaconu}
\author[\INFNNapoli,\Napoli]{G.~De Lellis}
\author[\LPSC]{T.~Descombes}
\author[\INFNNapoli]{A.~Di Crescenzo}
\author[\INFNLNGS]{N.~Di Marco}
\author[\RHUL]{G.~Druitt}
\author[\RHUL]{R.~Eggleston}
\author[\CEA]{E.~Ferrer-Ribas}
\author[\sagaUniversity]{T.~Fusayasu}
\author[\UNIZAR]{J.~Gal\'{a}n}
\author[\INFNNapoli,\Napoli]{G.~Galati}
\author[\UNIZAR]{J.~A.~Garc\'{i}a}
\author[\UNIZAR]{J.~G.~Garza}
\author[\INFNGranSasso]{V.~Gentile}
\author[\LBNL]{M.~Garcia-Sciveres}
\author[\CEA]{Y.~Giomataris}
\author[\RHUL,\MITLNS]{N.~Guerrero}
\author[\LPSC]{O.~Guillaudin}
\author[\CSU]{J.~Harton}
\author[\KobeUniversity]{T.~Hashimoto}
\author[\HAWAII]{M.~T.~Hedges}
\author[\UNIZAR]{F.~Iguaz}
\author[\KobeUniversity]{T.~Ikeda}
\author[\UFL]{I.~Jaegle} 
\author[\LBNL]{J.~A.~Kadyk}
\author[\Nagoya]{T.~Katsuragawa}
\author[\kyotouniversity]{S.~Komura}
\author[\kyotouniversity]{H.~Kubo}
\author[\Chiba]{K.~Kuge}
\author[\LPSC]{J.~Lamblin}
\author[\INFNNapoli,\Napoli]{A.~Lauria}
\author[\UNM]{E.~R.~Lee}
\author[\HAWAII]{P.~Lewis}
\author[\RHUL,\MITLNS]{M.~Leyton}
\author[\UNM]{D.~Loomba}
\author[\MITLNS,\UCOLORADO]{J.~P.~Lopez}
\author[\UNIZAR]{G.~Luz\'{o}n}
\author[\LPSC]{F.~Mayet}
\author[\UNIZAR]{H.~Mirallas}
\author[\KobeUniversity]{K.~Miuchi} 
\author[\kyotouniversity]{T.~Mizumoto}
\author[\kyotouniversity]{Y.~Mizumura}
\author[\INFNRoma]{P.~Monacelli} 
\author[\RHUL,\kekTsukuba]{J.~Monroe}
\author[\INFNNapoli,\Napoli]{M.~C.~Montesi}
\author[\Nagoya]{T.~Naka} 
\author[\kyotouniversity]{K.~Nakamura}
\author[\kyotouniversity]{H.~Nishimura}
\author[\KobeUniversity]{A.~Ochi} 
\author[\CEA]{T.~Papevangelou}
\author[\NSTC]{J.D.~Parker}
\author[\UNM]{N.~S.~Phan}
\author[\INFNLNGS]{F.~Pupilli}
\author[\LPSC]{J.~P.~Richer}
\author[\PARIS]{Q.~Riffard}
\author[\Roma,\INFNRoma]{G.~Rosa}
\author[\LPSC]{D.~Santos}
\author[\kyotouniversity]{T.~Sawano}
\author[\kamiokaObservatory]{H.~Sekiya}
\author[\HAWAII]{I.~S.~Seong}
\author[\OXY]{D.~P.~Snowden-Ifft}
\author[\Sheffield]{N.~J.~C.~Spooner}
\author[\sagaUniversity]{A.~Sugiyama}
\author[\KobeUniversity]{R.~Taishaku} 
\author[\kyotouniversity]{A.~Takada}
\author[\kamiokaObservatory]{A.~Takeda}
\author[\kekTsukuba]{M.~Tanaka}
\author[\kyotouniversity]{T.~Tanimori}
\author[\HAWAII]{T.~N.~Thorpe}
\author[\INFNNapoli]{V.~Tioukov}
\author[\MITLNS,\SILVERSIDE]{H.~Tomita}
\author[\Nagoya]{A.~Umemoto}
\author[\HAWAII]{S.~E.~Vahsen}
\author[\KobeUniversity]{Y.~Yamaguchi}
\author[\Nagoya]{M.~Yoshimoto}
\author[\MITLNS]{and E.~Zayas}

\address[\Wellesley]{Department of Physics, Wellesley College, 106 Central Street, Wellesley, MA 02481, USA}
\address[\UNIZAR]{Grupo de F\'{i}sica Nuclear y Astropart\'{i}culas, Departamento de F\'{i}sica Te\'{o}rica, Universidad de Zaragoza, 50009 Zaragoza, Spain}
\address[\INFNNapoli]{Istituto Nazionale di Fisica Nucleare, Sezione di Napoli, Naples, Italy}
\address[\METU]{METU Middle East Technical University, TR-06531 Ankara, Turkey} 
\address[\Nagoya]{Nagoya University, J-464-8602 Nagoya, Japan}
\address[\INFNLNF]{Istituto Nazionale di Fisica Nucleare, Laboratori Nazionali di Frascati, Frascati, Italy}
\address[\LPSC]{Laboratoire de Physique Subatomique et de Cosmologie, Universit\'e Grenoble Alpes, CNRS/IN2P3, 53, avenue des Martyrs, Grenoble, France}
\address[\IPNL]{IPNL, Universit\'{e} de Lyon, Universit\'{e} Lyon 1, CNRS/IN2P3, 4 rue E. Fermi 69622 Villeurbanne cedex, France}
\address[\Napoli]{Dipartimento di Fisica, Universit\`{a} Federico II, Napoli, I-80125 Napoli, Italy} 
\address[\MITLNS]{Laboratory for Nuclear Science, Massachusetts Institute of Technology, Cambridge, MA 02139, USA}
\address[\CALTECH]{Caltech Division of Physics, Mathematics, and Astronomy, Pasadena, CA, USA}
\address[\CEA]{IRFU, CEA, Universit\'{e} Paris-Saclay, Gif-sur-Yvette, France}
\address[\INFNLNGS]{Istituto Nazionale di Fisica Nucleare, Laboratori Nazionali del Gran Sasso, Assergi, Italy}
\address[\UCHICAGONOW]{University of Chicago, Kavli Institute for Cosmological Physics, Chicago, IL, USA}
\address[\RHUL]{Department of Physics, Royal Holloway, University of London, Egham Hill, Surrey, TW20 0EX, UK}
\address[\sagaUniversity]{Department of Physics, Faculty of Science and Engineering, Saga University, Saga, 840-8502, Japan }
\address[\INFNGranSasso]{Gran Sasso Science Institute (INFN), L'Aquila, Italy}
\address[\LBNL]{Lawrence Berkeley National Laboratory, 1 Cyclotron Road, Berkeley, CA 94720, USA}
\address[\CSU]{Department of Physics, Colorado State University, Fort Collins, CO 80523-1875, USA}
\address[\KobeUniversity]{Department of Physics, Kobe University, Rokodai, Nada-ku Kobe-shi, Hyogo, 657-8501, Japan}
\address[\HAWAII]{University of Hawaii, 2505 Correa Road, Honolulu, HI 96822, USA}
\address[\UFL]{University of Florida, Department of Physics, P.O. Box 118440, Gainesville, FL 32611, USA}
\address[\kyotouniversity]{Department of Physics, Kyoto University, Oiwakecho, Sakyo-ku Kyoto-shi, Kyoto, 606-8502, Japan}
\address[\Chiba]{Chiba University, J-263-8522, Chiba, Japan} 
\address[\UNM]{Physics \& Astronomy Department, University of New Mexico, 1919 Lomas Blvd NE, Albuquerque, NM 87131, USA}
\address[\UCOLORADO]{University of Colorado, Department of Physics, Boulder, CO, USA}
\address[\INFNRoma]{Istituto Nazionale di Fisica Nucleare, Sezione di Roma, Rome, Italy}
\address[\kekTsukuba]{Institute of Particle and Nuclear Studies, KEK, 1-1 Oho,Tsukuba, Ibaragi, 305-0801, Japan}
\address[\NSTC]{Neutron Science and Technology Center, Comprehensive Research Organization for Science and Society, Tokai-mura, Ibaraki, 319-1106, Japan}
\address[\PARIS]{APC, Universit\'e Paris Diderot, CNRS/IN2P3, CEA/Irfu, Obs de Paris, Sorbonne Paris Cit\'e, 75205 Paris, France}
\address[\Roma]{Dipertmento di Fisica dell'Universit a di Roma Sapienza, I-00185 Roma, Italy}
\address[\kamiokaObservatory]{Kamioka Observatory, ICRR, The University of Tokyo, Gifu, 506-1205 Japan}
\address[\OXY]{Department of Physics, Occidental College, Los Angeles, CA 90041, USA}
\address[\Sheffield]{Department of Physics and Astronomy, University of Sheffield, S3 7RH, UK}
\address[\SILVERSIDE]{Silverside Detectors Inc., Cambridge, MA, USA}

\begin{abstract}
  The measurement of the direction of WIMP-induced nuclear
  recoils is a compelling but technologically challenging strategy to
  provide an unambiguous signature of the detection of Galactic dark
  matter.  Most directional detectors aim to reconstruct the
  dark-matter-induced nuclear recoil tracks, either in gas or solid
  targets.  The main challenge with directional detection is the need
  for high spatial resolution over large volumes, which puts strong
  requirements on the readout technologies. In this paper we review
  the various detector readout technologies used by directional
  detectors.  In particular, we summarize the challenges, advantages
  and drawbacks of each approach, and discuss future prospects for
  these technologies.
\end{abstract}


\end{frontmatter}



\setcounter{secnumdepth}{4}

\section{Introduction}
\label{sec:introduction}
It is now widely accepted that a large fraction ($\sim$26\%) of the matter in the Universe is in the form of non-baryonic cold Dark Matter \cite{Bertone:2004pz}.  The Weakly Interacting Massive Particle (WIMP) is a leading Dark Matter particle candidate.  WIMPs arise in well-motivated extensions of the Standard Model, especially those including Supersymmetry \cite{Jungman:1995df,Feng:2010gw,Bertone:2004pz}.  A large, and growing, experimental effort employs diverse detection strategies to detect and characterize WIMPs. In particular, searches are underway to detect WIMPs produced at particle colliders \cite{Kane:2008gb}, for WIMP annihilation products via astrophysical searches for gamma rays and cosmic particles \cite{Jungman:1995df}, and for the direct interactions between WIMPs and target nuclei in the laboratory \cite{Goodman:1984dc,Gaitskell:2004gd}.

Direct detection experiments search for rare interactions between galactic halo WIMPs and nuclei in a detector. The scattering is generally assumed to be elastic \cite{Goodman:1984dc,Wasserman:1986hh}, though models of inelastic interactions have been proposed \cite{TuckerSmith:2001hy}.  With WIMP velocities on the order of $10^{-3}c$, the interaction is non-relativistic, and for typical WIMP and target nuclei masses, the recoil energies are small ($\lesssim$ 100 keV). The main challenge in direct detection is to positively identify rare WIMP interactions amid a diverse collection of backgrounds that can mimic the signal of interest.  As such, much effort is directed at reducing detector internal backgrounds, shielding from external backgrounds, and improving signal-background discrimination.  Additionally, experimenters seek a Dark Matter signature with high discriminating power -- \ie{} those that are only weakly correlated with backgrounds.
In 1988, Spergel pointed out that the motion of the Earth through the Galactic halo of WIMP Dark Matter would produce a forward-backward asymmetry in the recoil rates in the Galactic reference frame \cite{Spergel:1987kx}. At present, no known background can mimic this signal, and so the directional signal is widely held to be the cleanest signature of Galactic Dark Matter.
For reviews of direct Dark Matter detection, we refer the reader to an abundant literature \cite{Jungman:1995df,Feng:2010gw}.  Additionally, for an overview of the motivation for and discovery potential of {\em directional} Dark Matter detection, we point to Ref.~\cite{Mayet2016}, and for an overview of directional Dark Matter search experiments, we recommend Ref.~\cite{Ahlen:2009ev}. 

Directional Dark Matter detection aims to reconstruct both the energy and the track of a recoiling nucleus following a WIMP scattering. In contrast, direction-insensitive detectors typically measure only the (time-dependent) energy spectrum.  Direction-insensitive experiments have operated underground for several decades, with masses now approaching the ton-scale. This detection strategy has lead to a wealth of results \cite{Cushman:2013zza}, with exclusion limits that have begun to approach the neutrino floor \cite{Billard:2013qya,Akerib:2015rjg}.  Directional detection is a next-generation strategy that offers a unique opportunity to conclusively identify WIMP events, even in the presence of backgrounds \cite{Billard:2009mf}.  Indeed, the motion of the Solar System through the Galaxy causes a strong angular anisotropy in the WIMP velocity distribution (as observed in the Earth frame).  WIMP-induced nuclear recoils will, in turn, exhibit a dipole feature \cite{Spergel:1987kx}. On the contrary, the background distribution \cite{Mei:2005gm} is expected to be isotropic in the galactic rest frame. In fact, several directional features provide an unambiguous discriminant between backgrounds and WIMPs, \eg{} dipole~\cite{Spergel:1987kx}, ring-like\footnote{The maximum of the recoil rate lies in a ring around the mean recoil direction.}  \cite{Bozorgnia:2011vc} and aberration\footnote{The annual variation of the mean recoil direction.} \cite{Bozorgnia:2012eg}.  Depending on the unknown WIMP-nucleon cross-section, directional detection may be used to: exclude Dark Matter \cite{Billard:2010gp,Henderson:2008bn}, discover galactic Dark Matter with a high significance \cite{Billard:2009mf,Billard:2011zj,Green:2010zm} or constrain WIMP and halo properties \cite{Billard:2010jh,Lee:2012pf,O'Hare:2014oxa,Alves:2012ay,Lee:2014cpa}. Directional detection also holds the promise of achieving sensitivity below the neutrino floor \cite{Grothaus:2014hja,O'Hare:2015mda}.  In the absence of directional sensitivity, results from experiments using different targets must be combined to surpass the neutrino limit \cite{Ruppin:2014bra}.

The challenge lies in the construction of a detector that is sensitive to directional signatures.  To measure the direction of WIMP-induced nuclear recoils, one can reconstruct the 3D nuclear recoil track (or the projection of the track along one or two dimensions).  
Alternatively, one can employ a detector with anisotropic response to nuclear recoils to infer information about the recoil track direction, without the need to reconstruct the track. While several such ideas have been put forward \cite{Belli:1992zb,Spooner:1996gr,Shimizu:2002ik}, none have been shown yet to provide enough directional sensitivity for a WIMP search.  In the rest of this review, we focus on experimental techniques to reconstruct the nuclear recoil track.  In order to do so, the detector must have high spatial granularity over a large volume. This poses a significant technological challenge to the readout used for track reconstruction. The focus of this review is to provide a critical assessment of the diverse readout technologies currently in use to reconstruct tracks in directional detectors.

In Section~\ref{sec:techchallenges} we describe some of the technological challenges of building a directional Dark Matter detector.  The remainder of this work (Sections~\ref{sec:mwpc} through \ref{sec:emulsions}) describes each readout technology in turn: Multi-Wire Proportional Chambers (MWPCs), Micro Pattern Gaseous Detectors (MPGDs), optical readouts, and nuclear emulsions.

\section{Technological challenges}
\label{sec:techchallenges}
The required spatial resolution for directional detection is set by the length of a WIMP-induced nuclear recoil, which, in turn, depends on the recoil energy and density of the target material. For example, a nucleus recoiling with energy $\lesssim100$~keV travels $\lesssim 100$~nm in a solid. Readouts capable of measuring 100~nm-long tracks in solids do exist in the form of microscopic inspection of nuclear emulsions.  These are described in Section~\ref{sec:emulsions}. Most directional experiments, however, opt to use a lower-density target (gas at $\sim 0.1$~atm), in which recoil tracks have millimeter extent. This relaxes the required detector spatial resolution, albeit at the expense of detector mass per unit volume.  Gas detectors typically employ a Time Projection Chamber (TPC) and a readout with sub-millimeter spatial resolution in one, two, or three dimensions, enabling either partial or full reconstruction of the recoil track geometry.  A significant technological challenge for TPC detectors is the construction of large-volume detectors ($\sim10^3$~m$^3$) with high spatial resolution and high radiopurity.  

A WIMP-induced nuclear recoil may produce as few as $10^2$--$10^3$ primary electron-ion pairs in the gas.  The small signal strength is due both to the low energy of the nuclear recoil, and the small quenching factors in the gas.
The quenching factor is given as the ratio of the electron-equivalent energy (usually expressed as \kevee) to the recoil energy (\kevr).  For example, consider a 100\,GeV\,c$^{-2}$ WIMP traveling at 10$^{-3}\,c$, incident on a detector filled with \cff{} gas.  The resulting fluorine recoil will have a maximum recoil energy of 40\,\kevr.  At that energy, the quenching factor of fluorine in \cff{} is 0.45 \cite{Guillaudin:2011hu}, and so the electron-equivalent energy of the recoil is 18\,\kevee.  The W-value of \cff{} (the average energy required to ionize a gas molecule) is 34.3\,eV \cite{Reinking1986JAP}, and so the recoil will produce only 530\,electron-ion pairs.

To enhance this weak signal, an electron amplification device is used. 
This device can take many geometries, but in all cases it consists of a region of high electric field through which primary electrons are accelerated to sufficient energies to impact-ionize the surrounding gas molecules. This results in an exponential growth of ionization electrons -- an avalanche. The conventional choice for an amplification device is the multi-wire proportional counter (MWPC, see Section \ref{sec:mwpc}). More recently, micro-pattern gas detectors (MPGD) such as Micromegas (Section \ref{sec:micromegas}), GEMs, and $\mu$PICs (Section \ref{sec:mupic}) have been used. They achieve the necessary amplification field through patterned electrodes on planar substrates, and are constructed through lithography techniques, much like printed circuit boards. The use of MPGDs may bring advantages in terms of simplicity of construction, and high spatial granularity. The signal of interest is either the induced electrical signal on the amplification electrode in the case of the MWPC, Micromegas, and $\mu$PIC, or on separate {\em sense} electrodes in the case of GEMs. In addition, a target gas with strong scintillation photon yield can be used to produce optical signatures during the amplification process.  This scintillation light can then be imaged, as described in Section \ref{sec:optical}.

Each of these readout techniques is described in the following sections, together with a summary of the status of ongoing R\&D on these technologies. All of the technologies are under active development to study how well they meet the challenges of directional detection.

An ideal directional detector would be capable of reconstructing the nuclear recoil track in three dimensions (3D), with high spatial granularity and angular resolution.  It would also be sensitive to the vector direction of a recoil (sense-recognition), not just the axis of the track.  The readout would not introduce backgrounds from radioimpurities, and it would be able to reconstruct the absolute position of an event vertex, allowing for a full-volume detector fiducialization.  Although directional Dark Matter detection can tolerate a sizable background contamination \cite{Billard:2009mf} because of the intrinsic difference between the background-induced and WIMP-induced angular spectra, the discrimination of background electron recoils from nuclear recoils can be achieved through track topology (see \eg{} \cite{Billard:2012dy}).  Additionally, the readout would be robust to high-voltage operation over long periods of time (years), and would be scalable to tens or hundreds of cubic meters of volume.  

At present, no technology satisfies all of these requirements, and indeed these design criteria do not carry equal weight.  We describe the relative merits of the design criteria listed above.

To reconstruct WIMP-induced nuclear recoil tracks, the spatial granularity of the detector readout must be finer than the track length ($\sim$mm in low-pressure gases, $\sim$0.1$\mu$m in solids).  The spatial resolution can be achieved by segmenting the readout plane, or by measuring the temporal profile of the recoil signal.
Some readouts can provide both fine spatial granularity and high temporal resolution.  For example, pixel chip readouts (Section~\ref{sec:pixelchip}) sample the \xy{} plane, providing a 2D projection of the track, but do so at a sampling rate that is fast enough to enable full 3D track reconstruction.  Other readouts are restricted to 2D, but can be used in alongside another technology to achieve 3D tracking.

For example, CCD readouts (Section~\ref{sec:optical}) measure the \xy{} projection of a track, but could be combined with the timing resolution of photomultiplier tubes to recover the \z{} projection of the track.  Alternatively, it may be sufficient to operate with only 2D, or even 1D tracking capability.  Green and Morgan \cite{Green:2006cb} have shown that for a 100\,\gevcc{} WIMP and \cstwo{} target, a 2D detector requires twice the exposure than a 3D detector to observe a WIMP signature.  Meanwhile, Billard \cite{Billard:2014ewa} showed that in terms of WIMP discovery potential, a 1D detector is only three times less effective than a 3D detector, though in order to characterize the galactic WIMP velocity distribution, a 3D readout is required \cite{Billard:2010jh,Lee:2012pf,Lee:2014cpa}.  This study did not contemplate background discrimination/rejection, however, and experimental studies have shown that the discrimination threshold for a detector with 1D track reconstruction is three times worse than for a detector with 2D reconstruction \cite{Phan:2015pda}.

Whether 1D, 2D, or 3D track reconstruction is achieved, a second, powerful recoil signature is the vector direction (sense) of the track.  For a WIMP mass of 100\,\gevcc, a 3D detector with recoil sense recognition can identify a WIMP signature with an order of magnitude smaller exposure than a detector without sense recognition \cite{Green:2006cb}.  For 2D readouts, the difference is two orders of magnitude.  At low WIMP masses (10\,\gevcc), however, sense recognition has almost no effect on required exposure \cite{Billard:2011zj}.

The recoil sense is encoded in the track in two ways.  First, the nuclear recoils produce more ioniation per unit length at the start than at the end of the track.\footnote{WIMP-induced nuclear recoils have energies well below the Bragg peak, and so dE/dx decreases as the recoiling nucleus loses energy.} Second, because large-angle scatters are more likely at low recoil energy, the beginning of a recoil track is straighter than the end \cite{Billard:2012bk}.  Both of these signatures argue for readout granularity much finer than the track length, and the first requires sensitivity to the ionization density along the track.

A further challenge is to reconstruct the track axis and sense at low recoil energies.  The relevant energy threshold for directional detectors is the directionality threshold, meaning the lowest energy for which recoil track geometries can be reconstructed.  Nuclear recoil events may be detectable below the directionality threshold, but without geometric information. 

Directional experiments must pay close attention to signal-background discrimination.  Background rejection and mitigation requires attention to the choice of experimental site, detector target material, and operating conditions, in addition to constructing a readout with low intrinsic backgrounds.  In this work, we focus on the latter, describing ongoing R\&D on background mitigation and discrimination for each readout technology.

A clear challenge facing directional Dark Matter detectors is how to scale to large volume at reasonable cost, while preserving stable operation and track reconstruction capability.  The scalability of each readout technology is addressed in this paper.

Although in the present review we focus on the readout-specific features,
issues like the type of gas, pressure, drift distance, charge vs. light readout, electron vs. negative-ion drift, among others, will affect the final experimental parameters \cite{Ahlen:2009ev}, and therefore the constraints on the readout will vary accordingly.

\section{Multi-Wire Proportional Chambers}
\label{sec:mwpc}
Invented by Georges Charpak in 1968 and honored with a Nobel Prize in 1992, the Multi-Wire Proportion Chamber (MWPC) has a long and venerable history \cite{Charpak:1968kd}.  MWPCs continue to be a workhorse for large particle physics experiments.  MWPCs are the main tracking systems for the world's largest TPCs, including ALICE (88\,m$^3$) \cite{Alme:2010ke} and STAR (4.2\,m long, 4\,m diameter) \cite{Anderson:2003ur}, as well as next-generation liquid argon neutrino experiments such as DUNE \cite{Acciarri:2015uup}.
In the realm of Dark Matter detection, the Directional Recoil Identification From Tracks (DRIFT) experiment has used MWPCs for over a decade in a low-pressure gas TPC \cite{Battat:2014van}.  Thorough descriptions of MWPCs and their use in particle detection are provided in Refs. \cite{Sauli:1977mt,Blum:2008zza,Sauli:2014cyf}.  Here, we discuss MWPCs in the context of directional Dark Matter detection, including their use in DRIFT.

\subsection{General features of MWPCs}
\begin{figure}
  \centering
  \includegraphics[width=.4\textwidth]{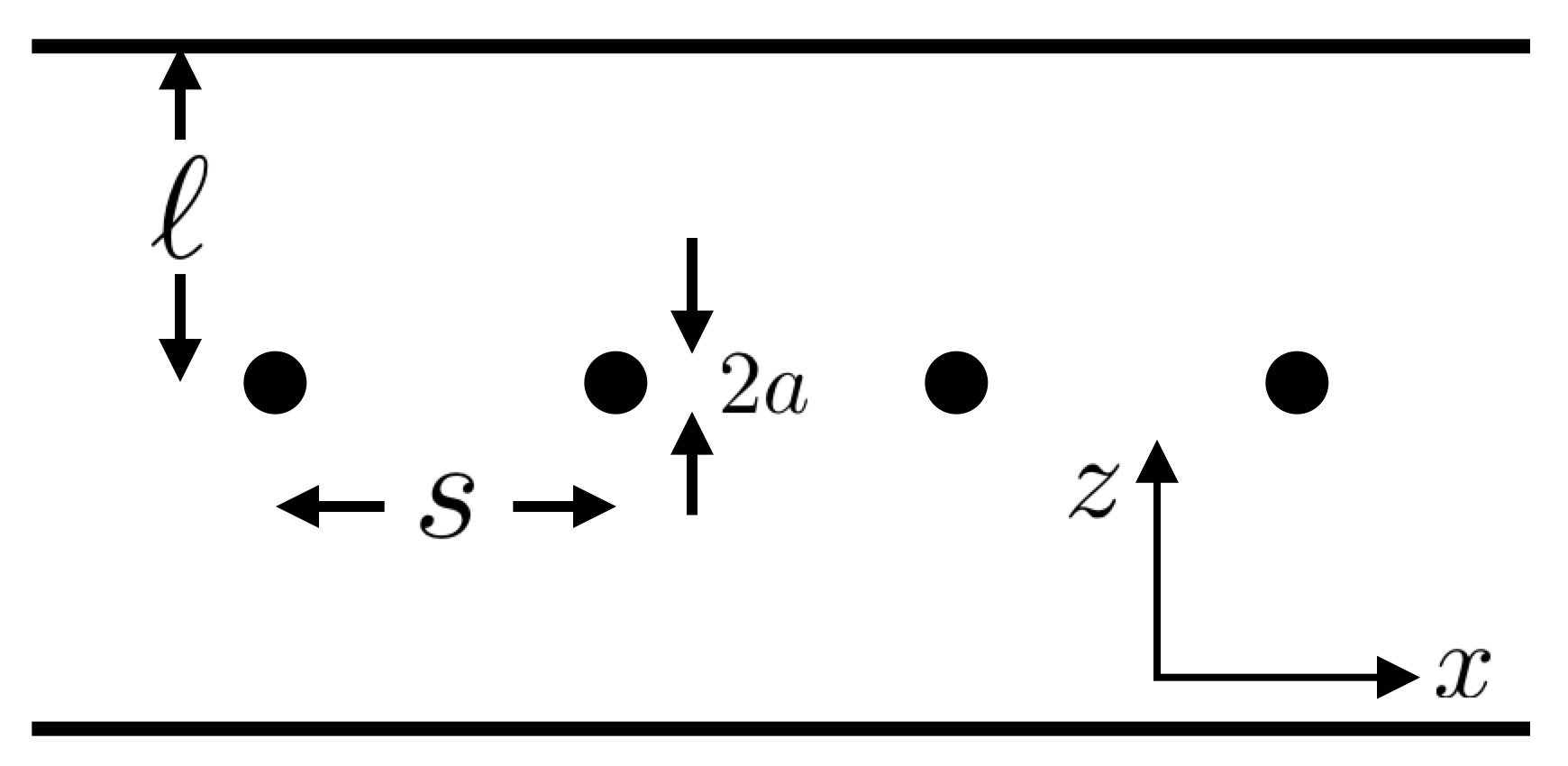}
\caption{\label{fig:mwpcGeometry}
MWPC geometry showing the anode wires (filled circles) of radius $a$ and pitch $s$, and cathode planes a distance $\ell$ away.}
\end{figure}

\begin{figure}
  \centering
  \includegraphics[width=.4\textwidth]{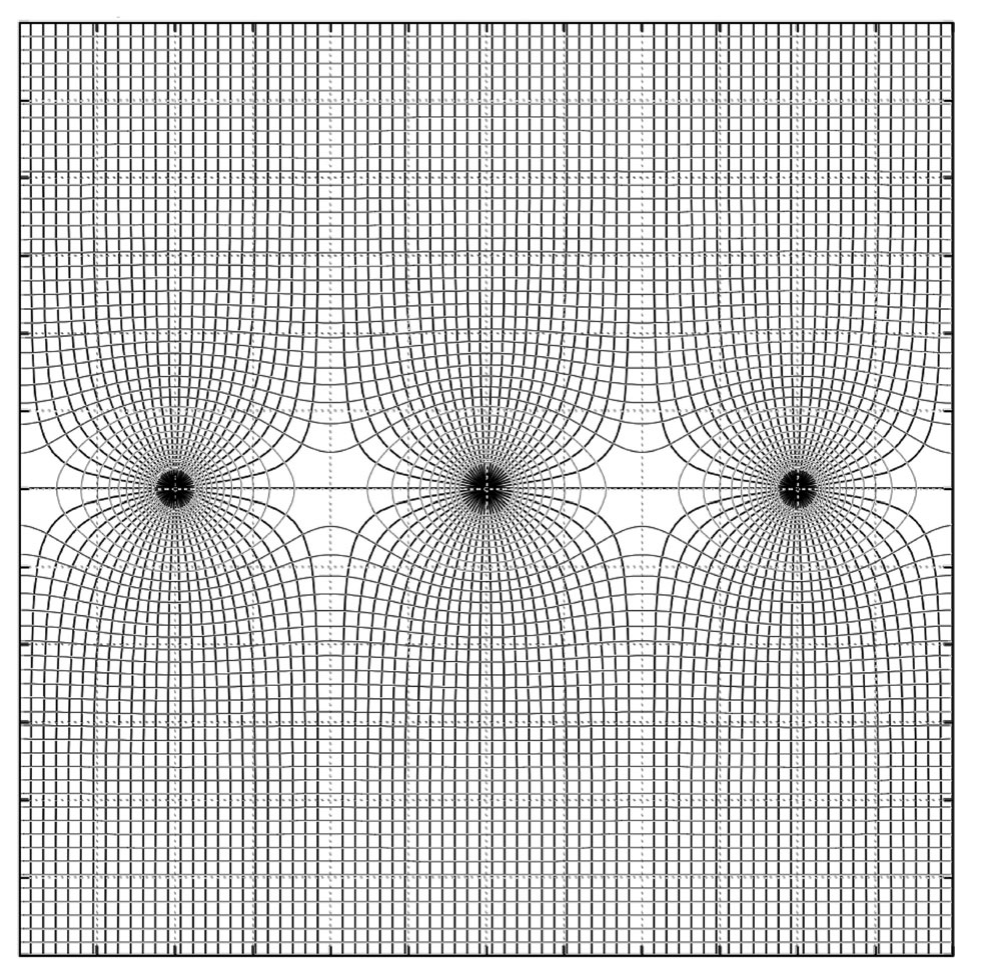}
\caption{\label{fig:mwpcFieldLines}
Field lines in the MWPC are uniform near the cathodes, and approximately radial near the anode wires.  Image from Ref. {\protect\cite{Sauli:2014cyf}}.}
\end{figure}

In the MWPC, a set of parallel, equally-spaced anode wires (typical spacing 1 to 5\,mm) are situated symmetrically between two cathode planes (see Figure\,\ref{fig:mwpcGeometry}).
The anode-cathode gap $\ell$ is typically many times larger than the anode wire spacing.  The cathode plane is held at a common, negative potential $-V_0$, and the anode wires are grounded.  The resulting electric field has two regimes (see Figure\,\ref{fig:mwpcFieldLines}).  Far from the anode wires, the field is largely uniform and perpendicular to the cathode plane, and electrons drift toward the anode wires while positive ions drift toward the cathode.  The field strength there is well approximated by
\begin{equation}\label{eqn:mwpcEfieldUniform}E \approx \frac{CV_0}{2\epsilon_0 s}, \end{equation}
where $C$ is the capacitance per unit length of the wire, $s$ is the anode wire pitch (see Figure~\ref{fig:mwpcGeometry}), and $\epsilon_0=8.854\times 10^{-12}$\,F/m is the permittivity of free space.

Close to an anode wire, the field is largely radial, as in a proportional tube, with approximate strength
\begin{equation}\label{eqn:mwpcEfieldRadial} E \approx \frac{CV_0}{2\pi\epsilon_0}\frac{1}{r}. \end{equation}
As electrons approach the anode wires, they experience a field that increases rapidly, and they gain enough energy to excite and ionize surrounding gas molecules.  The ionization leads to avalanche multiplication.  
Gains of $10^5$ are readily achievable, though in the context of directional Dark Matter detection where low gas pressures are used to ensure long recoil tracks, gains of $10^3$ are more typical.

A useful MWPC design parameter is $C$, the capacitance per unit wire length, given by
\begin{equation}\label{eqn:mwpcCapacitance} C = \frac{2\pi\epsilon_0}{\pi\ell/s - \ln\left(2\pi a /s \right)},\end{equation}
where $a$ is the radius of the anode wire.  The capacitance is a weak function of the anode wire diameter, but is approximately proportional to wire spacing.

Dark Matter searches require large detection volumes, so the MWPC is
used as an endcap detector attached to a large conversion/drift volume
(\ie{} a TPC or a drift chamber).  For example, in the DRIFT
experiment, an MWPC with $\ell=1$\,cm mates to a 50-cm-long
conversion/drift region.  The MWPC cathode is a wire plane, making it
transparent to electrons.

\subsection{MWPC geometry and size limitations}
\label{sec:mwpcGeomSize}
Practical limitations exist on both the anode wire pitch (and therefore the spatial resolution), and the anode wire length (and therefore the maximum MWPC size).  As a rule of thumb, the minimum anode pitch is $\sim 2$\,mm, and the maximum wire length is $\sim 2$\,m \cite{Sauli:1977mt}.

The spatial resolution of the MWPC is of order the wire pitch (though Section\,\ref{sec:mwpcSpatialResolution} describes the use of a segmented cathode to achieve a spatial resolution well below the anode wire pitch).  Difficulties, in the form of discharges, arise if one attempts to increase the spatial resolution by reducing the wire pitch.  In particular, the gas multiplication factor $M$ depends exponentially on the charge per unit wire length $C\,V_0$.  For constant detector gain, one must preserve the quantity $C\,V_0$.  But reducing the anode wire pitch will decrease $C$ (see Equation~\ref{eqn:mwpcCapacitance}), and so $V_0$ must be increased, often substantially.  For example, the DRIFT MWPC has $\ell=1$\,cm, $2a=20\,\mu$m, and $s=2$\,mm, which gives $C=2.9$\,pF/m.  If the anode pitch were reduced to 1\,mm, then $V_0$ would need to be increased by a factor of 1.8 in order to preserve $M$.  This would exceed the sparking threshold in the detector.  While it is possible to increase $M$ by decreasing the anode wire diameter, this introduces mechanical challenges, as described next.

The maximum size of an MWPC is limited by the achievable mechanical anode wire tension.  In the nominal MWPC configuration, the anode wires are in an unstable equilibrium, with the wire tension balancing the electrostatic attraction between anode and cathode.  The required anode wire tension grows with the square of the wire length.  A large wire tension is therefore desired, but the tension, and therefore the maximum wire length, is limited by the ultimate tensile strength of the wire.  If we define $T_M$ to be the maximum wire tension before deformation, then the maximum stable anode wire length $L_M$ is given by
\begin{equation}\label{eqn:mwpcMaxLength} L_M = \frac{s}{CV_0}\sqrt{4\pi \epsilon_0 T_M}. \end{equation}
%
For example, the DRIFT detector uses stainless steel wires with $2a=20\,\mu$m, so $T_M \approx 0.15$\,N. During standard DRIFT operation, $V_0 = 2884\,$V, with a corresponding maximum stable anode wire length of $L_M = 1$\,m.  The use of thinner wires (\eg{} as a way to increase gain while decreasing the anode wire pitch) would reduce $T_M$ and therefore restrict the maximum size of the MWPC.  

Techniques exist to construct MWPCs larger than allowed by the wire tension restriction described above.  For example, mechanical supports for the wires can be inserted along the MWPC, as was done in the NA24 experiment at CERN \cite{DePalma:1983kj}.  These supports contact the anode wires and must be insulating.  They will produce a local electric field perturbation that will impact the particle detection efficiency \cite{Schilly:1971gi}, though this effect can be largely mitigated \cite{Charpak:1971ku}.  Additionally, MWPCs can be tiled to form a larger readout plane, as was done in ALICE \cite{Alme:2010ke} and STAR \cite{Anderson:2003ur}.  A description of various MWPC construction methods is given in Ref.~\cite{Veres:1978nf}. The upcoming use of kiloton liquid argon (LAr) detectors with MWPC readouts for neutrino experiments has also motivated the construction of MWPCs with areas larger than $1\times1$\,m$^2$ \cite{Karagiorgi:2013cwa,Acciarri:2015uup}.  Section~\ref{sec:mwpcLAr} discusses some synergies between LAr-based neutrino detectors and directional Dark Matter detection.

\subsection{Spatial resolution}
\label{sec:mwpcSpatialResolution}
A recoil track in the TPC produces charge carriers (\eg{} electrons or negative ions) that diffuse in both the transverse (\xy) and longitudinal (\z) directions as they travel toward the readout plane.  Negative ions suffer less diffusion than electrons, often at the thermal limit \cite{SnowdenIfft:2013iy}:
\begin{equation}
\sigma_{thermal} = \sqrt{\frac{2kTL}{eE}} \approx 0.72\,\mbox{mm}\,\sqrt{\left(\frac{L}{1\,\mbox{m}}\right)\left(\frac{1\,\mbox{kV cm}^{-1}}{E}\right)}
\end{equation}
where $k$ is Boltzmann's constant, $T$ is the physical temperature of the gas, $L$ is the drift distance of the charge carrier, $e$ is the fundamental electric charge and $E$ is the magnitude of the drift electric field.
Assuming that no confinement techniques are implemented (\eg{} a magnetic field parallel to the drift direction would suppress diffusion in the \xy{} direction), diffusion can limit the achievable spatial resolution in all three dimensions.  

For an MWPC with solid cathode planes, as described in Section~\ref{sec:mwpcGeomSize}, the spatial resolution in the \x{}-direction is of order the anode wire pitch $s$.  There is no spatial information in the $y$-direction (parallel to the anode wires).  Early on, however, it was recognized that measuring the induced charge on a segmented cathode plane allowed for spatial reconstruction in both the $x$ and $y$ directions.  In fact, point resolutions much finer than the wire spacing have been demonstrated in both the $x$ and $y$ dimensions.  For example, Charpak \etal{} \cite{Charpak:1977sv} achieved a point resolution of $\sigma_x=0.15$\,mm and $\sigma_y=0.035$\,mm for soft \xrays{} in an MWPC with $s=2$\,mm.

Spatial resolution along the drift direction \z{} is achieved via timing (pulse shape) information, and generally provides higher spatial resolution than in either the \x{} or \y{} directions.  As described below, DRIFT uses an electronegative gas (\cstwo) that allows for the drift of negative ions, with charge carrier drift speeds of $\sim5$\,cm/ms.  Readout electronics sample the resulting anode electrical signal at 1\,MHz, which corresponds to $\Delta z = 50\,\mu$m.  Both the track extent along \z{} and the longitudinal diffusion of charge carriers during drift affect the pulse shape.  A recent discovery of minority carriers in negative ion gas \cite{Snowden-Ifft:2014taa} makes it possible to measure the absolute \z{} coordinate of an event in the drift/conversion region.  This allows for \threed{} detector fiducialization, as well as a correction for the diffusion contribution to the track size.

\subsection{Radiopurity}
For WIMP Dark Matter searches, the largest background found for wire chambers has been radon-progeny decays producing radon-progeny recoils (RPRs) and, through energy degradation, low energy alphas (LEAs) \cite{Battat:2015rna}.  Unlike the micropatterned gaseous detectors described in Section~\ref{sec:mpgdIntro}, MWPCs can be made to have only metallic wires in the $x$-$y$ readout plane (with some support structure around the perimeter).  The wires can be made very low background via material selection, and subsequent treatment (\eg{} electropolishing \cite{Schnee:2014eea} or nitric acid etching \cite{Battat:2014oqa}) can further improve radiopurity by removing a thin surface layer that is rich in daughters from the \isotope{Rn}{222} decay chain.  For example, removing $<1$\,$\mu$m of material via electropolishing has been shown to reduce the contamination by a factor of $>100$ \cite{Schnee:2014eea}.

Detector fiducialization in \x{} and \y{} (by applying an edge-crossing cut in the analysis), as well as by fiducializing in the \z{} direction using \eg{} minority carriers \cite{Snowden-Ifft:2014taa} allows for the rejection of remaining RPRs and LEAs.  Using this technique, the DRIFT collaboration has demonstrated zero-background operation during a 46.3 live-day exposure \cite{Battat:2014van}.

MWPCs have also been used for low-background material screening.  For example, the BetaCage project \cite{Bunker:2014bea} uses an MWPC with \z-fiducialization via charge induction fraction, and they anticipate a sensitivity of $\sim 0.1$\,alphas\,m$^{-2}$\,day$^{-1}$ and 0.1\,betas\,keV$^{-1}$\,m$^{-2}$\,day$^{-1}$.

\subsection{MWPCs in the DRIFT Dark Matter detector}

The use of MWPCs for directional Dark Matter experiments was pioneered by the DRIFT collaboration \cite{SnowdenIfft:1999hz}.  Utilizing negative ion drift technology, the DRIFT collaboration has been operating cubic-meter-scale TPCs underground in the Boulby mine (England) since 2001.  For a description of the currently active DRIFT-IId detector, see Ref.~\cite{Alner:2005xp}.  Recently the collaboration published the leading WIMP cross-section limits from a directional detctor \cite{Battat:2014van}.  A brief summary of the DRIFT-IId detector as it was operated for the limits run is provided here for convenience.

\begin{figure}
\centering
\includegraphics[width=.4\textwidth]{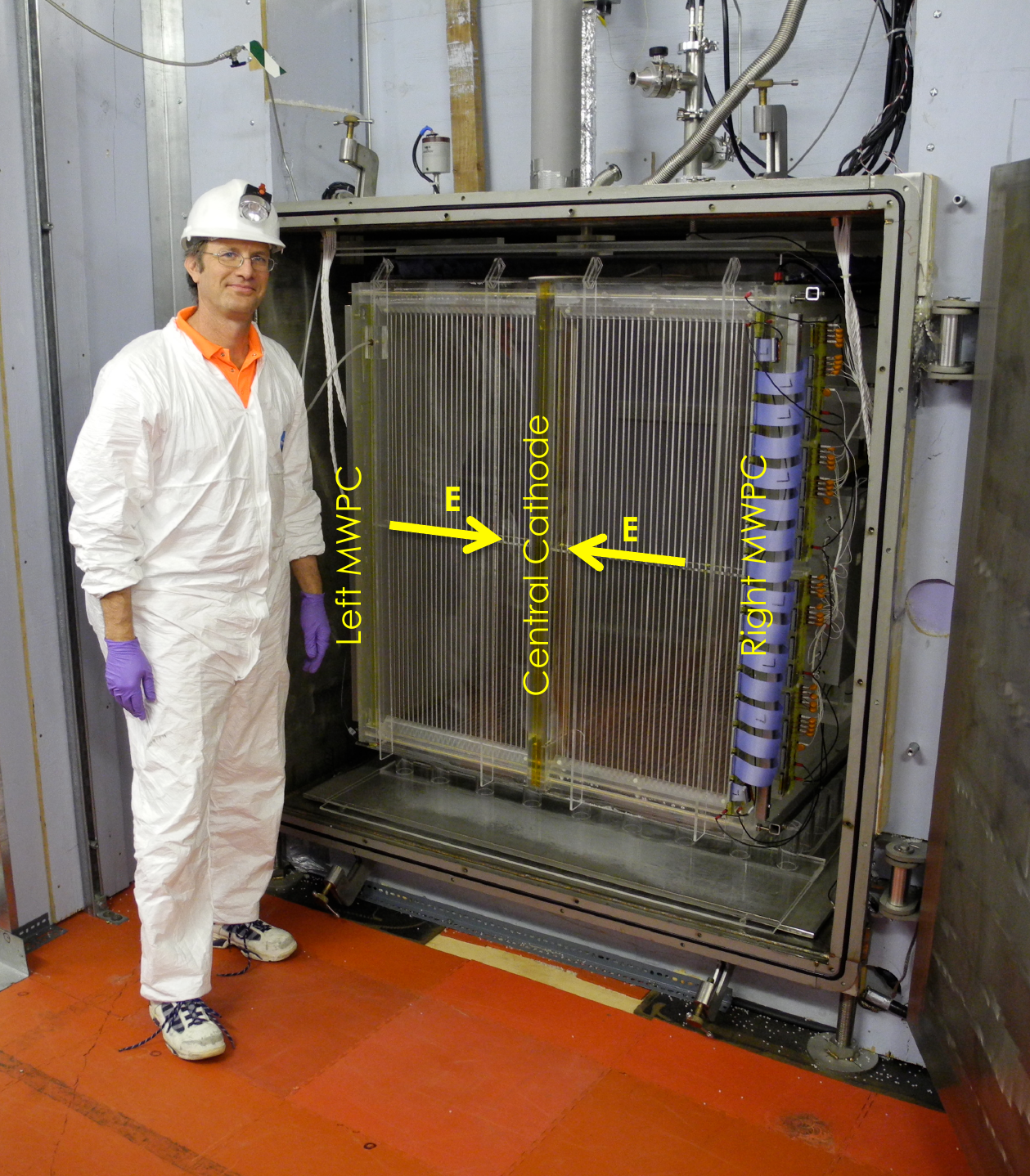}
\caption{\label{fig:DRIFTIIdPic.pptx.pdf}
The DRIFT-IId detector in the Boulby Mine.}
\end{figure}

A 1.53\,m$^3$ low-background stainless steel vacuum vessel provided containment for the gas, which was 30+10+1\,Torr CS$_2$+CF$_4$+O$_2$. Within the vacuum vessel were two back-to-back TPCs with a shared, vertical, central wire cathode cathode (Figure~\ref{fig:DRIFTIIdPic.pptx.pdf}). Two field cages, located on either side of the central cathode, defined two drift regions of 50\,cm length in which recoil tracks could be detected. Charge readout of tracks was provided by two MWPCs each comprised of an anode plane (20~$\mu$m stainless steel wires with 2\,mm pitch) sandwiched with a 1\,cm gap between two perpendicular grid planes (100 $\mu$m stainless steel wires also with 2\,mm pitch). The potential difference between the grids and the grounded anode planes was -2884\,V.  An acrylic strong back provided mechanical stability against the wire tension. The central cathode voltage, at -30\,kV, produced a drift field of 580\,V/cm. For each MWPC, 448 grid wires (measuring the \y-direction) were grouped down to 8 sense lines that were then pre-amplified, shaped and digitized. The anode sense lines (measuring the \x-direction) were treated identically. Eight adjacent readout lines (either anode or grid) therefore sampled a distance of 16\,mm in \x{} and \y. Voltages on the grid and anode lines were sampled at 1\,MHz providing information about the event in the drift direction $z$.  With a negative ion drift speed of $\sim$5\,cm/ms, this sampling rate corresponds to 50\,$\mu$m spatial resolution in \z. The 52(41) of the remaining wires form a grid(anode) veto on each side of this fiducial area against ionizing radiation from the sides of the detector, for each MWPC. Triggering of the data acquisition system occurred on individual anode lines. All lines were digitized from -3\,ms to +7\,ms relative to the trigger with 12-bit digitizers. The region bounded by the vetoes and the inner MWPC grid planes formed a fiducial volume of 0.803~m$^3$. Each side of the detector was instrumented with an automated, retractable, $\sim$100\,$\mu$Ci $^{55}$Fe calibration sources, which allowed regular monitoring of detector gain and performance.

\subsubsection{Directionality}
DRIFT detectors have been shown to be directional.  A range component signature, based on the measurements of the components of the range of the recoils, $\Delta x$, $\Delta y$ and $\Delta z$ was studied in \cite{Burgos:2008mv}.  Despite utilizing induced signals on the grid wires, no useful directional information was found in the $\Delta y$ component.  But the amplitude of the variation of the ratio $\Delta x/ \Delta z$ was found to increase with energy from $\sim$50\,\kevr, which at that time was the operational threshold (the threshold has since been lowered to 30\,\kevr).  DRIFT's use of negative \cstwo{} anion drift suppresses diffusion, and therefore reduces the directional energy threshold relative to electron drift gases.

\begin{figure}
\centering
\includegraphics[width=.55\textwidth]{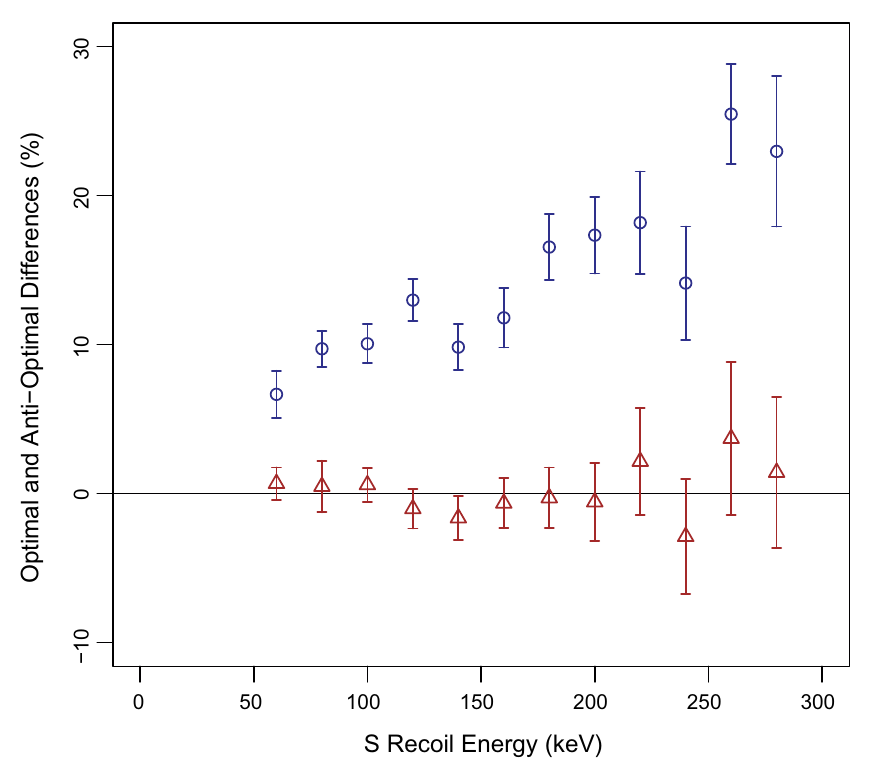}
\caption{\label{fig:mwpc_drift_ht}
Measurements of the recoil sense in DRIFT-II as a function of energy.  The \y-axis quantifies the recoil sense reconstruction, with zero corresponding to correct sense reconstruction only half of the time (\ie{} no head-tail sensitivity).  Blue circles correspond to data taken with a neutron source aligned with the \z-axis.  DRIFT has the best spatial resolution along the \z{} direction.  The measurements show a recoil sense sensitivity all the way down to the detector threshold of 50\,\kevr{} (the threshold has since been improved to 30\,\kevr).  The red triangles correspond to a combined data set with the neutron source aligned along the \x{} and \y{} axes, producing recoils that should exhibit no recoil sense asymmetry in the \z{} direction.  As expected, these measurements are consistent with zero.  Figure from \protect\cite{Burgos:2008jm}.}
\end{figure}

The head-tail signature (sense recognition) was also
studied \cite{Burgos:2008jm}.  This signature was found to be four
times stronger than the range component signature, as expected.  As
shown in Figure~\ref{fig:mwpc_drift_ht}, recoil sense recognition was
achieved all the way down to the threshold (at the time) of
50\,\kevr{}.  An extrapolation of the data suggest it is non-zero down
to energies below threshold.

\subsubsection{Cost and reliability}
The DRIFT collaboration has built seven cubic-meter-scale devices, and has operated MWPCs underground for more than a decade.  In each case the cost of the MWPC was insignificant compared to the electronic readout and vacuum vessel.  In recent years the stability of operation has improved to the point where the collaboration can reliably run for more than 6 months at a time.  The most vexing problem for DRIFT has been broken anode wires.  Repair times for broken wires, however, are only several days.

\subsubsection{Backgrounds}
As shown by DRIFT \cite{Daw:2010ud}, RPRs are expected to be the dominant background to directional WIMP searches in gas.  The RPRs were initially found to be related to radon emanation in the DRIFT detector \cite{Burgos:2007gv}.  An extensive effort to remove radon emanating materials was undertaken with the result that radon emanation no longer limited DRIFT \cite{Battat:2014oqa}.  It was then found that \isotope{Pb}{210} was coating the wires of both the central cathode and the MWPCs and creating RPRs not related to radon emanation.  Both the central cathode and MWPCs were etched in nitric acid, which eliminated these backgrounds.  DRIFT has since replaced the central cathode with a clean, texturized, 0.9\,$\mu$m-thick aluminized mylar film.  This has reduced the RPR background rate to $\sim$1/day \cite{Battat:2015rna}.  DRIFT has been able to tag and cut those events using $z$-fiducialization via minority carriers, producing a recent background-free limit \cite{Battat:2014van}.

\subsection{Future prospects for large-area MWPCs}

\subsubsection{Very large TPCs with MWPC Readout}
\label{sec:mwpcLAr}
The use of liquid argon detectors for the next-generation long-baseline neutrino experiments has led to substantial developments in MWPC technology and associated readout electronics.
For example, the DUNE collaboration is designing large anode plane arrays (APAs)  for the proposed 10\,kton liquid argon TPC modules to be installed underground at the Homestake site over the next decade \cite{Acciarri:2015uup}.  These APAs are 6\,m $\times$ 2.3\,m in size with three planes of MWPC readout.  Many hundreds of these APAs will be needed for the final assembly.  Each DUNE module has a proposed active volume of 12m $\times$ 14.5\,m $\times$ 58\,m.  As a point of reference, such a TPC if filled with gas at the operating pressure of DRIFT would hold about 2\,tons of target.  The DUNE TPCs are not designed for low background operation, and the MWPC readout has a wire spacing of 3\,mm.  Nevertheless this design could be modified for directional detection.

Perhaps the biggest challenge when scaling up the MWPC design is the electronics to read out the correspondingly large number of channels.  The synergy with the liquid argon neutrino detectors will help here as well.  The drift speed of electrons in liquid argon is closely matched to that of negative ions in a low-pressure gas, and so the electronics developed for DUNE and other related experiments can be used in directional Dark Matter detection.  In particular, Brookhaven National Laboratory (BNL) has developed custom ASIC chips for charge shaping and analog-to-digital conversion (ADC).  The BNL ASICs are already in use on several LAr experiments including MicroBooNE \cite{microbooneTDR}, LArIAT
\cite{2014arXiv1406.5560P}, and the 35T LBNE prototype detector \cite{ChenPersonal}, and will be used to readout 1.5 million channels for the full DUNE far detector.  The DRIFT collaboration has already successfully demonstrated the use of the charge shaping ASICs on a DRIFT detector \cite{ifftPrivate}, and work is underway to use the ADC ASICs as well \cite{battatPrivate}.

\subsubsection{Hybrid MWPCs}

In DRIFT, the MWPC provides both gas amplification and charge readout.  In doing so, the anode wire pitch is limited to 2\,mm because of inter-wire electrostatic forces.  It is interesting to consider alternative strategies that use wires only for charge readout, and a different device for gas amplification (\eg{} a gas electron multiplier -- GEM).  In doing so, the wires would not be at high-voltage, and the wire pitch could be reduced substantially (though not as fine a pitch as in Micro-Patterned Gas Detectors -- MPGDs).  In addition to the finer spatial resolution, such a hybrid scheme is worthwhile to explore because of the low capacitance per unit length of the wires, and the absence of radio-impure substrates in the active volume.  An added benefit of this approach is that a GEM (or a set of multiple GEMs) can potentially provide more gas amplification than a traditional MWPC because it produces a longer region of high-field in which to develop the avalanche.  Preliminary simulation work is underway to explore this GEM+wire possibility \cite{spoonerPersonalCommunication}.

\subsection{Conclusions}
MWPCs have played a central role in leading particle physics experiments for the last 40 years.  Their ease of construction and low radioactivity make them an interesting readout techology for directional Dark Matter detection.  Fast timing readout electronics provides spatial sampling equivalent to $50\,\mu$m in the \z{} direction, but the spatial granularity in the \x{} and \y{} directions is limited to $>1$\,mm.  Track reconstruction is therefore limited to 1D or 2D, though the recoil sense can be measured along the \z{} direction.  MWPCs are readily scalable to large volumes, and readout electronics with high channel density are under active development to handle channel counts in excess of $10^{6}$.  A synergy between the needs of directional Dark Matter searches in negative ion gases and liquid argon neutrino detectors is particularly interesting, and may provide enabling technologies to scale up directional Dark Matter TPCs with MWPC readout.

\section{Micro Pattern Gaseous Detectors (MPGDs)}
\label{sec:mpgdIntro}
In recent years, an increasing number of tracking detectors have utilized sophisticated amplifying structures made from printed circuit-like substrates. Such structures are generically referred to as micro-pattern gaseous detectors (MPGDs), and have a number of potential advantages over MWPCs, such as finer detector granularity, higher maximum rate capability, improved mechanical robustness, and greater ease of producing large detector planes. This recent trend is mainly driven by high energy physics experiments. The origin of MPGDs is attributed to Anton Oed, who in 1988 invented the microstrip gas chamber (MSGC) \cite{Oed:1988jh}, a plane of thin metallic strips imprinted on a plastic substrate, around which the amplification takes place. 
 
Although the MSGC was not widely adopted, due to its ageing effects and the appearance of destructive discharges (mainly due to the geometrical proximity of the plastic substrate to the amplifying avalanche), this was the start of a new, and soon very active, line of development. In order to overcome the limitations of the first MSGC, a number of alternative amplifying structures were developed, based on gap, hole, well, wire, or dot-based geometries, with different (sometimes complementary) degree of success \cite{Titov:2007fm}.  Although a number of MPGDs have been successfully employed in specific applications, two generic MPGD types are now widely used in particle physics experiments: the micro-mesh gas structure (Micromegas \cite{Giomataris:1995fq}), and the gas electron amplifier (GEM \cite{Sauli:1997qp}). They are considered the most successful MPGDs, and they also form the basis for more recent derived concepts  like the ThickGEM and RETGEM detectors (derived from the GEM) and the bulk and microbulk detectors (derived from the Micromegas). 

In addition, combining MPGD amplification structures with highly integrated readout electronics allows for gas-detector systems with channel densities comparable to those of modern silicon detectors. New types of MPGDs have been built using CMOS pixel ASICs directly below GEM or Micromegas amplification structures. In 2008, the RD51 collaboration at CERN was established to further advance the development of MPGDs and associated readout electronics for applications in basic and applied research \cite{rd51}. Also in recent years, more specific studies have investigated the suitability of MPGDs for rare event searches \cite{Micromegasrareevents}. Here, the focus is on more specific features, such as radiopurity, energy resolution,
tracking performance such as angular resolution and particle identification
of particular interest for the field.

The advantages offered by MPGDs are of interest for directional Dark Matter detection, and a number of MPGD development and prototyping initiatives are underway. The following sections review the status of MPGDs, including current results and future prospects. In Section~\ref{sec:micromegas}, the development of Micromegas readouts is described, including a number of technical R\&D results of relevance to low background applications as well as the specific experience of the MIMAC directional Dark Matter detection experiment.  In Section~\ref{sec:mupic}, results on the $\mu$-PIC readout used in the NEWAGE experiment are described. Finally, recent efforts on pixel ASIC readouts are described in Section~\ref{sec:pixelchip}.

\subsection{MPGD: Micromegas}
\label{sec:micromegas}
With the advent of MPGDs, the possibility to design and built precise
readout planes allowing the reconstruction of low energy (few keV)
recoil tracks of a few mm length is simpler and easier than with
conventional wire planes. In particular, Micromegas detectors have
shown these qualities for rare event
searches~\cite{Micromegasrareevents,cast1,Aune:2013nza,tpchp,T2Kdetect,mimac,Irastorza:2015dcb,Irastorza:2015geo}.
Micromegas are double-gap MPGD consisting of a metallic micromesh
suspended over a pixelized anode plane by insulating pillars. The
mesh-anode gap defines an amplification region (usually with width in
the range of 25-256\,$\rm{\mu}$m). The drifting electrons go through
the micromesh holes and trigger an avalanche inside the gap, inducing
detectable signals both in the anode pixels and in the mesh.

The CERN Axion Solar Telescope (CAST) experiment has been a pioneer in
the use of Micromegas detectors for rare event searches since
2002. The context is the search of solar axions converted into 1--10\,keV
photons in the magnetic field of the experiment.  In 2007, the
shielded TPC with a multi-wire proportional counter as a readout
structure covering two detector emplacements of the experiment was replaced by two
Micromegas detectors.  The background level achieved in the 2012 data
campaign showed a two-order-of-magnitude improvement with respect to
the Micromegas detectors used at the beginning of the
experiment \cite{Aune:2013nza}. This background level of $1.5\times10^{-6}$\,
keV$^{-1}$cm$^{-2}$s$^{-1}$ is also two orders of magnitude better than
the best level achieved with the CAST TPC \cite{tpc}. Since 2007, the
amplification structure used is a microbulk Micromegas detector made
of kapton and copper exhibiting very low levels of radioactivity per
unit surface area \cite{microbulkRadiopurity,Irastorza:2015dcb}. The pattern of the
Micromegas anode, an array of highly granular pixels interconnected in
the $x$ and $y$ directions, as well as the recording of the micromesh
pulse are features that are used, for instance, in the MIMAC
directional Dark Matter detector \cite{mimac}.

In the following, we review the suite of Micromegas technologies,
highlighting their advantages and drawbacks for directional Dark
Matter search (Section~\ref{sec:mmtech}) as well as the current R\&D
efforts relevant to directional Dark Matter search
(Section~\ref{sec:retd}).  We then present the first use of a
Micromegas-based detector for directional Dark Matter detection,
namely in the MIMAC experiment (Section~\ref{sec:mimac}).

%
%
\subsubsection{Micromegas technologies relevant to directional Dark Matter search}
\label{sec:mmtech}
Micromegas technologies can be classified according to the type of
mesh that is used and the way the insulator spacers are manufactured.
In the following, we review the various Micromegas technologies and
highlight their advantages and drawbacks for use in directional Dark
Matter searches with requirements outlined in
Section~\ref{sec:techchallenges}.  We note that, at present, the choice of
the Micromegas technology that is best suited to a directional Dark
Matter search may require some compromises.  For instance the
production of large area Micromegas with a good radiopurity has not
yet been achieved.

\paragraph{Standard Micromegas} 
In the first generation of Micromegas, the amplification gap spacers
were glued onto the anode plane \cite{Giomataris:1995fq}, and the micromesh was
suspended on top of the spacers. Besides, the spacers were implemented
on the anode plane using standard lithography techniques \cite{Charpak:2001tp}.
Electroformed micromeshes have also been produced with attached
pillars \cite{Abbon:2007ug}. A significant drawback of this approach is that the
parallelism between the anode and the mesh depended greatly on the
manufacturing of the mechanics of the detector, and on the expertise
of the user.  This technology has been used for the first Micromegas
applications in the early 2000s, such as neutron time-of-flight (nTOF) \cite{Pancin:2004vja},
CAST \cite{CASTMM} and COMPASS \cite{Kunne:2003qe}.  This standard
Micromegas has been more or less abandoned after the advent of bulk
and microbulk technologies in 2006--2007, as described in
Secs.~\ref{sec:bulkmm} and \ref{sec:microbulkmm}. However, there is a
renewed interest in standard Micromegas as they have been shown to be
the most favorable for the production of large size and large number
of detectors \cite{NSWTDR,Wotschack}. In particular, the upgrade of
the muon chambers of the ATLAS New Small Wheel \cite{NSWTDR} within
the framework of the luminosity upgrade of the LHC in 2019, requires a
surface of 1200\,m$^2$, with more than 2\, million electronic
channels. This scalability is a key advantage of standard Micromegas
for use in a large-scale directional Dark Matter search.

\paragraph{Bulk Micromegas}
\label{sec:bulkmm}
In the bulk technology \cite{bulk}, the amplification structure is
manufactured as single entity, instead of as two separate components
(mesh and anode plane).  This enables the production of large, robust
and inexpensive modules.  In bulk Micromegas a commercial woven mesh
is encapsulated in insulating pillars by a standard printed circuit
board (PCB) process. The main steps in the fabrication process consist
of placing a stretched mesh between insulating layers (often two on
the bottom and one on the top) directly on a readout plane. A mask
with the pattern of the pillars is placed on top of this stack and is
illuminated by ultraviolet light. A chemical bath then removes the
non-illuminated regions.  The final structure is a readout plane with
an encapsulated mesh between insulating pillars. The bulk Micromegas
is very robust and easy to manufacture. This type of detectors has
been produced by CEA (IRFU/SEDI) and CERN. As the raw material of the
insulator is 64\,$\mu$m thick, the amplification gap of a bulk
Micromegas detector is a multiple of this value.  Amplification gaps
of 64, 192, 256, 512\,$\mu$m have already been manufactured with a
35\,$\mu$m thick woven mesh.  Bulk Micromegas are used for different
applications, \eg for the largest MPGD TPC (9\,m$^2$) in the
T2K \cite{T2K} experiment, and the MINOS TPC \cite{Minos} tested
successfully in 2013.  The CLAS12 \cite{CLAS12Mod,Charles:2013cvy}
experiment is also finalising the design and construction of bulk
Micromegas detectors to operate in the near future in a magnetic
field.

A key advantage of the bulk technology for directional Dark Matter
detection is the possibility to produce Micromegas with a broad range
of well-controlled amplification gaps. Indeed, larger gaps are
preferred at the low gas pressures used in directional searches.  For
example, the MIMAC bi-modules operated in the underground laboratory
of Modane \cite{Riffard:2013psa}, are using bulk detectors with a
256\,$\mu$m amplification gap, allowing operation at 50 mbar. An
effort is underway to select radiopure construction materials for the
bulk Micromegas, in particular the micromesh and the PCB readout plane.

\paragraph{Microbulk Micromegas}
\label{sec:microbulkmm}
As with the bulk technology, the final structure for the microbulk
technology is a single entity containing the micromesh and the anode
plane.  However, the main constituents are different: the raw material
is a thin, flexible kapton foil with a 5\,$\mu$m copper layer on each side.  The possible amplification gaps are 50, 25 and 12.5\,$\mu$m.  The
manufacturing process is based on lithography \cite{Andriamonje:2010zz}.

Due to their thin mesh and amplification gap homogeneity, the energy
resolution demonstrated with microbulk Micromegas is extremely good
for a gaseous detector, $\sim$11\% (FWHM) at 5.9\,keV in an Argon +
5\% Isobutane gas mixture \cite{Iguaz:2012ur}, \ie{} close to the theoretical
statistic-limited value of 10.8\%.  Microbulk detectors are known to
be less robust to sparks than bulk detectors. However, they have shown
remarkable stability in the case of the CAST experiment for long data
taking periods, and in the nTOF experiment as a neutron beam
profiler \cite{ntof1,ntof3}. These readouts are also used in the non-directional Dark Matter search TREX-DM  \cite{Iguaz:2015myh}. Microbulk detectors are currently
being developed in the context of double beta decay searches, for which good energy resolution and operation at high pressure in Xe mixtures is required. 
Energy resolutions of 7.3\% (9.6\%) FWHM in 1(10) bar at 22\,keV  in Xe and trimethilamine has been demonstrated \cite{MicrobulkTMA,NEXTHector,NEXTDiego}.

Although energy resolution is not a central requirement of directional Dark Matter detection \cite{Billard:2011zj} 
it is connected with important aspects like homogeneity and stability of response, as well as the quality of the 
topological information of the readout. The main advantage of the microbulk option remain its excellent intrinsic radiopurity \cite{microbulkRadiopurity,Irastorza:2015dcb},
 as it is being object of active R\&D (see Section~\ref{sec:trex}).

\paragraph{Ingrid Micromegas}
Gridpix is a detector integrating a Micromegas grid with a pixel
readout chip as the signal collecting anode manufactured by
microelectronics techniques \cite{PhDMax}. This technique has been
developed by the MESA industry and the University of Twente. The
precision of the fabrication Micromegas grid (ingrid) on silicon
wafers (holes controlled to a 1\,$\mu$m precision and gaps varying
less than 1\%) results in an excellent resolution $\sim$11\% (FWHM) at
5.9\,keV in an Ar+10\% Isobutane mixtures with high enough gains to
detect single primary electrons.  An example of a first implementation of this type of detector in a low background experiment is the Ingrid 
CAST detector \cite{Krieger:2014pea} where the detector has been operated successfully exhibiting a very low energy threshold (300 eV).

Despite its low threshold and excellent energy resolution, this
technology does not seem very appropriate for directional Dark Matter
detection as only small surfaces (largest 12\,cm$^2$ with mosaic techniques \cite{Lupberger:2013lfa}) can be manufactured at present.

\paragraph{Resistive bulk Micromegas}
In order to decrease the total number of electronics channels, it has
been proposed to increase the charge dispersion by means of a
resistive coating. This allows to keep a good spatial resolution while
reducing the total cost of the detector \cite{Dixit}.  The first
Micromegas with a resistive coating have been developed in the context
of the R\&D for the ILC-TPC. With this strategy, the sparks limit is
surpassed and sparks are aborted. Hence, resistive Micromegas
detectors constitute an interesting solution for high flux
environments where the consequences of sparking need to be reduced. In
the last years an active program of R\&D has been developed for
resistive strip readout in particular within the framework of of the
upgrade of the muons detectors of the New Small Wheel (NSW) for the
High Luminosity LHC (HL-HLC) \cite{alexpoulos-kobe}.  Resistive
Micromegas detectors will also be used for the forward tracker of the
CLAS12 tracker \cite{CLAS12Mod}.

The reduction of the total number of channels that can be obtained
with resistive coatings is an appealing feature in the context of
large TPC for direction Dark Matter detection.  However, the
radiopurity of the coatings needs to be studied. Moreover, dedicated
studies should be carried out in order to optimise the resistivity
value of the resistive film so that the directional information is not
lost.

%
\subsubsection{Current R\&D efforts on Micromegas technologies}
\label{sec:retd}
There is currently a vigorous R\&D program to refine and expand the
Micromegas technologies.  In this section, we highlight the efforts
that pertain to directional Dark Matter searches. In particular, we
focus on strategies to reduce the required number of electronic
readout channels to enable a cost-effective scaling of the Micromegas
technology to large area readouts, as well as techniques to localize
charge to facilitate 2D/3D track reconstruction.

\paragraph{Genetic Multiplexing} 
This is an innovative technique to reduce the number of electronic
channels.  It assumes that a signal is deposited on at least 2
neighbouring strips. Spatial resolution around 100 microns can be
achieved with meter size detectors equipped with only 64
channels. With this technique, the degree of multiplexing can be
easily adjusted to the incident flux of particles to solve the
ambiguities. The technique has been tested on a large,
$50\times50$\,cm$^2$ Micromegas prototype equipped with 1024 strips
and read with only 61 channels \cite{genetic}.

\paragraph{Segmented Mesh Microbulk}
The localization of charge is a key issue for directional Dark Matter
detection as it is related to the 3D reconstruction of the recoiling
nucleus \cite{Billard:2012bk}, providing the electron drift velocity in
the gas mixture is known \cite{Billard:2013cxa}.

To cover medium or large surfaces with a 2D pattern, the strategy used
up to now is an $x$-$y$ structure out of electrically connected pads in
the diagonal direction through metallised holes. This readout strategy
reduces the number of channels with a fine granularity covering a
large anode. In CAST, these anode signals are combined and are
correlated to the integrated signal of the mesh in the discrimination
algorithms. The mesh signal can also be used as a trigger signal. This
strategy has been very successful and has allowed impressive
background rejections. One of the issues of this structure is that, in
the theoretical limit of no gas diffusion, the charge would only be
collected in one pixel and the 2D capabilities would be lost. Even if
in practice this does not happen, the charge collection in the two
directions is not completely equivalent. Another major disadvantage is
the complexity of manufacturing this multilayer stack. Moreover,
when coupled to the microbulk technology, this readout structure
complicates the manufacturing of the detector and decreases the
production yield.  A new concept has been developed to manufacture a Microbulk Micromegas detector  with 1D strips in the readout plane and a stripped micromesh in the orthogonal direction. The so-called ``segmented mesh Microbulk'' provides two opposite fully correlated signals induced both on the anode and on the mesh strips giving intrinsically spatial information in two directions. Furthermore this structure coupled to autotrigger electronics provides a very low threshold detector compared for instance to the CAST detectors, where the trigger is provided by the whole mesh taking into account the total detector capacitance.  The manufacturing process is similar to that of the Microbulk but simpler. First detectors have been manufactured and  operated successfully as  neutron beam profilers in the context of the nTOF Experiment.  

\paragraph{Piggyback resistive Micromegas}
In most Micromegas applications the design of the detector vessel and
the readout plane are completely linked. A way of decoupling these
two components would be by separating the amplification structure and
detector volume from the readout plane and electronics.  This is
achieved with the so-called ``Piggyback'' Micromegas
detectors \cite{Piggyback-2013}. The signal is then transmitted by
capacitive coupling to the readout pads. This opens up new
possibilities of application in terms of adaptability to new
electronics. In particular, Piggyback resistive Micromegas can be
easily coupled to modern pixel array electronic ASICs.  The novelty is
the way in which the resistive layer is deposited on an insulator
substrate instead of being directly deposited on the anode plane. The
insulator is then posed on the readout plane. The Micromegas detector
operates as usual in the proportional avalanche mode inducing signals
on the resistive anode plane. The structure needs to be optimised in
such a way that the electronic signal is not lost through the
resistive layer but is propagated to the readout plane \ie{} the
capacitance of the insulator needs to be much larger than the
capacitance of the amplification gap. Materials with large dielectric
constants are favoured ($\gg$10). The first experimental tests have
been performed with a bulk Micromegas with an amplification gap, $
t_\mathrm{amp}$, of $128\,\mu$m and a ceramic layer with
$t_\mathrm{insu}$ of 300\,$\mu$m. For the resistive layer, ruthenium
oxide (RuO$_2$) has been chosen for its robustness, stability and wide
range of resistivity values available.
This readout structure is being explored for \xray{} Polarimetry and the system needs 
to be optimised to perform spectro-polarimetry using different gas mixtures at low 
pressures \cite{Attie:2015uoa,Serrano:2016pfh}. The interesting feature is the isolation 
of the amplification gap from the readout plane, allowing to reduce the cost of the electronics 
and probably facilitating the installation of the readout electronics outside of the chamber. 
However, the radiopurity of the structure needs to be measured and
will have to be optimised for rare event detection experiments.
 
\subsubsection{Micromegas for low background applications }
\label{sec:trex}
 
An active R\&D program on Micromegas for low background applications is ongoing, namely the T-REX project \cite{trexwebpage}. This development is mostly focused on microbulk Micromegas (see Section~\ref{sec:microbulkmm}), due to their good intrinsic radiopurity. Indeed, specific measurements of both raw material and fully processed microbulk readouts with high purity Ge detectors \cite{microbulkRadiopurity,Irastorza:2015dcb} have demonstrated extremely low levels of radiopurity, below 0.1\,$\mu$Bq/cm$^2$ corresponding to a negligible component in the radioactive budget.
Microbulks are limited in size by their current fabrication technique and this may be a drawback in their scalability. One way to surpass this is to design efficient  tiling or ``mosaic'' strategies to reach large surface of readouts with relatively small single elements. Some ideas have already been developed allowing this in an almost dead-zone-less way, and with efficient extraction of a high number of channels. Currently the largest surface of microbulk in operation is the NEXT-MM prototype built at University of Zaragoza, with a circular readout of 30 cm diameter, composed of 4 circular sectors planes (themselves the largest single microbulk made so far) \cite{NEXTHector,NEXTDiego}. In parallel, work is ongoing to build radiopure bulk readouts, as an alternative strategy to reach large surfaces. Work is also ongoing to understand radiopurity limitations in components other than the readouts typically associated with Micromegas setups \cite{Aznar:2013jwa}, as well as into shielding strategies, understanding of the microphysics of the avalanche and the optimal treatment of the topological information of the pixelised readout towards background rejection.  The most recent series of highly-granular microbulk detectors in CAST have benefited from this R\&D and have achieved background levels down to $\sim10^{-6}$\, keV$^{-1}$cm$^{-2}$s$^{-1}$  at surface ($\sim10^{-7}$\,keV$^{-1}$cm$^{-2}$s$^{-1}$  in a special underground setup \cite{Aune:2013pna,Aune:2013nza}) a factor more than 100 (1000) lower than the first generation CAST Micromegas detector. Many of the elements in these detectors are directly translatable to an application to Dark Matter directionality. A larger, highly radiopure, Micromegas TPC prototype called TREX-DM is being commissioned at the moment to test these concepts, and to measure background rates and energy threshold.  This study will inform the design of a more massive detector, with potential applications in a non-directional search for low-mass WIMPs \cite{Iguaz:2015myh}.
 
%
%
\subsubsection{Micromegas in the MIMAC experiment}
\label{sec:mimac}
The MIMAC collaboration \cite{mimac} is currently building a large TPC devoted to Dark Matter directional detection for which 
three dimensional track reconstruction is recognized as a major issue \cite{Billard:2012bk}.
The following detection strategy has been chosen.
The nuclear recoil produced by a WIMP in the TPC produces electron-ion pairs in the conversion gap of the Micromegas detector. 
The electrons drift towards the amplication gap 
(128$\,\mu\mathrm{m}$ or 256$\,\mu\mathrm{m}$ in this case) where they produce an avalanche that 
induce signals in the  pixelized \xy{} anode and in the mesh. 
The third dimension \z{} of the recoil is reconstructed by a dedicated  self-triggered electronics specifically designed 
for this project \cite{mimacelec,Richer:2011pe,Bourrion:2011vk,Bourrion:2011it} that is able to perform a full anode 
sampling at a frequency of 50\,MHz. 
The concept has been verified by the construction of  a  10 $\times$ 10\,cm$^2$ detector to validate the 
feasability of a large TPC for directional detection with a realistic size prototype.

\paragraph{Design of the MIMAC readout}
\label{sec:mimac.readout}
The design of the bulk Micromegas was guided by 
the requirements on the granularity of the anode as well as by the operation conditions (various gas mixtures and various pressures). Simulation studies showed that the granularity of the readout plane needed strips of 200\,$\mu\mathrm{m}$ size.  The design of the bulk Micromegas end cap  takes into account these requirements. Moreover the end cap was designed to be versatile as the detector was 
first to be validated by the T2K electronics \cite{T2Kelec1,T2Kelec2} before the final conclusive test with the specifically designed MIMAC electronics \cite{mimacelec,Richer:2011pe,Bourrion:2011vk,Bourrion:2011it}.

\begin{figure}
\centering%
\begin{tabular}{cc}
\includegraphics[width=0.5\textwidth]{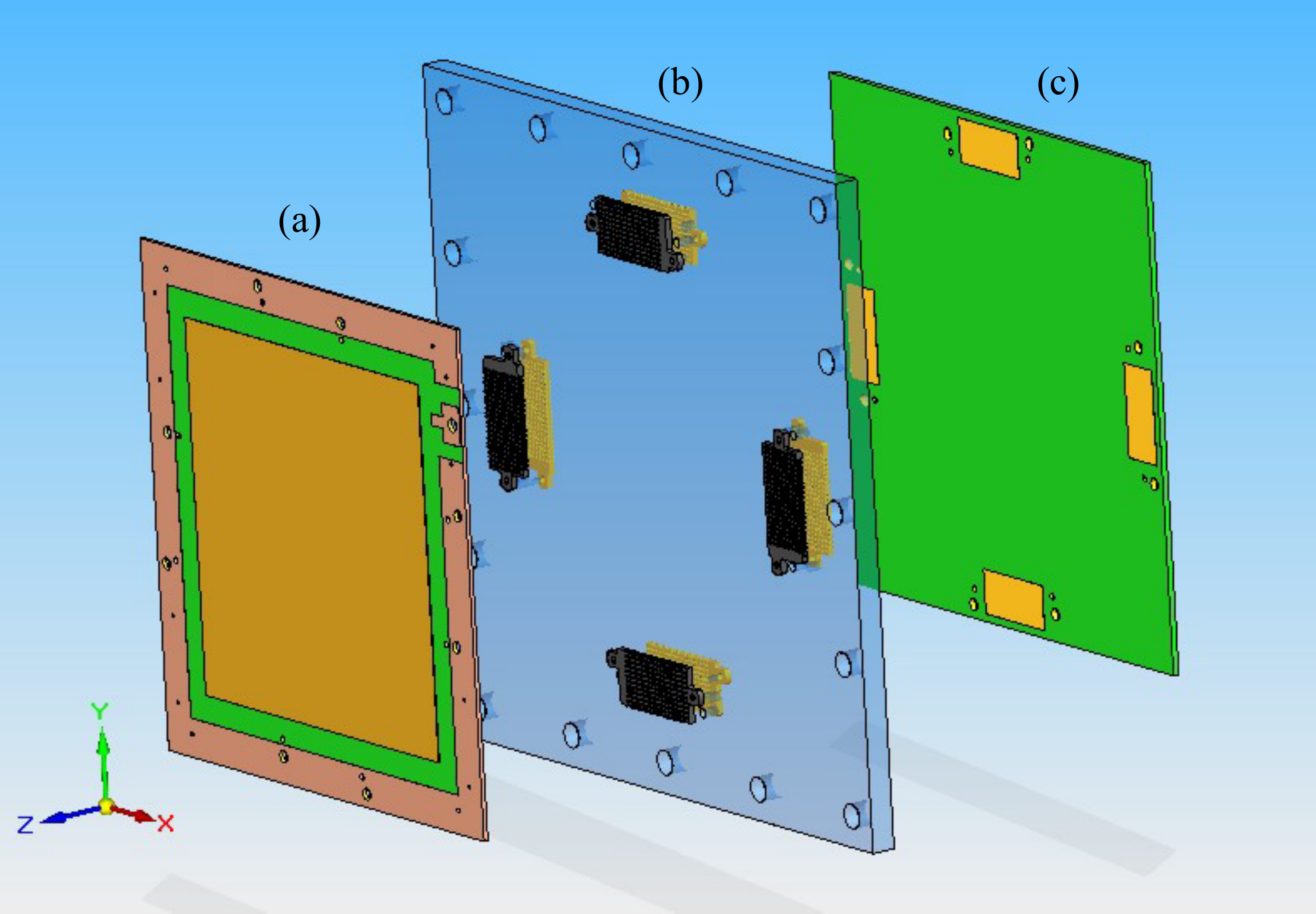}
\end{tabular}
\caption{Zoom of the Readout PCB (\emph{a} in the image), Leak-tight PCB (\emph{b}) and the Interface PCB (\emph{c}). }
\label{fig:assembly}
\end{figure}

\begin{figure}
\centering%
\begin{tabular}{cc}
\includegraphics[width=0.45\textwidth]{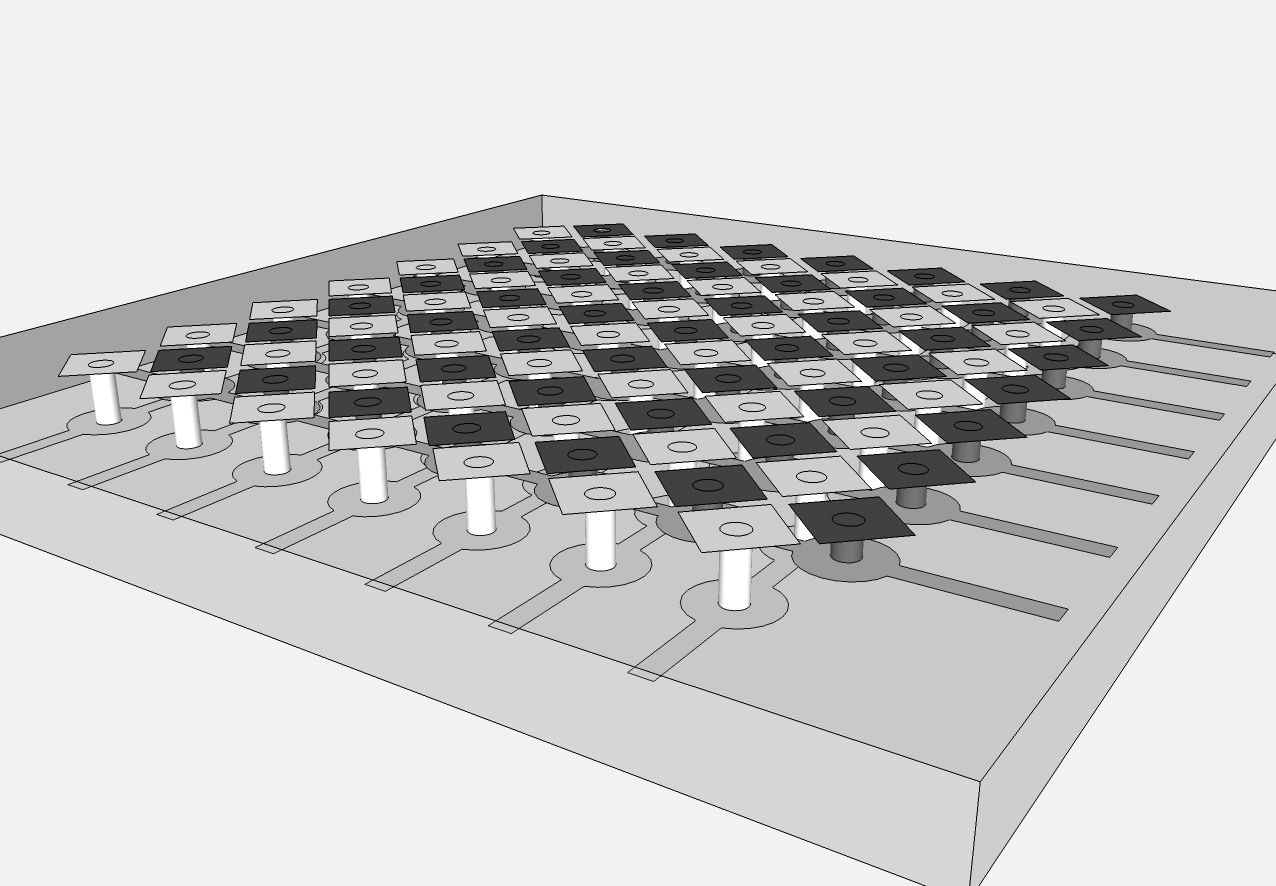} &
\includegraphics[width=0.42\textwidth]{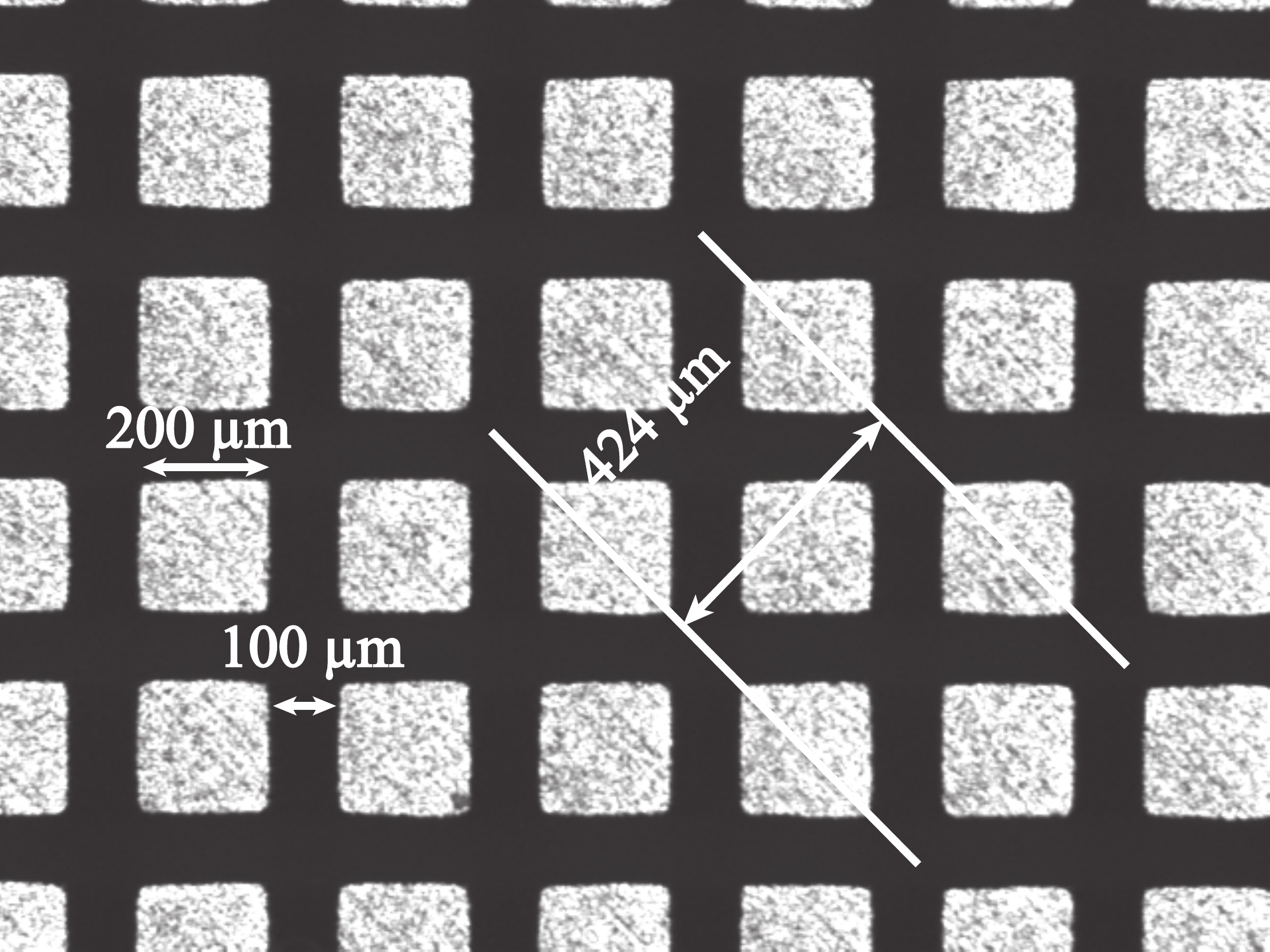}
\end{tabular}
\caption{Left: Sketch of the 2D readout used. Right: Microscope photograph of the 2D readout.}
\label{fig:2d}
\end{figure}

The detector and the characterisation tests have been described in detail elsewhere \cite{MIMAC-detector}. Here we will only describe the 
main features. The system consists of 
a   leak-tight readout plane on a 2\,cm aluminium cap.  A general sketch of the mechanical assembly is given in 
Figure~\ref{fig:assembly}. The bulk Micromegas is fabricated on a Printed Circuit Board (PCB), called \emph{Readout PCB}, of 1.6\,mm thickness (\emph{a} in Figure~\ref{fig:assembly}).  The signal connections from one board to another are done by means of  connectors  that are placed and screwed between the two boards (\emph{b} in Figure~\ref{fig:assembly}). On the outside of the vessel an \emph{Interface card} distributes the signals to the desired electronics (\emph{c} in Figure~\ref{fig:assembly}). 
The active surface is of 10.8 $\times$ 10.8\,cm$^2$ with 256 strips per direction. The charge collection strips 
make-up an \xy{} structure out of electrically connected pads in the diagonal direction through metallized 
holes as can be seen in Figure~\ref{fig:2d} (left). This readout strategy  reduces the number of channels 
with a fine granularity covering a large anode surface. The pads are 200\,$\mu\mathrm{m}$ large with an isolation 
of  100~$\mu\mathrm{m}$ resulting into a strip pitch of 424~$\mu\mathrm{m}$. The surface quality of the 
readout plane can be observed in Figure~\ref{fig:2d} (right). The 100\,$\mu\mathrm{m}$ diameter metallized holes 
have been fully fillled, hence yielding a completely uniform surface.  
This fact is a prerequisite to  obtain a uniform 
performance of a bulk Micromegas detector. To enable the use of two different electronic system, two versions of the interface PCB to the electronics were 
fabricated.  This design offers several advantages: 
a simple, compact and leak-tight way for the signal connections and a versatility for two different types of electronics. 

Bulk Micromegas with two different amplification gaps (128\, and 256 $\mu$m) were produced in order to choose the best gap for different
running pressure conditions.  The detectors were tested in a dedicated vessel in an  Argon (95\%)-Isobutane (5\%) mixture at atmospheric pressure with a
  $^{55}$Fe source (5.9\,keV X-rays).  The tested detectors exhibit typical bulk Micromegas behaviour in terms of gain and resolutions:  gains greater  than $2 \times 10^4$ before sparking for both amplification gaps, and energy resolutions between 16\% and 21\% (FWHM) at 5.9\,keV.

\paragraph{Micromegas in the MIMAC operation conditions}
The current MIMAC prototype is a dual-TPC, composed of two TPC sharing a common cathode. 
Each TPC is 25\,cm long and equipped with the pixelized micromegas detector ($10 \times 10 \ {\rm cm^2}$). 
Each strip of pixels is monitored by a current preamplifier and the fired pixel coordinate is obtained by using the
coincidence between the \x{} and \y{} strips, \ie{} with a signal on the grid greater than a given threshold, 
are stored. The end of the track event is defined by a series of 8 empty samples. 
Each strip of pixels has its own threshold determined 
by an autocalibration algorithm. The criterium used to define the thresholds is the minimal 
value with no electronic noise signal within  15 seconds. In order to reconstruct the third dimension of the recoil, the \z{} coordinate (along the drift axis), 
 a self-triggered electronics has been developed \cite{Richer:2011pe,Bourrion:2011vk,Bourrion:2011it}.
 It allows anode plane sampling at a frequency of 50\,MHz. 
 Providing the electron drift velocity is known, either by simulation \cite{magboltz} or by measurement \cite{Billard:2013cxa}, the track
 may then be 3D reconstructed \cite{Billard:2012bk}. The ionization energy is digitized at 50\,MHz 
 by the flash ADC placed after the charge preamplifier used to read the grid.
 
The stainless steel chamber is equipped with field rings to ensure a good
homogeneity of the drift field. The leak-tight interface shown in Figure~\ref{fig:assembly} provides a connection with the electronic board located outside the chamber.
 To ensure a good charge collection, the gas is permanently recirculated through a gas control system including a 1-bar buffer and 
 a   filter  to remove H$_2$O and O$_2$ impurities.
 
\begin{figure}
\centering
\includegraphics[width=0.7\textwidth]{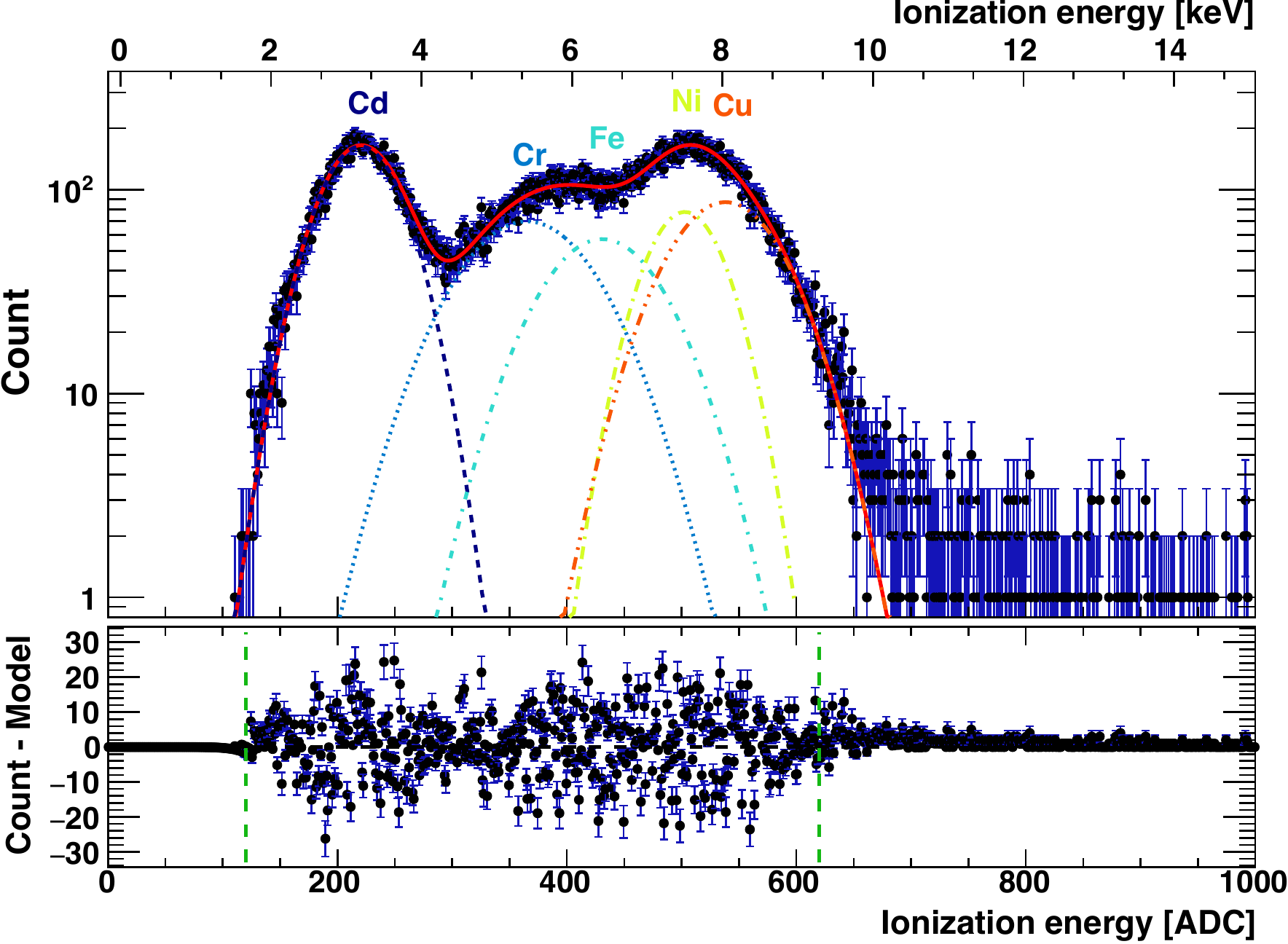}
\caption{Left: Calibration spectrum and best fit obtained from the minimization of the likelihood function. The lower plot presents the residuals between data and model. Figure from~\cite{riffardThesis}.}
\label{fig:calib}
\end{figure}

The detector is calibrated using electronic recoils induced by X-rays from the fluorescence of copper, iron and cadmium foils, 
produced with an \xray{} tube. Note that the orientation of the \xray{} tube is chosen in such a way 
that primary X-rays cannot enter the detector volume. 
Figure \ref{fig:calib} presents a typical calibration spectrum measured in the dual-TPC during underground operation. The spectrum features 
 X-rays from the fluorescence of the foils, as well as of the stainless steel chamber (Co, Fe).
 These X-rays are used for
 the energy calibration of the MIMAC detector up to 10\,\kevee, thanks to the minimization of a likelihood function~\cite{riffardThesis}.

The gas mixture chosen for Dark Matter search with MIMAC is CF$_4$ (70\%), CHF$_3$ (28\%) and C$_4$H$_{10}$ (2\%).
CF$_4$ is usually considered as the standard gas for directional detection \cite{cygnus2011} due to its electron 
transport properties \cite{Caldwell:2009si}, and the sensitivity to spin dependent interaction \cite{AlbornozVasquez:2012px} thanks to 
the non-vanishing spin of $\rm ^{19}$F. CHF$_3$ is added  to lower the electron drift velocity 
while keeping almost the same fluorine content of the gas mixture \cite{Billard:2013cxa}. 
In the case of the MIMAC detector, a fraction of 30\% was found to be  
adequate as it allows to significantly enhance the 3D track reconstruction while
conserving sufficiently dense primary electron clouds in order to keep a high nuclear recoil track detection efficiency.
Eventually, C$_4$H$_{10}$ is used as a quencher to improve the avalanche process in the Micromegas gap.  
Since Summer 2012, the detector has sucessfully operated underground at the Laboratoire Souterrain de Modane \cite{Riffard:2015rga}.

The next step of the MIMAC project is a 1\,m$^3$ detector that will be a demonstrator for the  
large TPC devoted to Dark Matter directional search. The design is a matrix of dual-TPC chambers. 
Upgrades are required in order to fulfill technical specifications required for
such a detector. In particular, it must include a larger Micromegas detector ($35\times 35 \ {\rm cm}^2$), together with  
dedicated electronics (1960 channels), synchronized by a global clock in order to veto multiple-chamber events.  

\subsubsection{Conclusions} 
Micromegas detectors are a mature technological choice for rare event detection. 
The strengths of the Micromegas technology like energy resolution, homogeneity of the 
gain, stability of operation, good spatial resolution and radiopurity combined with 
ultra-low background techniques turn out in very competitive options for a directional Dark Matter search. 
Microbulk detectors are being used in the context of axion solar search  and of double beta  decay detector 
R\&D achieving impressive background levels at surface and energy resolution respectively. A dual-TPC prototype 
has been built and operated at low pressure  by the MIMAC Collaboration   in the Laboratoire Souterrain de Modane. The next steps will be to 
study and improve the achieved background levels  by improving the radiopurity of the detector and 
vessel construction materials while  scaling the detector to 1\,m$^3$.

\subsection{MPGD: $\mu$-PIC}
\label{sec:mupic}
The micro-pixel chamber ($\mu$-PIC), a type of MPGD, was first developed in 2000 by Tanimori and Ochi \cite{ref:3uPIC}.
The most unique feature of the {$\mu$-PIC} compared to other types of MPGDs is that the detector is a monolithic board.
This offers an advantage in developing large-scale detectors, which is essential for rare event search experiments. 
Another feature of the $\mu$-PIC is that the gas amplification structure (typically 50\,$\mu$m-diameter anode electrodes) also constitutes a
two-dimensional micro-pattern (typically 400\,$\mu$m pitch) readout. This simplifies the detector system, which could be another 
advantage in developing low-background detectors.

\begin{figure}
\includegraphics[width=.4\linewidth]{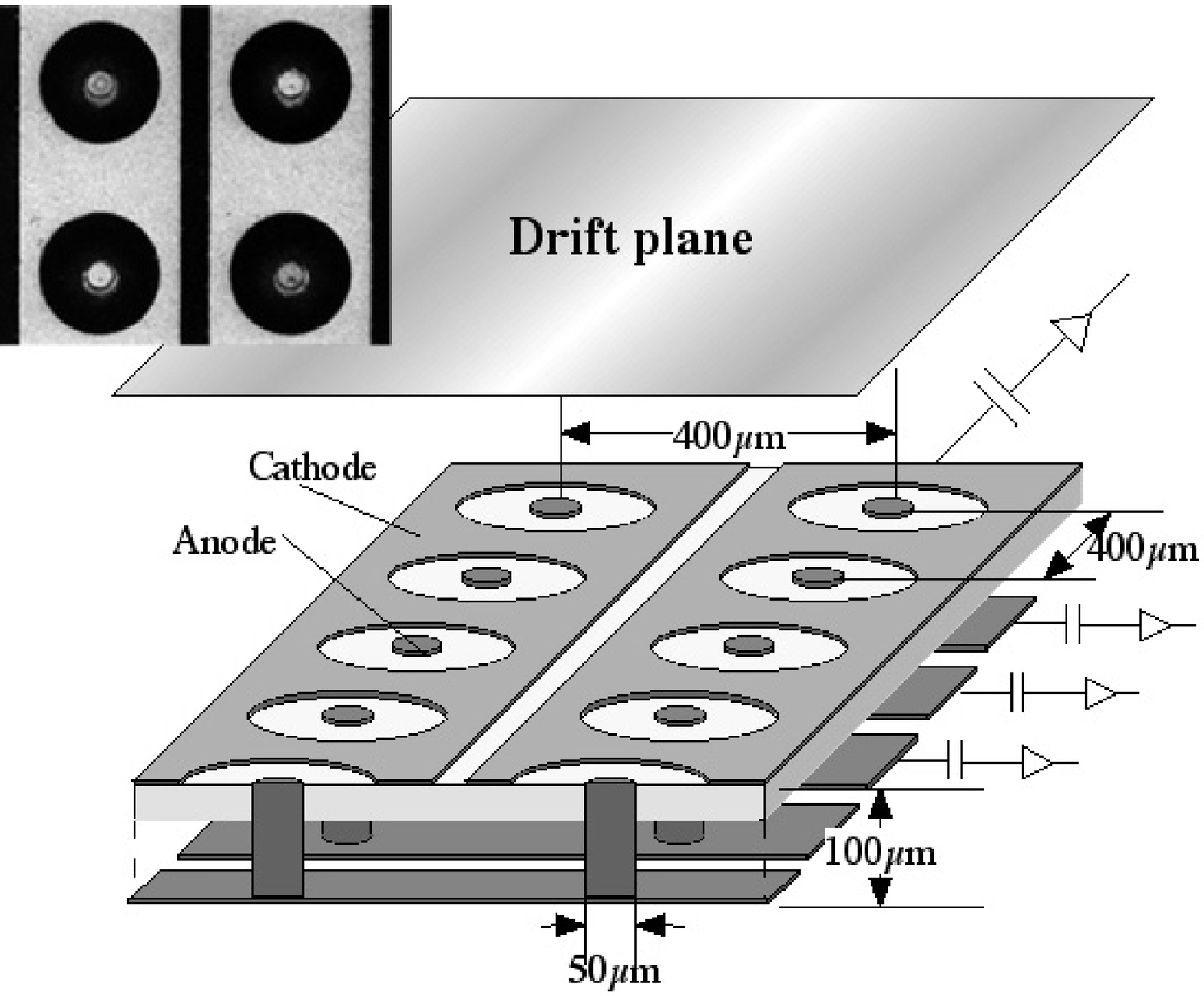}
\includegraphics[width=.4\linewidth]{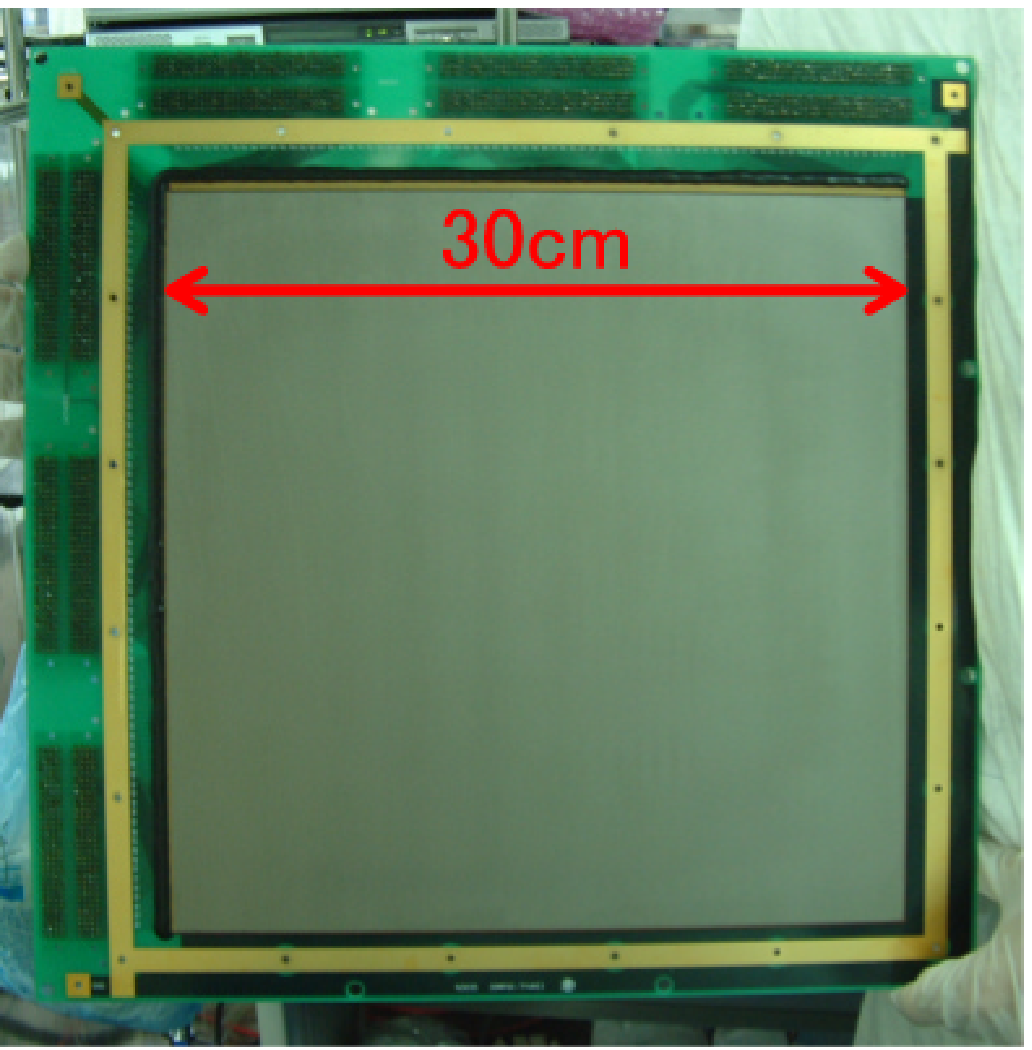}
\caption{\label{fig:uPICstruct}
Schematic (left) and photograph (right) of a $\mu$-PIC.}
\end{figure}

A schematic of the $\mu$-PIC is shown in the left panel of Figure~\ref{fig:uPICstruct}. 
Cathode electrodes are formed on the detection side of a substrate with a pitch of 400\,$\mu$m. The polyimide film substrate is 
100\,$\mu$m thick. The cathode strips have circular openings with 260\,$\mu$m diameter and 400\,$\mu$m pitch. Anode electrodes with diameters of 50\,$\mu$m are formed at the center of each opening.
Both the anode and cathode electrodes are made of copper.
The gas 
amplification occurs in the strong electric field near the anode electrode. The pixel-shaped electrodes 
were implemented so that the detectors do not suffer destructive discharges, which had been a problem 
in previous MSGCs \cite{ref:MSGC_tanimori}.
Anode electrodes are connected through the substrate to the anode readout strips on the rear side. 
The anode strips run orthogonal to cathode strips so that the two-dimensional position of an event can be determined from the 
coincidence of the anode and cathode-strip signals.

\subsubsection{Development of the $\mu$-PIC and related technologies}
The development of the $\mu$-PIC started in the year 2000 
with a small prototype that had a detection area of $3\times 3$~$ \rm cm^2$ \cite{ref:3uPIC}.
After this proof-of-concept, development of a practical-sized $\mu$-PIC with a detection area of $10\times 10$~$ \rm cm^2$ 
started in 2002 \cite{ref:10uPIC}. This practical-sized $\mu$-PIC is the ``standard'' $\mu$-PIC and many units have been produced since its inception.  
The quality of the standard $\mu$-PIC was improved in subsequent years, and by 2004,  
the performance satisfied the requirements for several intended applications \cite{ref:uPIC_neutronNIM2014,ref:uPIC_imaging2003Nagayoshi,ref:uPIC_imaging2003Ueno,ref:uPIC_imaging2003Orito}. Typical performance characteristics of a standard $\mu$-PIC include a position resolution of 120\,$\mu$m, a maximum gas gain of 15000, a gain uniformity of 5\% RMS,  an energy resolution of 23\% FWHM at 5.9\,keV, and stable operation for more than 1000 hours at a gain of $\sim$6000 \cite{ref:uPIC_NagayoshiThesis}.
Fabrication of large ($31 \times 31$\,cm$^2$) $\mu$-PICs 
commenced in 2004 \cite{ref:30uPIC}. A photograph of a large-sized $\mu$-PIC is shown in the right panel of Figure~\ref{fig:uPICstruct}.
The fabrication of the anode electrodes for large $\mu$-PICs has not yet been optimized. Therefore, at present, the gas gain during operation is restricted to 5000.  This is to be compared with the typical gain of 7000 for standard $\mu$-PICs.
$\mu$-PICs are currently being used in such fields as MeV gamma-ray astronomy \cite{ref:uPIC_MeVBalloon}, medical imaging \cite{ref:uPIC_MeVmedical}, neutron imaging \cite{ref:uPIC_neutron2013}, small-angle X-ray scattering \cite{ref:uPIC_SAX}, gas photo-electron-multipliers \cite{ref:uPIC_PMT}, space dosimetry \cite{ref:uPIC_PS-TEPC}, and directional Dark Matter searches \cite{ref:NEWAGE_PTEP2015}.

A dedicated readout system for the $\mu$-PIC detector has also been developed. An initial version was based on the CAMAC system \cite{ref:MSGC_tanimori}. Subsequently a more compact, FPGA-based, system was developed \cite{ref:uPIC_KuboPSD}; a revised version of which is now widely used for $\mu$-PIC readout. This system consists of an amplifier-shaper-discriminator chip (SONY CXA3183Q or CXA3653Q), which discriminates the signal from each strip, and FPGAs which encode information on the hit strips at a frequency of 100\,MHz.   
The original FPGA firmware was designed to use the coincidence of anode ($x$) and cathode ($y$) signals to realize a pipe-line process of high event-rate signals. 
The firmware and the data acquisition system have since been tailored for individual applications. The latest FPGA firmware outputs the time-over-threshold of each strip, which improves the spatial resolution and tracking performance in some applications \cite{ref:uPIC_neutron2013}.

Simulation studies with the finite-element method (FEM) have been indispensable 
for understanding the gas avalanche in the $\mu$-PIC and optimizing the detector structures.
Early-stage studies published in 2004 \cite{ref:uPIC_Nagayoshi_sim,ref:uPIC_Oleg_sim} were performed with 
Garfield \cite{ref:Garfield} and Maxwell-3D \cite{ref:Maxwell}. In these studies, experimental results were only partially explained 
by the simulation, but suggestions for further detector optimization
were obtained. The latest studies use Gmsh \cite{ref:Gmsh}, Elmer \cite{ref:Elmer}, and Garfield++ \cite{ref:Garfield++}.
There is now a good agreement between the simulated avalanche size and measured signals from the electronics. 
As a result, the detector simulation is now used to design the entire system \cite{ref:uPIC_Takada_sim}.

\subsubsection{$\mu$-PIC technologies for directional Dark Matter searches}
The application of the $\mu$-PIC detector and its data-acquisition system to 
directional Dark Matter detection began in the early 2000s, and the first proposal was  
published in 2004 \cite{ref:NEWAGE_PLB2004}.
The idea was to utilize a state-of-the-art $\mu$-PIC-based TPC 
for directional Dark Matter searches. Thus, a
proof-of-concept was important in the early stages.
Demonstrations of the three-dimensional detection of nuclear tracks
and of gamma-ray discrimination were given top priority among many R$\&$D items considered, 
as described in Ref.\,\cite{ref:NEWAGE_PLB2004}.
One technical issue particular to directional Dark Matter searches is the chamber gas. 
In most $\mu$-PIC applications, argon-based gas at atmospheric or greater pressure is used. 
However, a directional Dark Matter search experiment needs to use  
gas at lower pressure so that the recoiling nuclei travel a distance 
comparable to the detector pitch. 
The optimum pressures for a $\mu$-PIC chamber with $\rm CF_4$ gas were found to be 
20 and 30 Torr, for a detector with and without the ability to detect track sense, respectively \cite{ref:NEWAGE_PLB2004,Mayet2016}.  
Detector operation with a low-pressure gas is generally more difficult than at atmospheric pressure,
because the chamber tends to discharge more frequently.
As a first step, $\rm CF_4$ at 152\,Torr was used.
A Dark Matter search in a surface laboratory was performed 
to set direction-sensitive Dark Matter limits derived from the 
``sky-map'' of nuclear tracks \cite{ref:NEWAGE_PLB2007}. This measurement was performed to 
demonstrate that a three-dimensional gaseous tracking detector can set Dark Matter limits, 
although the specific limit obtained was much weaker than those set by non-directional detectors due to the limited exposure and the high number of background events.
After these new concepts had been demonstrated, the NEWAGE collaboration 
installed a $\mu$-PIC based TPC in an underground laboratory \cite{ref:NEWAGE_PLB2010}.
With the proof-of-concept accomplished, the development of $\mu$-PICs for 
directional Dark Matter searches has now progressed to the pursuit of low-background detectors. 

\subsubsection{$\mu$-PICs in NEWAGE}
The NEWAGE experiment uses $\mu$-PICs for directional Dark Matter searches. 
Three TPCs (NEWAGE-0.1a,b,c) with $\rm 10\times10$~$\rm cm ^2\, \mu$-PICs 
and two (NEWAGE-0.3a,b) with $\rm 31\times31 $~$\rm cm ^2 \mu$-PICs 
are operating in surface and underground laboratories, respectively.

\begin{figure}
\includegraphics[width=1.\linewidth]{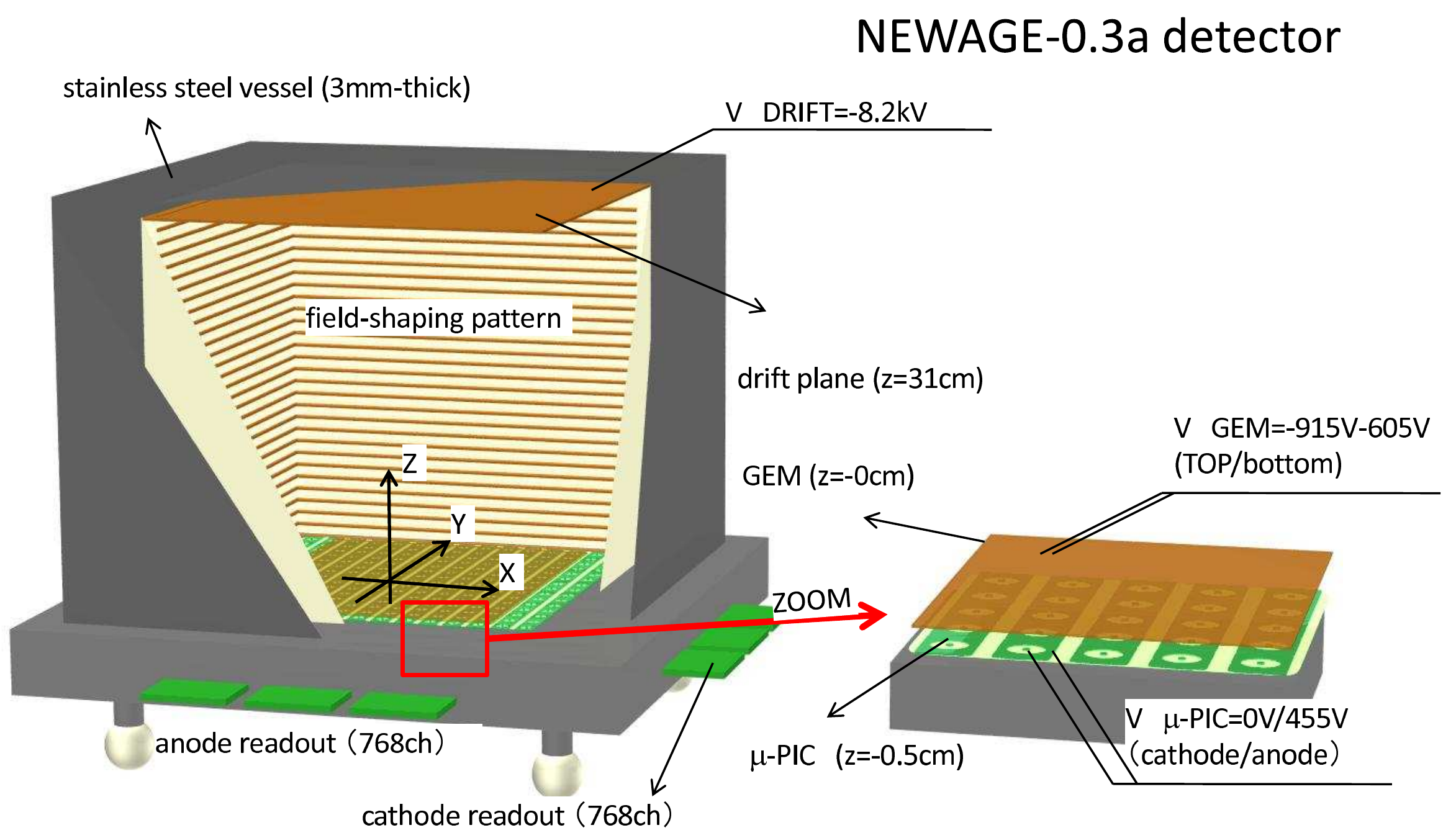}
\caption{\label{fig:NEWAGE-03a}
A schematic of the NEWAGE-0.3a detector.
}
\end{figure}

A schematic of a typical NEWAGE TPC is shown in Figure~\ref{fig:NEWAGE-03a}. 
The $\mu$-PIC located at the bottom of the vacuum vessel serves as a two-dimensional readout with the third dimension from timing. 
An electric field in the detection volume causes the 
primary electrons to drift toward the detection device.
A GEM located above the $\mu$-PIC provides additional gain.
The performance of the latest and largest detector, NEWAGE-0.3b', 
filled with \cff{} gas at 76\,Torr, has been characterized in detail \cite{ref:NEWAGE_PTEP2015}.
The directional energy threshold was 50\,keV electron equivalent; the angular resolution, nuclear track detection efficiency, and electron 
detection efficiency at the energy threshold were measured to be $\rm 40^\circ$, 40$\%$ and $2.5\times 10^{-5}$, respectively.  The latest underground directional Dark Matter search was performed in 2013 and a 
direction-sensitive spin-dependent limit of 557~pb for 200~$\rm GeV/c^2$ WIMPs
was obtained \cite{ref:NEWAGE_PTEP2015}.

\subsubsection{Conclusion}
The $\mu$-PIC and associated electronics are a well-studied system
for three-dimensional tracking in gaseous detectors. 
A TPC with a $\mu$-PIC pioneered the use of MPGDs for directional Dark Matter searches in the early 2000s. 
The next step in the development of $\mu$-PICs for 
directional Dark Matter searches is to create low-background detectors.

\subsection{MPGD: Pixel chip readout}
\label{sec:pixelchip}
Conventional readout techniques for gaseous detectors have been based on the use of wires (with few millimeter pitch), strips (with 100--200\,$\mu {\rm m}$ pitch), or pads (with area on the order of a square millimeter).  As discussed in Sec.~\ref{sec:mpgdIntro}, in recent years MPGD-based gas amplification has significantly improved gas detector performance over what can be achieved with wire-based approaches, so that nowadays the limiting factor has become the readout.
To match the spatial pitch of the MPGD amplification device, a finely pixellated readout structure could be used. Advances in commercial CMOS processes have led to the availability of highly segmented detectors with several million pixels each operating as an independent sensing element.
These pixel semiconductor detectors have become powerful tools for particle tracking, 
radiation detecting systems in high energy physics experiments, and fast imaging applications with penetrating radiation. 
MPGDs using pixel readout chips as anodes offer the possibility of high (single electron) charge-collection efficiency, three-dimensional reconstruction with position-dependent ionization measurement, negligible noise at room temperature, small demands on DAQ, and low cost per channel.  If pixel chip readout can be combined with an amplification stage stable at gas pressures of order $\sim$10\,Torr (like a single thick GEM or 3-4 thin GEMs), then this technology may be suitable for imaging nuclear recoils with high signal-to-noise ratio, excellent position and energy resolutions and low energy threshold, making it very attractive for directional Dark Matter searches.

To date, work on these technologies has focused on single-chip miniature TPC prototypes and gas pressures ranging from 
atmospheric pressure down to 100 torr. Three main families of CMOS-based chips have been used in R$\&$D targeting directional Dark Matter searches, namely the ATLAS, QPIX, and Timepix pixel chip families.
The chips can operate in different modes including time-of-arrival (TOA), time-over-threshold (TOT), charge integration (ADC), event counting (EC), and integral time-over-threshold (iTOT).  In the following we will discuss each of these R$\&$D efforts in turn, together with chip specifics (summarized in Table \ref{tab:pixel_table}), and future prospects.

\begin{table}
\centering
\caption{Summary of the main characteristics of the most advanced versions of pixel chips discussed in Sec.~\ref{sec:pixelchip}. The acronyms used in the last row (Modes) are described in the text.}
\begin{tabular}{l | c c c }
\hline\hline
                        & FE-I4b   	                  & QPIX-ver.1   	  & Timepix3 \\
\hline 
Pixel Size ($\mu$m$^2$) & 50 $\times$ 250                 & 200 $\times$ 200      & 55 $\times$ 55 \\
Pixel Matrix            & 80 $\times$ 336                 & 20 $\times$ 20	  & 256 $\times$ 256 \\
CMOS Process            & 130 nm	                  & 180 nm  	          & 130 nm \\
Clock Speed (MHz)       & 40		                  & 100		          & 100\textsuperscript{*} \\
Readout                 & Serial	                  & Parallel or serial	  & Parallel or serial \\
Trigger                 & Data driven or external         & External              & Data driven or external \\
Threshold (e$^-$)       & 2000\textsuperscript{$\dagger$} & 60000		  & 500\textsuperscript{$\ddagger$} \\
Noise (e$^-$)           & 100		                  & 2500		  & 100\\
Modes                   & TOA$+$TOT	                  & ADC$+$TOT$+$TOA       & TOA, EC$+$iTOT,  \\
 		        & 		                  & 			  & TOA$+$TOT  \\
\hline\hline
\multicolumn{4}{l}{\textsuperscript{*}\footnotesize{A time resolution of 1.6\,ns is possible, see Section~\ref{sec:TIMEPIX}.}}\\
\multicolumn{4}{l}{\textsuperscript{$\dagger$}\footnotesize{The threshold is adjustable.  This is a typical value.}}\\
\multicolumn{4}{l}{\textsuperscript{$\ddagger$}\footnotesize{The threshold is adjustable.  This is the minimum threshold for which $\leq$1 pixel will fire due to noise~\cite{PoikelaTimepix3}.}}
\end{tabular}
\label{tab:pixel_table} 
\end{table}%

\subsubsection{ATLAS pixel chips}
\label{sec:pixelchipAtlas}
The ATLAS pixel chip family has been developed in order to cope with the large radiation dose and high
particle flux seen by the innermost detector of the ATLAS experiment at the LHC. The FE-I3 and FE-I4b chips are fabricated in 250 nm and 130 nm CMOS technology, respectively, with pixel sizes 
of $50 \times 400~\mu{\rm m}^2$ and $50 \times 250~\mu{\rm m}^2$, and are both nominally clocked at 40\,MHz.
Each pixel contains an integrating amplifier, a discriminator, a shaper, and associated digital controls. 
Only a small fraction of each pixel's surface area, nominally used for bump bonding of pixelated silicon sensors in 
the ATLAS experiment, is conductive. When using these chips (without such silicon sensors) to detect amplified TPC drift charge, a pixelized metal layer is deposited onto the chip surface, to increase the conductive area of each pixel and hence the charge collection efficiency. Pixels are typically tuned to thresholds of 1500-3000 electrons, 
well above the typical pixel noise (100--200 electrons RMS at room temperature). As a result, in the absence of a true charge signal,
the rate of hits is typically negligible ($<10^{-9}$~Hz per pixel) and no data are output by the chip. 
When avalanche charge reaches the pixel chip and at least one pixel detects charge above threshold, 
a digital trigger signal is produced by the chip. This signal triggers the readout, resulting in a zero-suppressed digital 
serial stream that encodes the \twod{} position, arrival time, and amount of ionization observed in all pixels above threshold. 
Because the typical pixel thresholds (about 2000 electrons) are small compared to typical MPGD gains ($10^3 - 10^5$).
It is reasonable to expect readouts based on MPGD gain stages and pixel chip readout to be capable of detecting all primary ionization - even single electrons - 
with efficiency near unity and negligible rates of noise hits. This expectation was confirmed in the first measurements with double-GEM gain stages and ATLAS FE-I3 pixel chip readout at LBNL \cite{Kim:2008zzi}.

\begin{figure}
\centering
\includegraphics[width=8cm, trim=2cm 20cm 12cm 1cm, clip=true,]{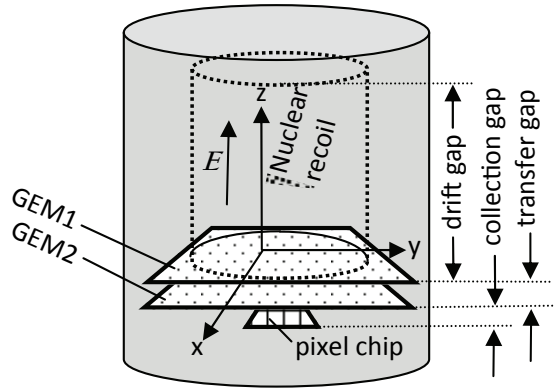}
\caption{Schematic of charge readout via double GEMs and a pixel chip.  \label{pix:tpcdrawing}}
\end{figure}

\begin{figure}
\centering
\includegraphics[height=8cm]{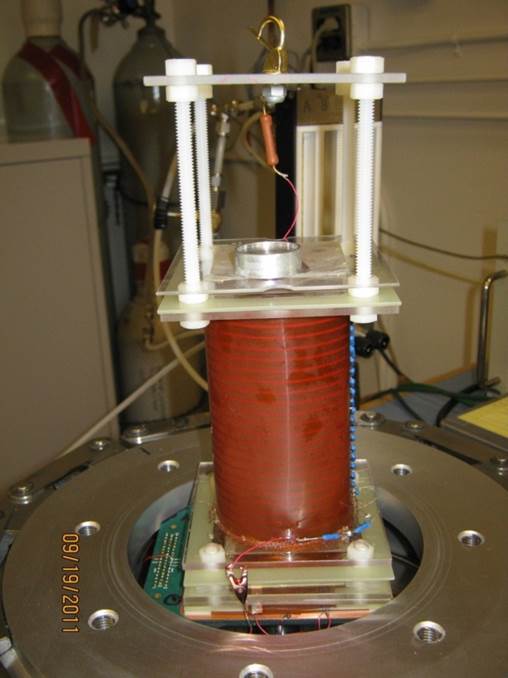}
\includegraphics[height=8cm]{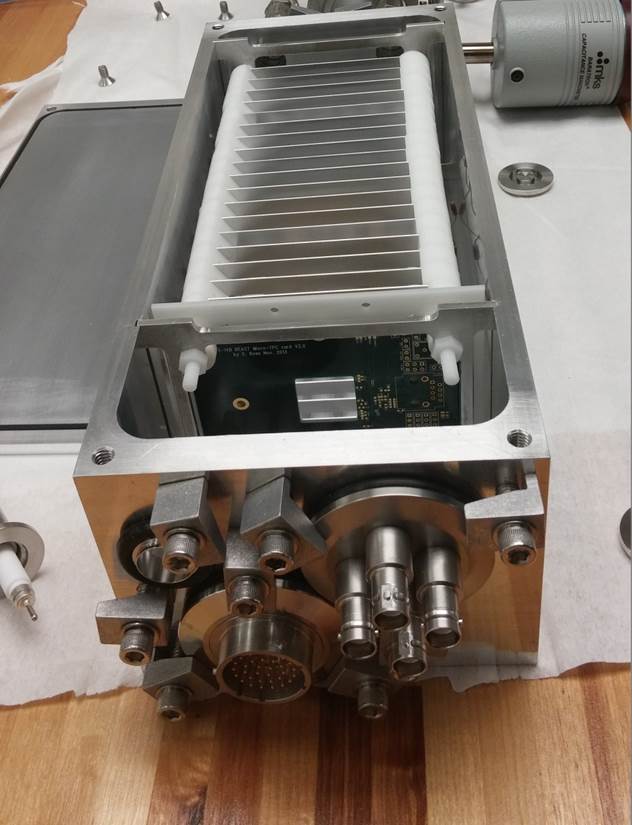}
\caption{Second-generation prototype TPCs with pixel readout constructed at LBNL (left) and the University of Hawaii (right). Both prototypes utilize a double layer of GEMs for charge amplification, and a single ATLAS FE-I4b pixel chip to detect the charge. The LBNL prototype utilizes a 12-cm cylindrical field cage constructed from a flexible Kapton sheet with copper field-shaping rings. The Hawaii prototype utilizes a 15-cm field cage consisting of aluminum rectangular field shaping electrodes, mounted onto four Delrin acetal support rods.\label{pix:tpcs}}
\end{figure}

\paragraph{Existing prototypes and performance}
Four miniature TPCs with fiducial volumes ranging from 0.5 to
60~cm$^3$ have been constructed to study TPC charge readout with ATLAS pixel chips, 
two at LBNL \cite{Kim:2008zzi,Vahsen:2014mca} and two at the University of
Hawaii \cite{Vahsen:2015oya,Jaegle:2015beasttpc}.  All four prototypes have
employed a double layer of CERN standard GEMs \cite{Sauli:1997qp} with $5
\times 5$~cm$^2$ active area and $140~{\rm \mu m}$ hole spacing for charge
amplification, and a single pixel chip for recording the resulting
avalanche charge, as depicted schematically in Figure~\ref{pix:tpcdrawing}. The pixel chip is glued to a circuit board and electrically connected with wire bonds, 
which are shielded against the electric field by a small metal overhang \cite{Kim:2008zzi}. The first prototypes 
at both sites used the ATLAS FE-I3 pixel chip \cite{Aad:2008zz}, while the two most recent 
ones (shown in Figure \ref{pix:tpcdrawing}) both use the ATLAS FE-I4b \cite{GarciaSciveres:2011zz}.

\begin{figure}
\centering
\includegraphics[width=12cm]{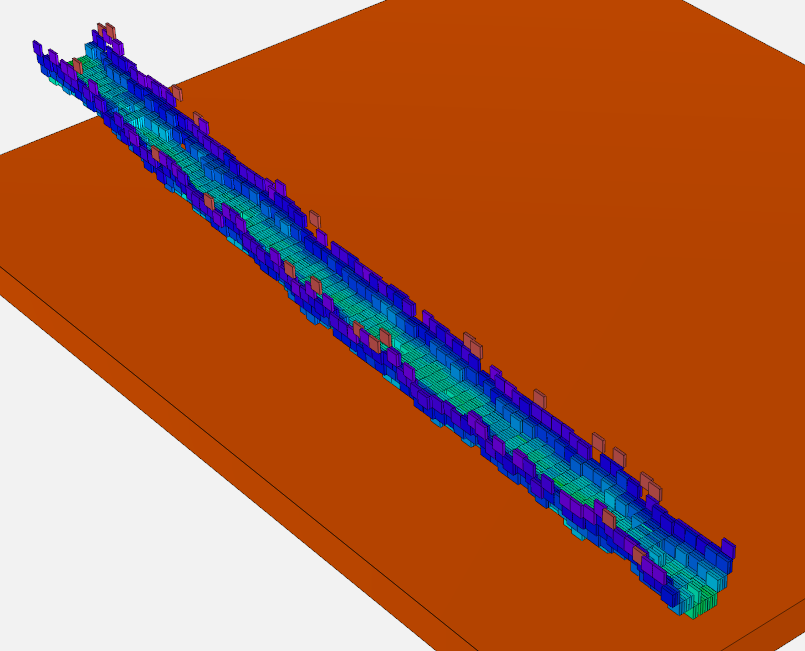}
\caption{Alpha-particle track recorded with Hawaii TPC prototype with pixel chip readout. Each block shown represents one $50 \times 250 \times 250\,\mu$m$^3$ unit of the readout volume. The color represents the ionization measured in that volume, and in the detector volume directly above it, as further explained in the text. Note the absence of noise hits.}
 \label{pixel_alpha}
\end{figure}

Since the charge detected in each pixel is measured via TOT, which typically has a time constant equal to multiple clock cycles (and hence time bins) of the pixel chip, for typical electron drift velocities and track shapes, each pixel provides a single hit per track, which is the location of charge closest to the chip. The measured charge assigned to that hit is then the integral of all charge in the track directly above the pixel. By using the known drift velocity in the gap, a \threed{} image of the surface of the track's charge cloud that is facing the pixel chip can be reconstructed. This provides a detailed measurement of the distribution of total ionization across and along the track, as shown in Figure \ref{pixel_alpha}, with a point resolution of order 100--200\,$\mu$m in all three dimensions. Detailed measurements of detector performance can be found in a series of published \cite{Kim:2008zzi,Vahsen:2014mca,Seong:2013nua,Vahsen:2015oya,Lewis:2014poa} and forthcoming \cite{Jaegle:2015beasttpc} papers. These studies have shown that the pixel chip contribution to the point resolution and energy resolution of the readout plane is small or negligible. 

Instead, the point resolution of the readout plane has been found to be limited by diffusion in the transfer and collection gaps (see Figure~\ref{pix:tpcdrawing}), which for any practical application are much smaller than diffusion in the drift gap. As a result, as is also visible in Figure \ref{pixel_alpha}, a TPC with pixel readout can record a large number of independent measurements across the typical diffusion width of ionization trails. This capability can be used to measure the diffusion contribution to charged tracks, including nuclear recoils. The LBNL and Hawaii groups recently demonstrated that by making use of detector-internal calibration sources, robust measurements of diffusion can be obtained. As a result, the absolute position of mm-sized track segments were measured to an accuracy of 1--2\,cm \cite{Lewis:2014poa}. This has enabled absolute position measurement and hence fiducialization in the drift direction, which is critical for low-background search experiments. It may also be possible to trade some of the better-than-needed readout plane resolution for larger area coverage via charge focusing \cite{Ross:2013bza}, as discussed also in Section~\ref{sec:scale}. 

The event-energy resolution of TPC readout with pixel chips is energy dependent, and typically limited by the gain resolution of the amplification stage, in this case the double GEMs. In 70:30 He:CO$_2$ and 70:30 Ar:CO$_2$ gas mixtures at atmospheric pressure, the gain resolution is asymptotic to $\sigma_E/E\approx 2$--$4\%$ at 1--2 MeV. For keV-scale energies, of interest for WIMP Dark Matter detection, it is of order $\sigma_E/E\approx 10\%$ at atmospheric pressure, limited by primary ionization statistics and statistics of the avalanche process.

Perhaps surprisingly, despite the high precision and sensitivity of the pixel chip readout, the data rate and demands on the DAQ system tend to be lower than in most competing approaches. This is because the discrimination against noise occurs far upstream in the data flow -- in the analog part of each individual pixel -- and because the chip can self-trigger. As a result, in the absence of ionization in the drift gap, there are no digital hits or triggers created, and no data is output by the chip.

Though other amplifications stages than GEMs could be considered, the GEMs and pixel chips have proven to be a robust combination of high gain, low noise, and low pixel threshold, that promises exceptional sensitivity in the context of Dark Matter detection. Initial concern that discharges from the GEMs would damage the sensitive pixel electronic have proven to be unfounded; for example, the first Hawaii prototype ($D^3$-Micro) \cite{Vahsen:2015oya} operated more or less continuously for two years with the same pixel chip, and that chip still works today. One drawback of thin GEMs is that they are rather vulnerable to dust, so that detector assembly in a clean room is recommended. 

At atmospheric pressure, two thin GEMs provide sufficient gain (of order $10^4$) compared with typical discriminator settings in the pixels (2000--3000 electrons), so that even single ionization electrons in the drift gap can be easily detected. This should, in principle, translate into the lowest possible discrimination and directionality threshold levels achievable when reconstructing ionization trails of WIMP recoils. However, a detector optimized for low energy threshold, such as for low-mass WIMP searches, might operate well below atmospheric pressure, probably in the 10-torr range. Though below-atmospheric pressure clearly reduces the target mass, it also lowers the directionality threshold, and thus can increase the directional sensitivity of a detector \cite{Vahsen:2011qx,Jaegle:2012sma,Mayet2016}.  A double thin-GEM layer would not be stable at such low pressures \cite{Vahsen:2014mca}, but thicker GEMs can provide stable, high gain even down below 1\,Torr \cite{Shalem:2005ix}. The drawback of thick GEMs is larger hole spacing and thus somewhat reduced spatial resolution. Another option might be to mix the low-pressure target gas with Helium at higher partial pressure \cite{Vahsen:2014mca}, which improves GEM stability and may allow operation with multiple thin GEMs, while largely preserving recoil track length. The Hawaii group is currently building a next-generation prototype to study these options in more detail.

\subsubsection{The Quasi-3D pixel chip (QPIX)}
When reading out TPC signals, TOT is a good estimator of the collected charge as long as
the charge collection time profile does not change significantly. Since, however, directional
Dark Matter detectors need a drift region as long as possible in order to maximize the active
volume, longitudinal diffusion modifies this parameter so that smaller-diffusion events  
possess shorter TOTs and larger-diffusion longer ones. Moreover, TOT is also affected
by the inclination of the track with respect to the detection plane, so that for a given charge
longer TOTs are expected for electron avalanches arriving perpendicular to, rather than parallel to the \xy{} readout plane.
Therefore, independent measurements of TOT and charge would provide additional 
information on the tracks such as diffusion and track direction. The measured diffusion would
in turn provide an estimate of the absolute \z{} position.  
Absolute \z{}, even if the resolution is several cm, would greatly help to reduce radioactive 
background from the cathode and the detection plane \cite{Billard:2012bk}.

It is with this idea in mind that a CMOS ASIC called Quasi-3d pixel chip (QPIX), with the capability of ADC measurement in addition to TOA and TOT in each pixel, is being currently developed \cite{ref:Khoa_Mth}. The development is still at the proof-of-concept stage, and cost and radioactive backgrounds from the readout system are currently under investigation. The QPIX, like other pixel-readout chips, is intended to be coupled with MPGDs such as GEMs.

\paragraph{The QPIX-ver.1 chip}
After R\&D on several types of Test Element Groups (TEGs),
QPIX-ver.1, the first prototype with a two-dimensional array, has 
been developed \cite{ref:NEWAGE_cygnus2011,ref:Fei_Dth}. 

\begin{figure}
\begin{center}
\includegraphics[width=0.7\textwidth]{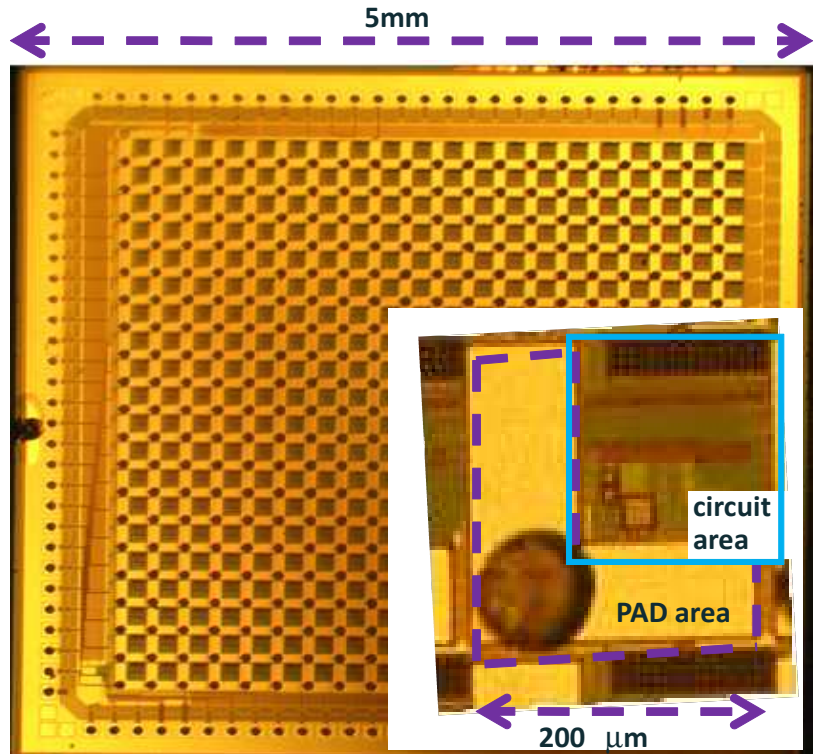}
\caption{\label{fig:QPIX-ver.1}A microscope photo of QPIX-ver.1. 20 $\times$ 20
pixels are seen. The inset shows a magnification of one pixel.}
\end{center}
\end{figure}

QPIX-ver.1 has 20$\times$20 pixels with a pitch of 200 $\mu$m. Each pixel has
a single register for 14-bit TOA, 8-bit TOT, and a 10-bit successive approximation register (SAR) ADC. The power consumption of each pixel is about 150~$\mu$W.
The chip was produced in the TSMC 0.18~$\mu$m process. A microscope photo of 20$\times$20 pixels is shown in Figure~\ref{fig:QPIX-ver.1}.
84 I/O pads are placed along three of the edges of the chip.
The inset shows the magnification of one pixel. A metal pad area is
indicated by the dashed line. This pad area can be used for
direct charge collection from the gas volume
or as a contact pad for bump bonding.
A trace of a bump bonding test is seen in the center. The circuit area is 130 $\times$ 140~$\rm\mu m^2$.
Further details are described in Refs. \cite{ref:NEWAGE_cygnus2011} and \cite{ref:Fei_Dth}. 

The electrical performance of QPIX-ver.1 was studied by injecting charges via the pads.
It was found that the TOA had good linearity up to 2\,$\mu$s while the ADC was linear up to 1.5\,pC. The comparator threshold for TOA and TOT measurement was 10~fC. The threshold, however, was about ten times higher than the design value. New TEGs with lower threshold of 1~fC and smaller dynamic range of 100~fC are being designed.

The chip was tested as a charged particle tracking device in a setup with four QPIX-ver.1 chips, a 2.8-cm drift gap, and three GEMs. Ion beams ($\rm Ne^{7+}$ at 260~MeV) at Takasaki JAEA were used and a position resolution of 85\,$\mu$m was obtained.

Once the electrical characteristics satisfy the requirements for QPIX to be used in directional Dark Matter experiment, background reduction and cost will become the next important R$\&$D items.

\label{sec:QPIX}
\subsubsection{The Medipix chip family}
Medipix2 and Timepix \cite{Plackett:2010zz} are the second generation of the Medipix chip family, fabricated in the IBM 250\,nm CMOS process, with a 256\,$\times$\,256 matrix of 55\,$\mu$m square pixels. The two chips share a similar architecture.
  
 Medipix2 was originally conceived as a photon counting, \xray{} imaging detector intended to exploit the superior noise performance and granularity of active pixel devices, compared to charge integrating systems based on conventional pad readout. Each Medipix2 pixel contains a charge-integrating preamplifier and a discriminator with a globally adjustable threshold, followed by mode control logic and a 14-bit counter. The chip is driven by an external shutter signal that determines if the pixel matrix is taking data or reading out. When the shutter is open, each pixel individually counts the number of particles above threshold and increments the counter. When the shutter is closed, the matrix is read out as a shift register with a dead time of $\sim$10\,ms. The Timepix chip is derived from Medipix2, where the second threshold of Medipix2 is replaced with a counting clock operating up to 100\,MHz (synchronized with the external clock reference) which is propagated to every pixel. This allows two new modes of operation, namely TOT and TOA, in addition to a particle counting mode. It is possible to program individual pixels to operate in different modes, and still read them out simultaneously. In TOT mode the energy resolution ($\Delta$TOT/TOT) is better than 5\% if the input charge is at least 1000\,e$^-$ above threshold. In TOA mode, the measured time-walk per pixel is $\leq$ 50\,ns. TOA is measured in 10\,ns due to limitations in the distribution of the (100\,MHz) clock across the pixel matrix. 

The main driving requirements for the development of Timepix3 \cite{Gromov:2011zz} (fabricated in 130\,nm CMOS) have been the simultaneous measurement of time (TOA) and charge (TOT), the minimization of dead time, the improvement of time resolution and improved monotonicity of TOT above $2\times 10^5$\,e$^-$ in detecting both polarities. Timepix3 has several modes of operation, namely TOA$+$TOT,  EventCount$+$Integral TOT, and TOA only. In the first mode, 10-bit TOT and 18-bit TOA are recorded simultaneously in each pixel. TOA is measured using a reference clock at 40\,MHz for the most significant 14 bits; the additional 4 bits of fine TOA measurement are provided by a Voltage-Controlled Oscillator (VCO) distributed across the pixel matrix, running at 640\,MHz and allowing to reach a time quantization of 1.6\,ns. 

Each pixel is equipped with a threshold-equalization DAC (Digital-to-Analog Converter), to correct for pixel-to-pixel variations. A Krummenacher scheme \cite{Krummenacher} guarantees a charge-sensitive preamplifier symmetric in both polarities, with a 500\,e$^-$ threshold. High gain (50\,mV/1000 e$^-$), and low noise ($75\,$e$^-$ RMS) ensure efficient hit identification. The circuit has fast peaking time ($\sim$10\,ns), limiting the time-walk. PMOS diodes are added to the Krummenacher feedback to provide monotonicity up to large values of positive holes ($\sim2\times 10^5$) and correct for the deviations observed in the original Timepix. The chip can be read out at the end of the shutter period as its predecessor (classical sequential readout), but also in data-driven mode. In the latter mode, Timepix3 sends out a 48-bit packet every time a pixel is hit; this optimizes the bandwidth of the system for the expected maximum hit rate of 40\,Mhits\,s$^{-1}$\,cm$^{-2}$. 

Preliminary tests demonstrated a time walk less than 25\,ns for pulses larger than 800\,e$^-$ \cite{DeGasperi}. The Timepix3 energy and track reconstruction performance has so far only been tested for hybrid chips bump bonded to 300\,$\mu$m-thick silicon sensors \cite{Frojdh:2015fta}. The energy resolution in TOT mode under normal operation conditions is found to be 4.07 keV FWHM at 59.5 keV (with possibility of improvement with stabilization of chip temperature). 

\paragraph{Medipix chips in TPCs}\label{sec:timepix_application}
The first use of the Medipix2 chip in combination with a MPGD occurred in 2005, and demonstrated single-electron detection with 90\% efficiency \cite{Campbell:2004ib}. Several realizations of this concept followed and brought to the development of the GridPix \cite{vanderGraaf:2007zz}, a Timepix coupled to a special type of Micromegas, the InGrid \cite{Lupberger:2014pba}, directly produced on the chip in a photolithographic process, with the holes of the grid aligned to the pixels. The GridPix, originally developed as an R$\&$D for the main tracker of the International Linear Collider, demonstrated a single point resolution of 41\,$\mu$m and 15\,$\mu$m  in the $x$ and $y$ directions respectively on the SPS 180\,GeV\,c$^{-1}$ muon beam~\cite{Koppert:2013lua}. The GridPix is also an excellent \xray{} detector, with demonstrated energy resolution of 3.85\% for 5.9\,keV photons (very close to the Fano factor of 3\%), and an energy threshold of 277 eV \cite{Krieger:2014wxa}, in the context of the CAST experiment.

A new detector called GEMPix has been developed within the Medipix collaboration and INFN where the amplification is provided by a new triple-GEM (active area 28 $\times$ 28 mm$^2$) to match exactly the area of a Quad-Timepix. Thanks to a specially designed high voltage power supply, and a carefully chosen GEM electrode layout, the GEMPix demonstrates good reliability and discharge resistance \cite{Murtas:2014zxa}. Two new R$\&$D projects called NITEC and DCant started in 2015 with a 5 cm drift TPC coupled to GEMPix in the context of directional Dark Matter searches, to test the performances in negative ion operation and the possible anisotropic response of carbon nanotubes \cite{Capparelli:2014lua}.

TPC readouts will benefit from new features available with Timepix3. The simultaneous TOT and TOA measurements will improve the spatial resolution (thanks to the possibility of sub-pixel resolution with the centroid of the hit), simplify pattern recognition, and help in single-ionization cluster detection.  Timepix3 also allows for the possibility of exploiting through silicon-vias instead of wirebonds at the chip edges, which can allow for large, high-density realizations of the readout (see Sec.~\ref{sec:vias}).

\label{sec:TIMEPIX}
\subsubsection{Prospects for realizing large-area pixel detectors}
In the path towards the development of ton-scale detectors with directional sensitivity, the data acquisition and readout structure is
likely to be both the cost driver (approximately  \$25/cm$^2$ of pixels) and the major scale-up issue. Since diffusion is limiting the size of the drift gap, a 1\,m$^3$ Dark Matter TPC would require 1-2 (negative ion drift) or three (electron drift) readout planes of 1\,m$^2$, with each plane resulting in a electronics cost of order \$250,000. 
Furthermore, inactive chip regions on at least one edge and the commonly used techniques for I/O connections typically limit the tiling possibilities of the devices to one double-row of chips. For these reasons, it is very difficult to seamlessly cover a large area detection plane, as is needed for a directional DM search.  Several approaches have recently been pursued to minimizing the dead region between chips and reducing the size and complexity of detection planes.

\begin{figure}
\begin{center}
\includegraphics[width=1.\linewidth]{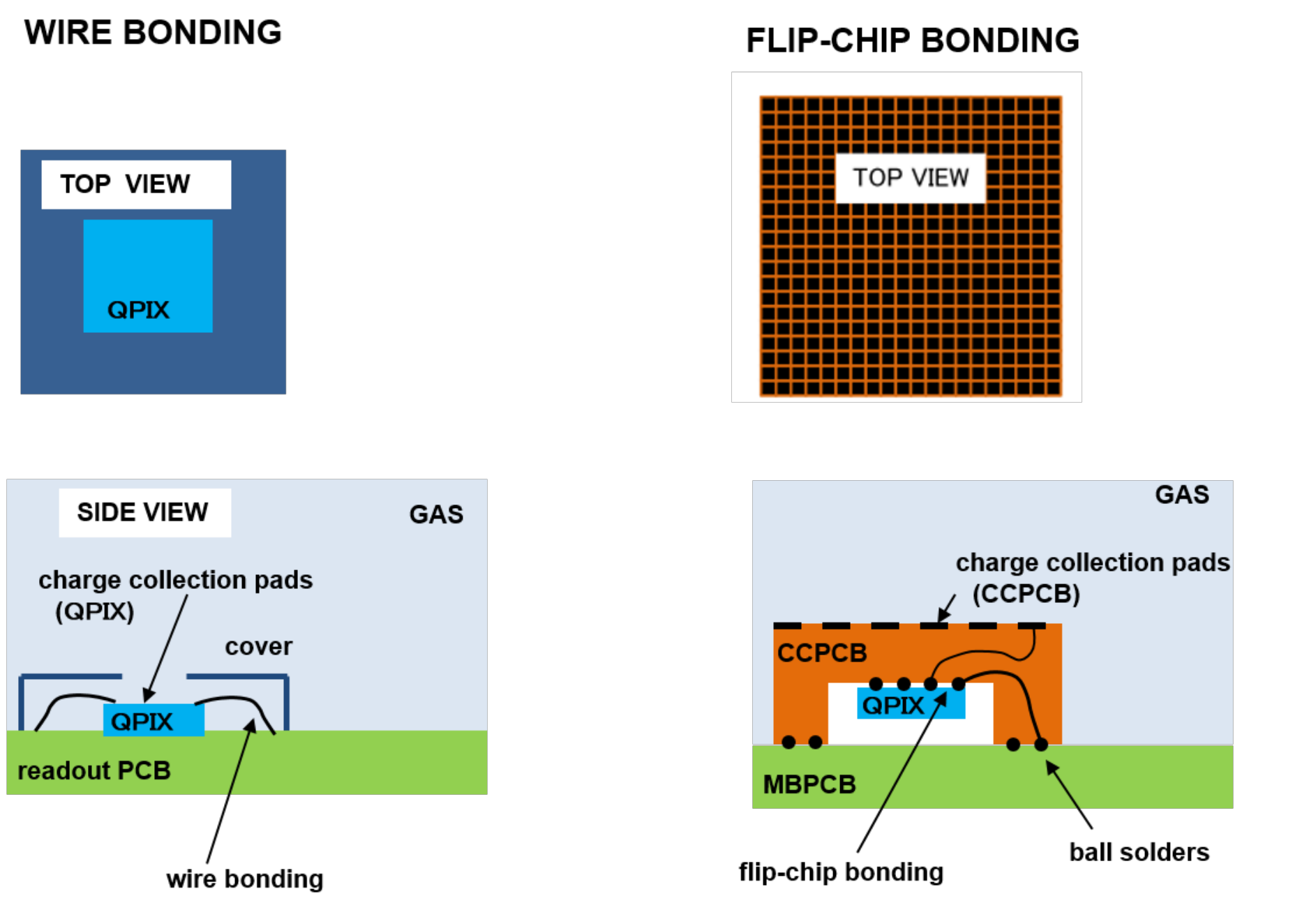}
\caption{\label{fig:QPIX_mountings}Two methods of QPIX mounting.
(Left) Wire-bond mounting.  (Right) Flip-chip mounting.}
\end{center}
\end{figure}

\paragraph{Flip-chip mounting}
Two mounting methods, wire-bonding and flip-chip mounting, depicted in Figure~\ref{fig:QPIX_mountings}, have been tried with the QPIX chip. The wire-bonding technique (left panel) is a well-studied and very reliable method. Wire-bonds were used to connect the I/O pads of QPIX-ver.1 to the readout PCB.
The problem with this approach is the presence of dead areas due to bond pads along at least one edge of the chip. In the case of the QPIX-ver.1, three edges were used for I/O, and a cover was used to shape the electric field. 

\begin{figure}
\begin{center}
\includegraphics[width=.5\linewidth]{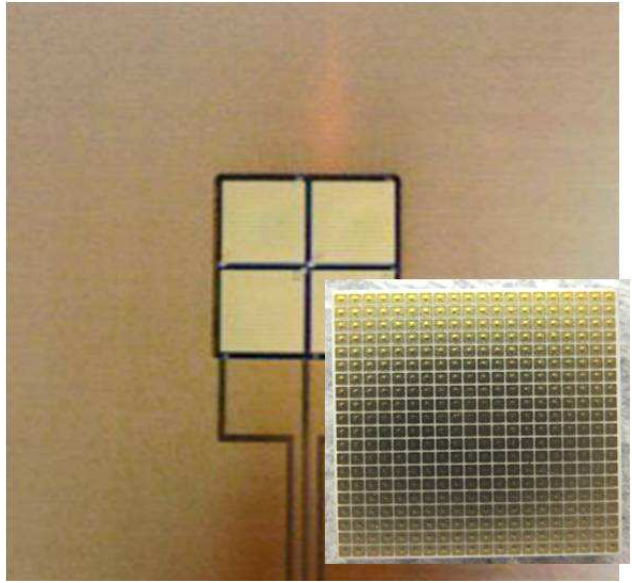}
\caption{\label{fig:QPIX_LTCC}Four CCPCBs mounted on an MBPCB. Inset shows a close-up of a CCPCB.}
\end{center}
\end{figure}

Flip-chip mounting (see Figure~\ref{fig:QPIX_mountings}) was attempted with QPIX in order to minimize dead area.  A charge collection PCB (CCPCB) was mounted on QPIX-ver.1 by flip-chip bump-bonds. This CCPCB has an array of $20 \times 20$ charge collection pads on the gas side, which are connected to the cavity underneath through the CCPCB.  QPIX-ver.1 is mounted in this cavity with flip-chip bump-bonds.  Similarly, I/O pads on the chip are also connected with flip-chip bump-bonds and to the
mother board PCB (MBPCB) through CCPCB. The CCPCB is larger than QPIX-ver.1 and can be mounted on the MBPCB with negligible dead areas. Four CCPCBs mounted on an MBPCB are shown in Figure~\ref{fig:QPIX_LTCC}. Mechanical mounting was successful, but electrical connectivity was not achieved because the surface of the CCPCB cavity was not sufficiently flat for flip-chip bump-bonding. An improved CCPCB is required for flip-chip mounting.  The performance of some QPIX-ver.1 units with wire-bonds and a triple-GEM amplification stage were studied \cite{ref:Fei_Dth} (see also Section~\ref{sec:QPIX}). Three out of four chips worked, while the rest had problems either in the ASIC production or the mounting.

\begin{figure}
\begin{center}
\includegraphics[width=0.8\textwidth]{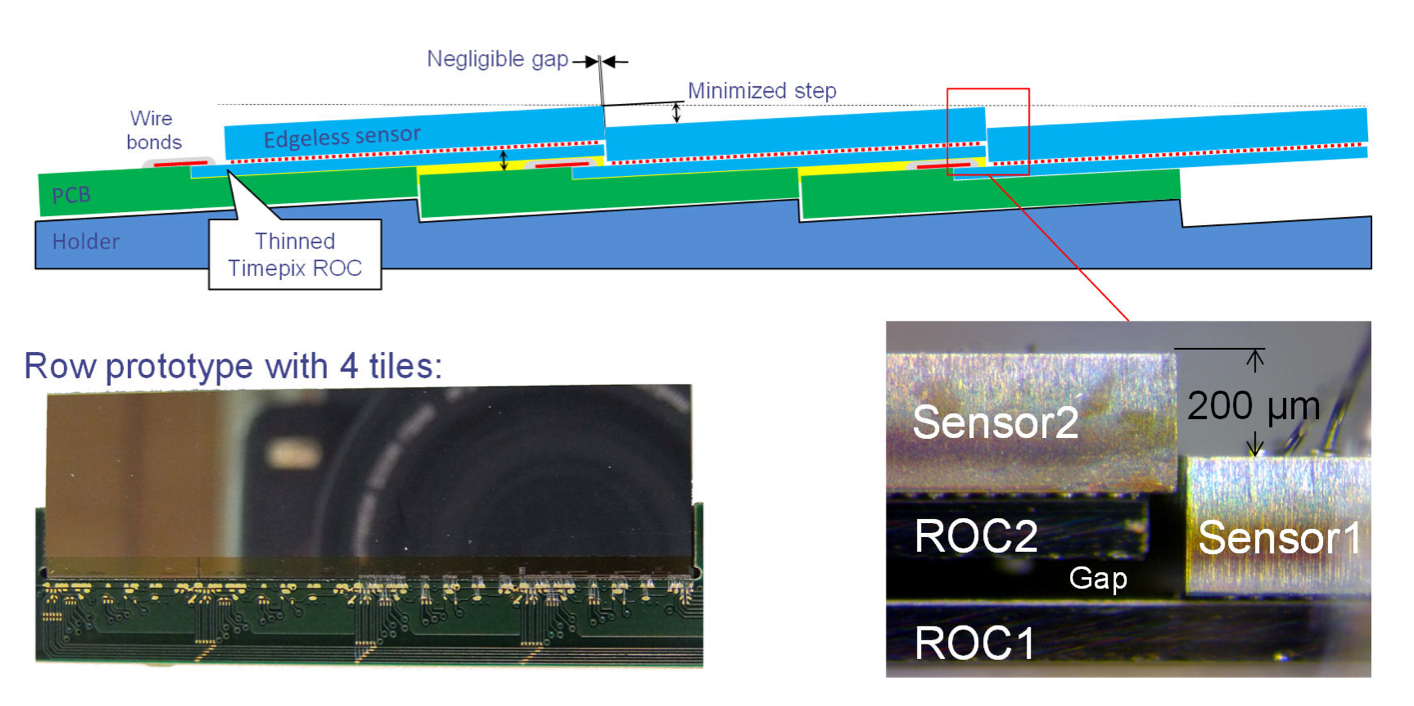}
\caption{\label{fig:tiling} The initial idea of tiling based on overlaying rows (top), microscopic photograph of real structure (bottom right) and the first prototype of single row with touching sensors (bottom left).
}
\end{center}
\end{figure}

\paragraph{Improved chip tiling}
Using an improved tiling technique, the Widepix company has produced the WIDEPIX 10 $\times$ 10 \cite{Jakubek:2014}, an array of $10 \times 10$ (silicon bumped) Timepix chips with a sensitive area of 14.3 $\times$ 14.3 cm$^2$, without dead space between chips. This is obtained by displacing the chips in a way that they are only partially supported by the PCB holder underneath. The overhanging part of the chips overlay the following PCB and is wire-bonded to it with a negligible gap to the next sensitive sensor (as can be seen in Figure~\ref{fig:tiling}). This creates single row of daisy-chained read-out chips providing a continuously sensitive area. This technology is fully scalable to virtually any size. The disadvantage of this approach is the necessity of tilting the rows, though the tilt angle can be minimized by reducing the chip thickness and the wire-bond area. For example, the backside-thinning of the Timepix chip, done after bump-bonding to sensors, achieved a minimal thickness of 50\,$\mu$m.  In the final prototype, a more conservative value of 120\,$\mu$m was chosen to decrease risks of damage during subsequent steps. The resulting tilting angle was 0.8\,degrees.  The tiles were connected and read-out along rows using the serial interface of the Timepix chips in a daisy chain. 

\begin{figure}
\begin{center}
\includegraphics[width=0.7\textwidth]{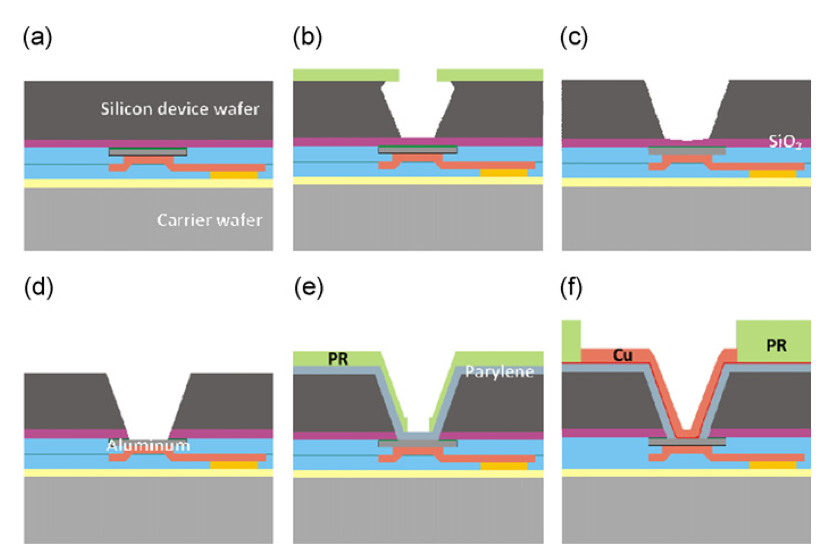}
\caption{\label{fig:TSV} Process flow of the through silicon via fabrication: (a) Thinned silicon device wafer (100 mm) is mounted face down on a glass carrier using wax as a glue layer. (b) The sloped via profile is achieved by reactive ion etching. (c) Maskless dry etch step is applied to remove negative angles near the via top and sidewall roughness. (d) Silicon oxide layer is etched away to expose the aluminum landing (bond) pad for electrical connection. (e) Parylene dielectric layer is deposited and opened at the bottom of the via by dry etching to reach the buried aluminum contact. (f) Seed layer sputtering and consecutive electroplating of the 5-20 mm thick copper layer which serves as the main conductive path between the silicon wafer front and back.
}
\end{center}
\end{figure}

\paragraph{Through-silicon via connections}
\label{sec:vias}
In order to overcome limitations associated with wire-bonding and flip-chip mounting, recent R\&D has focussed on the newly developed technology of through-silicon vias (TSV), which can be used to connect front and back sides of a chip \cite{Tezcan:2007}.
This technique can be performed at low temperature ($\leq 250^{\degr}$\,C) on standard CMOS processed wafers (see the process flow diagram in Figure~\ref{fig:TSV}). The interconnect is fabricated from the back-side of a thinned wafer.  Plasma etching is used to achieve a sloped profile, to allow the uniform deposition of the dielectric layer and the copper seed metallization. The vias are isolated from the substrate with Parylene, and a spray coating of photoresist is used to open the dielectric at the via bottom.
An electrical connection between the front and the back of the wafer is achieved by partially filling the via with copper. This copper layer is also used to redistribute all necessary I/O signals of the chip uniformly on the back-side. 
With this innovative connection and a new type of edgeless silicon sensor \cite{Bosma:2011zz} (where nearly all the surface is made sensitive thanks to the removal of the guard rings and the doping of the edge itself to make it actively participate in charge collection) the 
The RELAXd project (high REsolution Large Area X-ray Detector) \cite{Vykydal:2008} achieved four-side-tilable Timepix modules, with minimum dead space and able to cover an arbitrary large area. 

\paragraph{Charge focusing}
An alternative solution to the readout structure issue has been recently proposed by the Hawaii group of the D$^3$ experiment. 
By electrostatically focusing the TPC drift charge before it is detected, the size, cost, and complexity of construction of such detection planes may be reduced.  Simulations confirm such focusing can be achieved in low-pressure gas without introducing excessive diffusion due to the focusing stage. A first demonstration experiment has been performed, using a simplified setup at atmospheric pressure \cite{Ross:2013bza}. While further work is needed, preliminary data suggested a linear focusing factor of order 2.5 to 3.0 for a setup where the simulation predicted a factor of 2.2. The results are encouraging, and motivate further work to achieve uniform focusing at low gas pressure, and to determine the maximum feasible focusing factor. For example, linear focusing factors of 3 and 10 could lower the price (and manpower needs of construction) of a large-scale detector by a factor between 9 and 100. In optimistic scenarios, a 1-$\rm m^3$ detector could then be instrumented with a few (one to three) 1-$\rm m^2$ readout planes, each with 100 or less pixel chips, at a pixel chip cost of order \$2,500 per plane or less.
It must be noted that focusing is advantageous not only because it reduces cost, but also because (compared with using detectors with larger feature size) it reduces detector noise, which scales with the capacitance of each detector cell, which in turn scales with the cell area.
If charge focusing works well, it may enable large-volume, direction-sensitive WIMP searches at low cost.

\label{sec:scale}

\section{Optical readout}
\label{sec:optical}
In addition to ionization, the gas amplification process also produces
an abundance of photons \cite{Charpak:1987it}.  By imaging these
photons, one obtains a 2D projection (\xy) of the recoil track.  The
photon intensity depends on the ionization density, and so the surface
brightness of the recoil track encodes the recoil sense.  Because
vacuum viewports with high optical transparency are readily available,
the readout can be situated outside of the gas volume, which helps
minimize radioactive backgrounds.  Furthermore, recent advancements in
CCD and CMOS technology make optical imaging a particularly attractive
option for TPC readout.

In the early 1990s, Buckland's group pioneered optical readout of
low-pressure TPCs for directional Dark Matter
detection \cite{Buckland:1994gc,Lehner:1997fs}.  They built and
operated a TPC with 1\,m drift length and 100\,L volume, coupled to an
image intensifier and a CCD camera.  By placing the TPC in a 4.5\,kG
magnetic field, the transverse diffusion of electrons was suppressed
to under 1\,mm over the full 1\,m drift distance.

Since then, there have been many developments in optical readouts for diverse applications such as particle detection \cite{Breskin:1988sg,Gai:2011yb,Breskin:1988sd,Lightfoot:2008ig}, thermal neutron imaging \cite{Fraga:2002uc,FragaLuminescence:2003,Roccaro:2009tg}, and gas detector microphysics measurements \cite{Deaconu:2015vbk,Rubin:2013yua}, to name a few.  For direction-sensitive Dark Matter detection, these developments include studies of the photon yields of various gas mixtures,
measurements of the emission spectra as a function of gas mixture and
pressure, and discrimination and directionality studies using a
GEM-based detector with high spatial resolution and signal-to-noise in
low-pressure \cff{} gas.  At present, the Dark Matter Time Projection
Chamber (DMTPC) group is carrying out a WIMP search using CCDs coupled
to a low-pressure TPC filled with \cff{}.  We will describe the
various design considerations relevant to an optical TPC, and then
describe recent results from the GEM-based detector and the DMTPC
experiment.

\subsection{Design considerations for an optical TPC}

\subsubsection{Choice of optical sensor}
There are many choices for optical sensor, including CCD, intensified
CCD (ICCD), electron multiplying CCD (EMCCD), CMOS, as well as
non-imaging sensors such as PMTs and silicon photomultipliers (SiPM).
In this review, we focus on CCD sensors.

Because the readout of CCDs (as well as ICCD, EMCCD and CMOS
sensors) is slow compared to the temporal extent of a recoil track
along the \z{} direction, a CCD provides a 2D projection of the recoil
track.\footnote{As an example, the current state-of-the-art readout speeds allow for the full sensor to be read out in $\sim$1\,ms (with faster frame rates possible for a pre-selected sub-region of the sensor).  The drift speed of electrons in \cff{} is on the order of 10\,cm/$\mu$s, and so 1\,ms corresponds to 100\,m of drift distance.  In negative ion gases like \cstwo{} the drift speed is much slower, but the negative ions still travel 5\,cm in 1\,ms, a distance far larger than the millimeter length scale of the recoil track.}
In principle, the projected 2D range, combined with the energy
loss along the track ($dE/dx$) could be used to recover the track
angle relative to the drift direction, and this has been done for
MeV-scale alpha tracks
\cite{Fetal:2007zz}.  But for the low-energy recoils (tens of keV)
relevant for Dark Matter detection, this has not yet been successfully
demonstrated.  It may be possible to recover the third dimension of
the track by combining the CCD readout with a high-speed sensor.  For
example, in promising preliminary work the rise-times of the charge
signal \cite{Lopez:2013ah} and the PMT signal \cite{Fetal:2007zz} were
correlated with the \z{} extent of the recoil track.  Fraga \etal{} have further shown that with an array of PMTs alone (no CCD) it
is possible to reconstruct a 3D track, with better than 1\,mm point
resolution, albeit for 5\,MeV alpha particles from an
\isotope{Am}{241} source \cite{Fetal:2007zz}.

\subsubsection{Photon production}
\label{sec:photonyield}
The photon yield $\gamma/e^-$ during gas amplification (\ie{} the number of photons per electron after the avalanche) 
depends strongly on the choice of detector gas.  While a full review
of the optical properties of detector gas mixtures is beyond the scope
of this work, the emission properties of a few gases are mentioned
here.

Gas additives with small work functions such as tri-ethyl-amine vapor
(TEA) can enhance photon production \cite{Arnold:1992mv}.  In their
work with \chf+TEA and P-10+TEA, Buckland \etal{} found $\gamma/e^-\sim 1$
\cite{Buckland:1994gc}.  In
the context of optical TPCs for Dark Matter detection, \cff{} is a
particularly interesting gas because its large fluorine content
enhances the sensitivity to spin-dependent WIMP-proton interactions
\cite{Tovey:2000mm}, and it is a particularly efficient scintillator.  For
example, in pure \cff{} from 50 -- 600\,Torr, Pansky et
al. \cite{Pansky:1994zh} measured $\sim$300\,photons/MeV (primary
scintillation), compared with 0.06 for \chf.  In that work, they also
measured an avalanche-induced photon production rate of $\photperelec
= 0.3$ for $160 < \lambda < 600$\,nm.  Similarly, Kaboth \etal{}
\cite{Kaboth:2008mi} measured $\photperelec=0.34$ for $200 < \lambda <
800$\,nm using a single-wire proportional tube filled with pure \cff{}
gas at pressures of 140 -- 180\,Torr.

The emission spectrum of \cff{} is well-matched to the quantum
efficiency of silicon imagers.  Measurements at 180\,Torr
\cite{Kaboth:2008mi} show that the majority (60\%) of the emission is
in the visible range ($\lambda > 450$\,nm), with the rest between
250\,nm and 450\,nm.  Morozov \etal{} \cite{2010NIMPB.268.1456M}
found a pressure dependence of the emission spectrum of \cff{},
with the spectrum reddening with pressure from 1 to 5\,bar.

The temporal response of \cff{} scintillation has also been measured
\cite{Margato:2013gqa}.  Although not relevant for CCD readout, there
are schemes by which the third dimension of the track could be
recovered optically via timing (\eg{} using PMTs), in which case the
temporal response is of interest.  Avalanche-induced photon emission
in the visible (450--800\,nm) decays exponentially with a single time
constant of 15\,ns, independent of pressure.  The UV emission
(220--450\,nm) is dominated by a fast ($<10$\,ns) decay, while 10\% of
the UV emission decays more slowly (40\,ns).
Given that the typical drift speed of electrons in \cff{} is 10\,cm/$\mu$s, these scintillation time constants could complicate the reconstruction of the \z-coordinate of the track at the sub-millimeter level.

\subsubsection{Geometric acceptance}
\label{sec:opticalAcceptance}
The main challenge of optical readout for a TPC is the trade-off
between the imaged detector area and the photon throughput (the
fraction of photons produced in the amplification region that are
detected).  In practice, the most substantial loss comes from the
geometric acceptance (the solid angle of the lens as seen from the gas
amplification region).  As shown below, the geometric acceptance is
$\sim$10$^{-3}$ for prototype-scale detectors, and smaller for larger
detectors.

The photons generated during the gas amplification process are emitted
isotropically.
In principle, these photons could be collimated or focused near the
amplification plane (see \eg{} \cite{Ju:2007ra}), but in practice this has not been achieved for dark matter detection.
Instead, the photon flux decreases with the square of the distance from
the source to the lens, $s_o$.  The fraction of photons produced in
the amplification region that make it into the optical system
(ignoring any losses due to absorption) is called the geometric
acceptance $\eta$, and is simply the ratio of the lens cross-sectional
area to the surface area of a sphere of radius $s_o$: $\eta = D^2/(4
s_o)^2$, where $D$ is the lens aperture diameter, and we have assumed
that $s_o >> D$, which is the case in practice.

\begin{figure}
  \centering
   \includegraphics[width=0.5\textwidth]{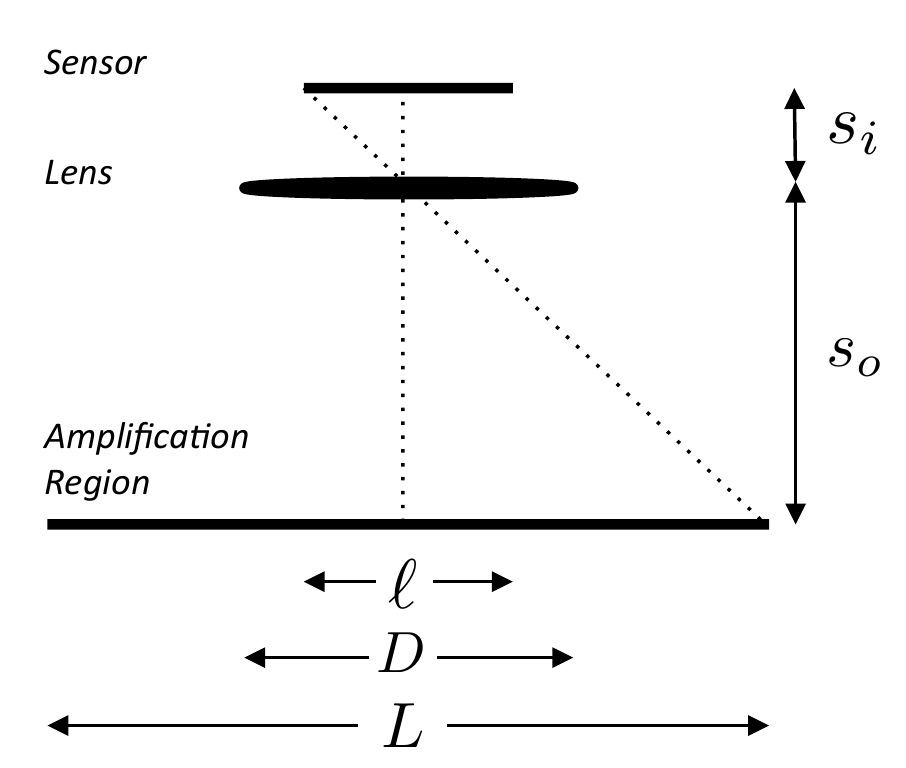}
   \caption{\label{fig:optical_schematic}
Schematic showing the side-view of an amplification region (object plane) of linear dimension $L$ imaged by a lens of diameter $D$ onto an optical sensor (image plane) of linear dimension $\ell$.  The distance between object plane and lens is $s_0$, and the distance between the lens and image plane is $s_i$.}
\end{figure}

The expression for $\eta$ is more useful when cast in terms of the
magnification $m$ of the optical system and the $f$-number of the
lens, which we now do.  The magnification is defined as the ratio of
the length scale $L$ of the amplification region to the length scale
$\ell$ of the sensor.
As seen in Figure~\ref{fig:optical_schematic}, the magnification can
equivalently be expressed as the ratio of the lens-object separation
$s_o$ and the lens-image separation $s_i$: $m=s_o/s_i$.  The
$f$-number of the lens, written $f/\#$, is defined as the ratio of the
lens focal length $f$ to the lens diameter: $f/\# \equiv f/D$.  Small
$f$-number lenses are called ``fast,'' and are advantageous in Dark
Matter detection (and in low-light photography) because they give a
larger geometric acceptance for a given field of view.\footnote{The
``lens speed'' nomenclature (\ie{} ``fast'' and ``slow'' for a small
and large value of $f/\#$, respectively) comes from photography and
refers to the exposure time required to achieve a given signal to
noise ratio.  The cinematographer Stanley Kubric famously filmed
scenes of the movie {\em Barry Lyndon} under natural candlelight using
a very fast $f/0.7$ lens developed by Carl Zeiss for NASA to film the
Apollo lunar landings.  Commercially available lenses are typically
slower than $f/0.95$.}

The final, and reasonable, constraint requires that the optical system
is focused.  In that case (and taking the thin-lens limit, which is
adequate for this discussion), the Lens Maker's equation states $1/f =
1/s_o + 1/s_i$.  By combining this with the expressions for the
magnification and $f$-number, we find
\begin{equation}
\eta = \frac{1}{16\,\left(m+1\right)^2 \left(f/\#\right)^2}.
\end{equation}
A fast lens (small value of $f/\#$), and a small magnification factor
(large sensor size) are preferred.

For example, a small gas amplification device (\eg{} $10 \times
10$\,cm$^{2}$), imaged by an $f/1$ lens onto a $2.5 \times
2.5$\,cm$^2$ CCD chip has a geometric acceptance of $3 \times
10^{-3}$.  The acceptance falls approximately quadratically with the
linear dimension of the amplification region, so if the same imager
was used for a $1 \times 1$\,m$^2$ amplification region, then the
acceptance would be $4 \times 10^{-5}$.

\subsubsection{Photon throughput}
In addition to the geometric acceptance, the signal strength depends
on the quantum efficiency $QE(\lambda)$ of the imaging sensor, as well
as the transparency $T(\lambda)$ of all elements between the gas
amplification region and the sensor.  Both of these depend on the
photon wavelength $\lambda$, and so it is convenient to compute an
average of these quantities $QE^\star$ and $T^\star$, weighted by the
photon emission spectrum (in Section~\ref{sec:photonyield}, we discussed the
emission spectra of various gas mixtures).  Thus
\begin{equation}
QE^{\star} \ = \ \frac{\int d\lambda \ QE(\lambda) \ I(\lambda) }{\int d\lambda \ I(\lambda) },
\end{equation}
and
\begin{equation}
T^{\star} \ = \ \frac{\int d\lambda \ T(\lambda) \ I(\lambda)}{\int d\lambda \ I(\lambda)},
\end{equation}
where $I(\lambda)$ is the number of photons emitted in the gas amplification region as a function of wavelength.

To enhance the quantum efficiency, a back-illuminated CCD (peak
$QE\sim0.95$) is preferred to front-illumination (peak $QE\sim0.7$),
although it is more costly \cite{JanesickBook2001}.  The transparency parameter $T(\lambda)$ is the
product of several independent contributions, the details of which
depend on the particular detector configuration.  As an example,
consider a micromegas gas amplification device, imaged by a CCD camera
through a vacuum viewport.  Then $T(\lambda)$ would have contributions
from the optical transparency of the micromegas mesh and cathode material
($\sim$0.7--0.9, depending on mesh pitch and wire thickness), the
vacuum viewport ($> 0.9$ for 250\,nm $< \lambda < 1\,\mu$m), and the
lens and window of the CCD ($\sim$0.7--0.9).

\subsubsection{Spatial resolution}
\label{sec:opticalSpatialResolution}
Once the amplification region and photon sensor chip size are chosen,
the spatial granularity is determined by the number of pixels along a
linear dimension of the sensor.  For example, if the $2.5 \times
2.5$\,cm$^2$ sensor in the example above were divided into $1024
\times 1024$ pixels, then each pixel would image a $100\times100\,\mu$m$^2$ portion of the amplification region.  Geometric track parameters (\eg{} the centroid) can be determined with resolution finer than the pixel scale~\cite{HowellBook2006}.

The granularity can be intentionally degraded by grouping adjacent
pixels into a single ``bin'' on-chip before readout.  This has the
added benefit of increasing the signal strength per bin without increasing the read noise, reducing the
data file size, and decreasing the total readout time, since in one
dimension, the binning is done in parallel (though experience shows
that for some sensors, the read noise increases moderately with
binning, see Section~\ref{sec:readnoise}).
The area of the amplification region imaged by a single sensor bin is
referred to as a ``vixel.''

\subsubsection{Signal strength}
The overall signal size per sensor pixel (or bin, if the sensor is binned) is given approximately
by
\begin{equation}
N_{signal} \ = \ \frac{E_{ion}}{W} \times G \times (\gamma/e^-) \times T^\star \times QE^{\star} \times\eta
\end{equation}
where $E_{ion}$ is the ionization energy deposited by a recoiling
nucleus over one vixel, $W$ is the work function of the gas (for
\cff, $W$ = 34\,eV \cite{Reinking1986JAP,DMTPC_wolfethesis}), and $G$ is
the gas gain, typically 10$^4$--10$^6$.

\subsubsection{Noise sources}
\label{sec:readnoise}
There are three main sources of noise to consider when using CCD
sensors: Poisson noise of the signal, read noise from the chip's
output amplifier and digitizing electronics, and thermally generated
electrons in the sensor, referred to as dark
noise \cite{JanesickBook2001}.

The Poisson noise of the signal is inescapable and is equal to the
square root of the number of photoelectrons generated in the CCD by
incoming light: $\sqrt{N_{signal}}$.  Ideally, one would reduce other
noise sources so that the observations are Poisson limited.  In
practice, this is rarely the case.

The read noise, $N_{read}$, is specified in terms of the number of
electrons (rms), with typical values of $\sim$3--10 e$^-$ rms for
cameras available off-the-shelf.  Self-heating of the readout
amplifier can increase the read noise \cite{HowellBook2006}, so it is
often advantageous to choose a slower digitization speed to reduce the
read noise (albeit at the cost of increased dead-time).  For example,
the FLI ProLine 9000 camera has two digitization speeds: 1\,MHz and
8\,MHz, with corresponding specified read noises of 10\,e$^-$ and
15\,e$^-$ rms.  The effect of read noise can be mitigated somewhat by
binning pixels before digitization, in which case the read noise
contributes only once to the bin, while the signal is increased by
combining individual pixels.  The DMTPC group, however, found that the
read noise increases with binning in one dimension.  This is likely
due to an effect called ``spurious charge'' that is described in
Janesick's comprehensive text on CCDs \cite{JanesickBook2001}.

Dark current refers to thermally generated electrons, and is specified
in terms of the number of thermal electrons generated per pixel per
time.  The dark current is an exponential function of temperature.
With typical room temperature dark rates $R(T)$ of
$10^4$\,electrons/pixel/second
\cite{HowellBook2006}, it is clear that the CCD must be cooled.
Typically, thermoelectric (Peltier) devices can cool the CCD chip to
$\sim$50$^{\degr}$\,C below ambient temperature, which may be sufficient
to keep the dark noise sub-dominant to the read noise.\footnote{This statement depends on the read noise and the exposure time for a given CCD.  For example, a Peltier-cooled chip can typically suppress the dark current to less than 0.1\,electron/pixel/second, and a typical read noise is $\sim$10 electrons rms, so for exposure times up to 10$^3$\,s, the dark noise is smaller than the read noise.}  If the dark rate is
too large, then cryogenic cooling could be employed.  While the mean
number of thermally generated electrons can be corrected by the
subtraction of a dark frame (a CCD exposure with closed shutter so no
external illumination is present), the noise in that correction is
given by the square root of the number of electrons.  The dark
noise per pixel is then $N_{dark} = \sqrt{R(T) \, t_{exp}}$, where
$t_{exp}$ is the exposure time of a pixel.

The three noise terms described above are independent, and they
combine in quadrature to give the rms noise per bin:
\begin{equation}
N_{noise} \ = \ \sqrt{N_{signal} + n_{pix} \, R(T) \, t_{exp} + N_{read}^2},
\end{equation}
where $n_{pix}$ is the number of CCD pixels that have been grouped
into a bin, and $N_{read}$ is the read noise of the bin.  If the read noise is dominant, then $N_{noise}\approx N_{read}$.

\subsubsection{Spatial and temporal uniformity}
CCD sensors typically have small inhomogeneities across the sensor
which come from the sensor material, or the manufacturing
process~\cite{JanesickBook2001}.  A highly-localized source of CCD
non-uniformity arises from ``hot'' pixels, in which the dark current
is significantly higher than their neighbors, by at least 50\%; this
is typically due to contamination embedded in the sensor. Similarly,
``cold'' pixels, in which the response is less than 75\% of the
average pixel, can arise from contamination on the surface of the
sensor.  These pixels can be identified and masked (or assigned
interpolated values) by comparing each pixel to the median pixel value
in an image, across a sequence of images.

Other inhomogeneities come from spatial variations in the quantum
efficiency, and optical throughput of the imaging system.  These
effects can be calibrated by imaging a source of uniform brightness.
In astronomy, this is done with a ``flat-field'' source.  In
directional Dark Matter detection, this can be done by uniform
irradiation of the TPC with a gamma calibration source (\eg{}
\isotope{Co}{57} or \isotope{Fe}{55}) \cite{DMTPC_4shins}.  This latter technique also
corrects for spatial variations in the gas gain of the amplification
region.

Stability of the CCD readout over time is also monitored and calibrated.
For example, the CCD bias level and dark rate may evolve over time.
This can be corrected with periodic dark frames, or with regions of
the image that contain no events (possible because in any given event,
the vast majority of the CCD pixels are not illuminated by a recoil
track).  One can measure the average pedestal value of the non-hit
pixels and subtract this image mean from all pixels in order to
correct for short-time-scale temporal variations~\cite{Ahlen:2010ub}.
 
\subsubsection{Practical challenges}
DMTPC has found that the main challenge to CCD robustness comes from
the shutter; on several cameras the shutter has failed mechanically
after of order 10$^6$ exposures.  If operating with 1\,s exposure
time, then such a shutter would fail within two weeks.
Because there is no ambient light when coupled to the vacuum chamber,
the CCD can in principle operate without a shutter.  The main drawback
is that a scintillation event that occurs during chip readout produces
image artifacts that must be handled during the analysis.  It would
likely be advantageous to develop a shutter with extended lifetime
(10$^8$ cycles).
To address this issue, DMTPC has implemented a mechanically separate shutter that is triggered electrically by a signal from the camera. This is a common solution for large-format astronomical sensors, and allows for changing a shutter without modifying the camera or accessing the low-pressure gas detector volume.

Another issue is the degradation of CCD readout electronics in
underground environments.  For example, at the Waste Isolation Pilot
Plant facility, the DMTPC collaboration has observed anomalous noise
behaviour in one of the sensors that had been operating underground in
a salty atmosphere for two years.  This issue is likely generic to all
electronics, but straightforward to address with filtered electronics
enclosures and environmental control.

\subsection{Backgrounds associated with optical readout}
Optical readouts have the advantage of being located outside of the
vacuum chamber.  This means that there is no concern about the readout outgassing into the target and modifing the gas properties or
producing recoiling radon progeny in or near the active detector volume.
For these reasons the radiopurity requirements on the readout are
relaxed, though attention must be paid to potential sources of
neutrons.\footnote{For example, in the COUPP 4\,kg detector, the piezo
  electric sensors, which were external to the detector volume, were a
  source of neutrons that limited the detector sensitivity
  \cite{Fustin_thesis}.}

However, there are a number of backgrounds associated with
radioactivity passing through the sensor, such as from cosmic rays or
from radioactive decays within the sensor housing, that can produce
direct interactions with the CCD~\cite{Ahlen:2010ub}.  These events
must be discriminated against, either using pattern recognition
analysis in the CCD data, or by requiring a coincidence between
optical information from the CCD and ionization information from,
\eg{} a charge readout channel on the TPC anode~\cite{Lopez:2013ah},
or PMT readout.

Another background population is residual bulk images (RBI), also
called ghost-images \cite{JanesickBook2001}.  When the CCD is exposed
to intense illumination (\eg{} from a spark in the detector), some
photons (especially ones with long wavelength) can penetrate deep into
the silicon to generate photoelectrons in the depletion region.  These
photoelectrons diffuse thermally, eventually reaching the potential
well of a pixel, and will appear in the digitized image as a ghost of
the original intense illumination.  The intensity of the RBI signal is
proportional to the exposure time, and fades from exposure to exposure
with a time constant of minutes to hours, depending on the chip
temperature.  RBI from sparks can mimic nuclear recoils, but these
image artifacts can be tagged by looking for a signal that is present
at the same location on the chip across a series of exposures.

\subsection{A high-resolution GEM-based TPC with CCD readout}
\label{sec:optical_unm}
An R\&D program was undertaken at the University of New Mexico \cite{Phan:2015pda} to study the properties of electron and nuclear recoil tracks in a CCD-based TPC, in regards to discrimination and directionality. For this work, both high spatial resolution and signal-to-noise were essential.  A cascade of Gas Electron Multipliers (GEMs) enabled both high gas gains at low pressures, and fine granularity for spatial resolution.

Gas gains of $>10^5$ were achieved in 100\,Torr \cff. The resulting signal-to-noise was high enough to image 5.9\,keV electron tracks from \isotope{Fe}{55} X-rays, and use them to obtain an energy spectrum. At even lower pressures (40\,Torr), comparably high gas gains were achieved using Thick GEMs (THGEMs). At these pressures, tracks from 5.9\,keV electrons were resolved.

\subsubsection{The GEM-based Optical Detector}
\begin{figure}
 \centering
\includegraphics[width=0.45\textwidth]{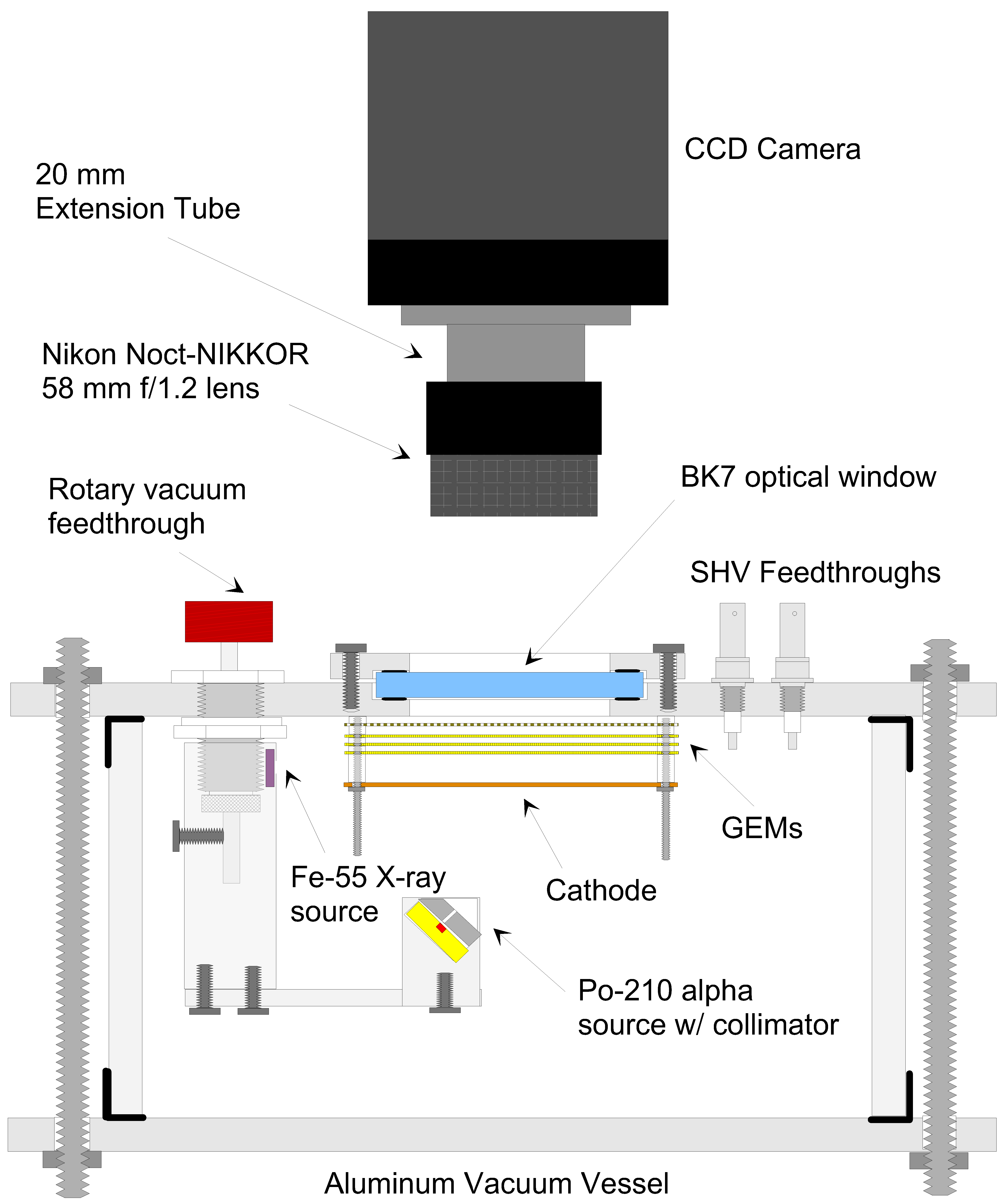}
   \hfill
   \includegraphics[width=0.45\textwidth]{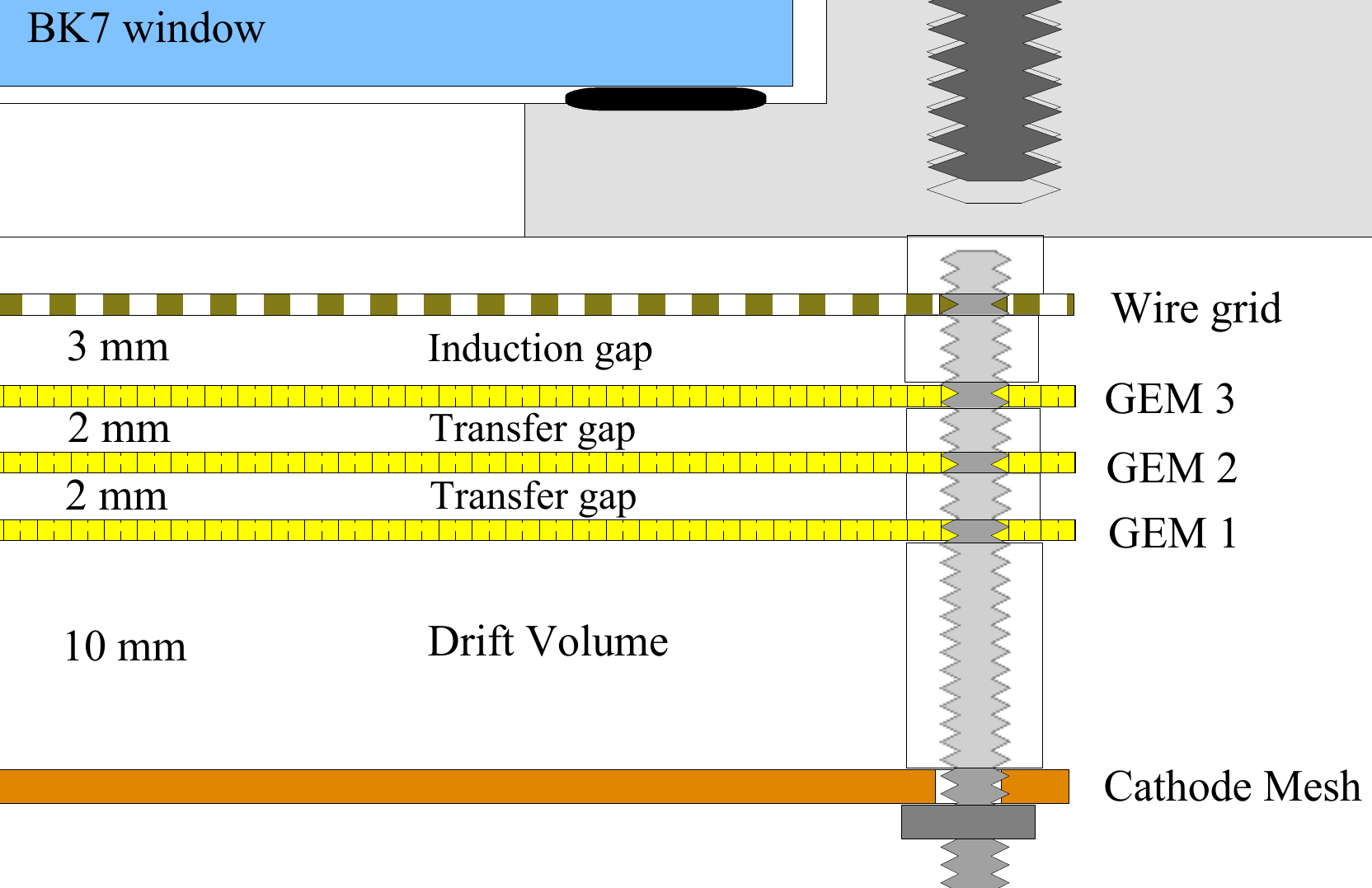}
   \caption[]{Left: Schematic of the CCD detector showing the relative positions of detector components.  The optical system, consisting of the CCD camera and lens, sits outside the vacuum vessel and images the central $2.8 \times 2.8$\,cm$^2$ region of the top-most GEM surface.  Right: Close-up view of the detection volume and GEM stack showing the relevant detector dimensions.}
\label{fig:unm_detector}
\end{figure}

A schematic of the detector is shown in Figure \ref{fig:unm_detector}, and a detailed description is provided in Ref.~\cite{Phan:2015pda}. For the 100\,Torr data the detector operated with a cascade of three standard, $7\times7$\,cm$^2$ thin GEMs manufactured at CERN with a 140\,$\mu$m pitch hexagonal hole pattern.
At lower pressures (down to 40\,Torr), a single or double stack of THGEMs (also manufactured at CERN) replaced the three standard GEMs.
Several THGEMs were evaluated, with thicknesses of 0.4\,mm and 1\,mm.  
The hexagonal-patterned holes had diameters and pitches of 0.3\,mm and 0.5\,mm, which was the smallest that CERN could make.
The induction gap was defined by a wire grid plane with 1\,mm pitch placed 3\,mm above the last GEM in the cascade (GEM3 in Figure~\ref{fig:unm_detector}). The grid was made from 20\,$\mu$m diameter gold-plated tungsten wires.
The cathode, placed 1\,cm below GEM1, was a $7\times7$\,cm$^2$ copper mesh made from 140\,$\mu$m diameter wires with a 320\,$\mu$m pitch.

The detector was housed inside an aluminum vacuum vessel and calibrated using \isotope{Fe}{55} (5.9\,keV X-rays) and \isotope{Po}{210} (5.3\,MeV alphas) sources, both mounted inside the vacuum vessel and each having the capability of being activated or disabled remotely.
The scintillation light from the final GEM (GEM3) was imaged through a BK-7 glass window positioned above the anode grid.

A back-illuminated CCD camera was mounted on top of the vacuum vessel and coupled to a fast 58\,mm $f/1.2$ Nikon Noct-NIKKOR lens.
The camera was a Finger Lakes
Instrumentation (FLI) MicroLine ML4710-1-MB, using a back-illuminated $1024 \times 1024$
array with $13\times 13$\,$\mu$m$^2$ square pixels made by E2V (CCD47-10-1-353).  
The peak quantum efficiency of the CCD was 96$\%$ at 560\,nm, and the read noise was 10 e$^-$ rms for a 700\,kHz digitization speed.
The CCD was cooled to $-38^{\degr}$C using a Peltier device, giving a dark current of $< 0.1$ e$^{-}$/pix/sec.
The camera imaged a $2.8$ $\times$ 2.8 cm$^2$ region of the top-most GEM surface, giving a magnification of $m=2.1$ and an geometric acceptance of $\eta=4.5\times 10^{-3}$.

\subsubsection{Gas gain at low pressure}
An \isotope{Fe}{55} \xray{} source was used to measure the effective gas gain and for energy calibration.
All gain measurements were made by reading out the signals from the last GEM electrode (GEM3 in Figure \ref{fig:unm_detector}, or the equivalent if THGEMs were used) with an ORTEC 142IH charge sensitive  preamplifier.  
At a pressure of 100\,Torr, with the triple-GEM configuration, effective gains as high as $3\times 10^5$ were achieved.
At 100\,Torr, a stable gain (no sparking or corona) of $\sim 1\times10^5$ was achieved.
At lower pressures of pure \cff, the triple-GEM detector was found to be unstable due to sparking and THGEMs were used instead. The THGEMs provided excellent gas gains:  $2\times10^5$ in 50\,Torr with a single THGEM, and in 35\,Torr with a double-THGEM stack.
Further details of the detector operation with THGEMs can be found in Ref.~\cite{nphan_ccd2}.

\begin{figure}
\centering
\includegraphics[width=0.45\textwidth]{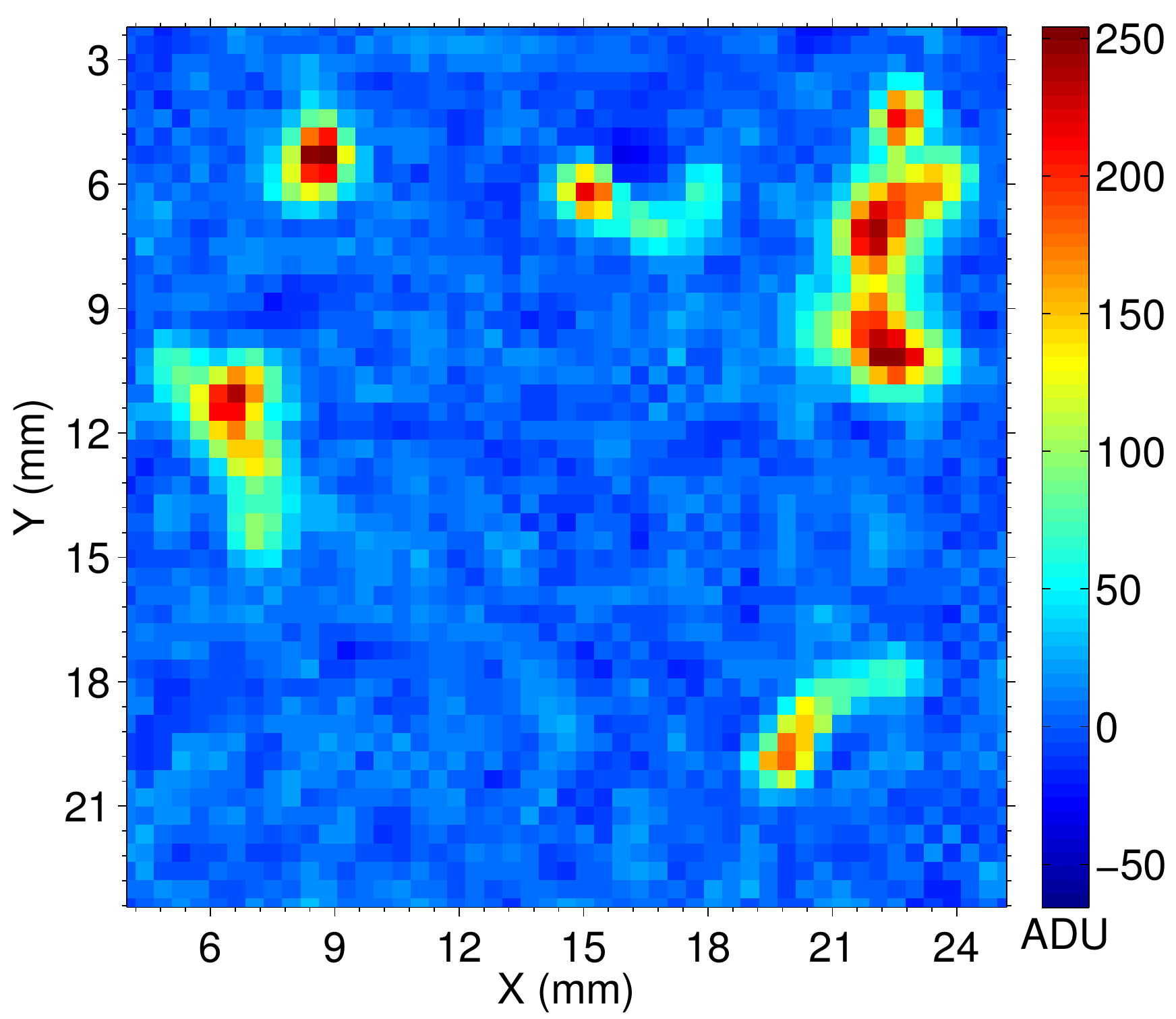}
   \hfill
   \includegraphics[width=0.45\textwidth]{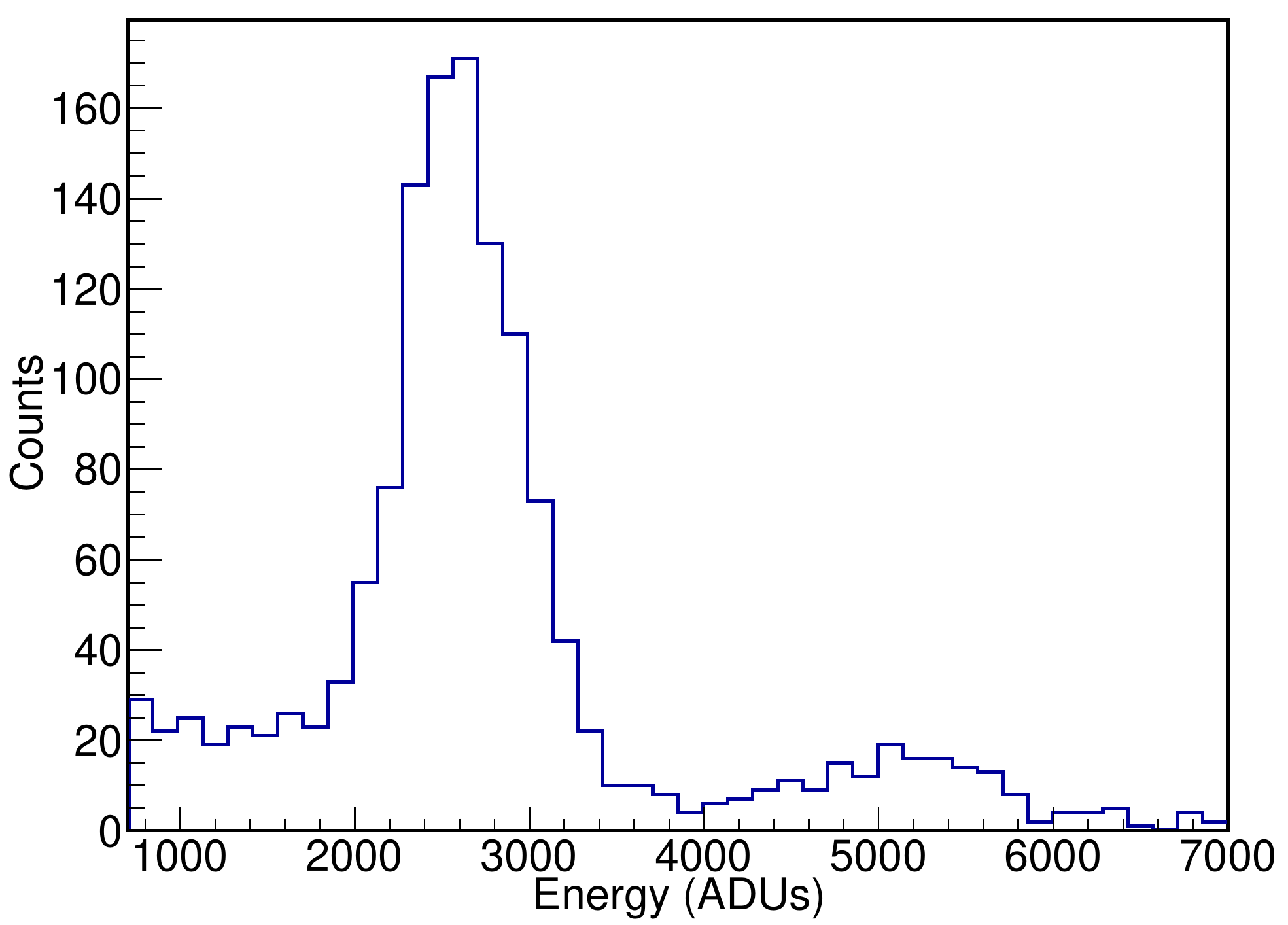}
	\caption[]{Left: Image of 5.9\,keV \isotope{Fe}{55} electronic recoil tracks in 50\,Torr \cff{} at $6 \times 6$ on-chip binning. An averaging filter with a 3 $\times$ 3 block size has been applied to the image to improve signal-to-noise without significantly degrading resolution. At this pressure, the tracks are well-resolved and fluctuations in energy loss and range straggling are also clearly seen. Right: An \isotope{Fe}{55} energy spectrum obtained {\em optically} from CCD imaging of electronic recoil tracks in 100\,Torr \cff.  The data was taken with a maximum stable gain of $\sim 2\times10^5$ and with $6 \times 6$ on-chip binning. The smaller secondary feature to the right of the primary peak is from event pile-up.}
	\label{fig:unmfe55}
\end{figure}

The extremely high ($>10^5$) gas gains achieved at the low, 35--100\,Torr pressures provided ample signal-to-noise to image electron tracks from \isotope{Fe}{55} X-rays. These tracks were used to derive {\em optical} \isotope{Fe}{55} spectra that were used to calibrate the energy scale across a large dynamic range, from low-energy electrons from Compton scattering (with a $^{60}$Co gamma source), to $\sim$5\,MeV alpha tracks. The detection threshold for the 100\,Torr data was found to be $\sim$2\,\kevee.  The energy resolution varied between 30\% and 40\% (FWHM/mean) in all pressures where it was measured.  Some examples of electron tracks from \isotope{Fe}{55} and the resulting energy spectra are shown in Figure~\ref{fig:unmfe55}.  There it can be seen that the 5.9\,keV electron tracks are clearly resolved. In fact, the tracks are resolved even at 100\,Torr, although those at the lower pressures show greater details of energy loss and fluctuations.  Resolving both electron and nuclear recoil tracks at these energies is an important result, both for discrimination and directionality, as discussed in \cite{Phan:2015pda}. For example, this could open up the window for directional low-mass WIMP searches.

\subsubsection{Discrimination and directional sensitivity}
The electron/nuclear recoil discrimination, as well as nuclear recoil track direction sensitivity were studied experimentally using gamma (\isotope{Co}{60}) and neutron (\isotope{Cf}{252}) calibration sources.  Using the reconstructed track properties of events in the detector, an electron recoil rejection factor of $\leq 3.9\times 10^{-5}$ was achieved above 25\,\kevr{} (10\kevee) in 100\,Torr \cff.  The high spatial resolution and high signal to noise were very important to tag electron recoils.  The large fluctuations in $dE/dx$ for an electron recoil means that at a lower signal to noise, only the portion of an electron recoil track with large $dE/dx$ may be visible, and that portion can mimic a nuclear recoil (for further details, see \cite{Phan:2015pda}).  Studies of the corresponding discrimination possible with only a 1D readout found that the discrimination threshold is a factor of three worse than for 2D.  Simulations of this detector suggest that with 3D track reconstruction, the discrimination threshold would be 35\% better than in the 2D case.  

The directional sensitivity of the detector to nuclear recoils was studied using neutrons from a \isotope{Cf}{252} source.  The track axis of nuclear recoils was resolvable down to 40\,\kevr{} (20\,\kevee), meaning that the axial directionality threshold was about a factor of two worse than the discrimination threshold.  The vector (sense) reconstruction threshold was 55\,\kevr, above which the sense is reconstructed correctly more than 50\% of the time.

\subsection{Optical readout in the DMTPC experiment}
The DMTPC group has developed optical readout of TPC detectors with
emphasis on measuring the direction and energy of 100\,GeV/$c^2$ WIMP
dark matter scatters in low-pressure (30--100 Torr) CF$_4$ gas.
Signals from recoiling nuclei are read out with both optical and
charge sensors \cite{Dujmic:2008ut,DMTPC_4shins}.  This section
describes the DMTPC detector R\&D work, and presents directional
sensitivity results from those experiments.

\subsubsection{DMTPC detector configuration}
\begin{figure}
  \centering
  \includegraphics[width=0.45\textwidth]{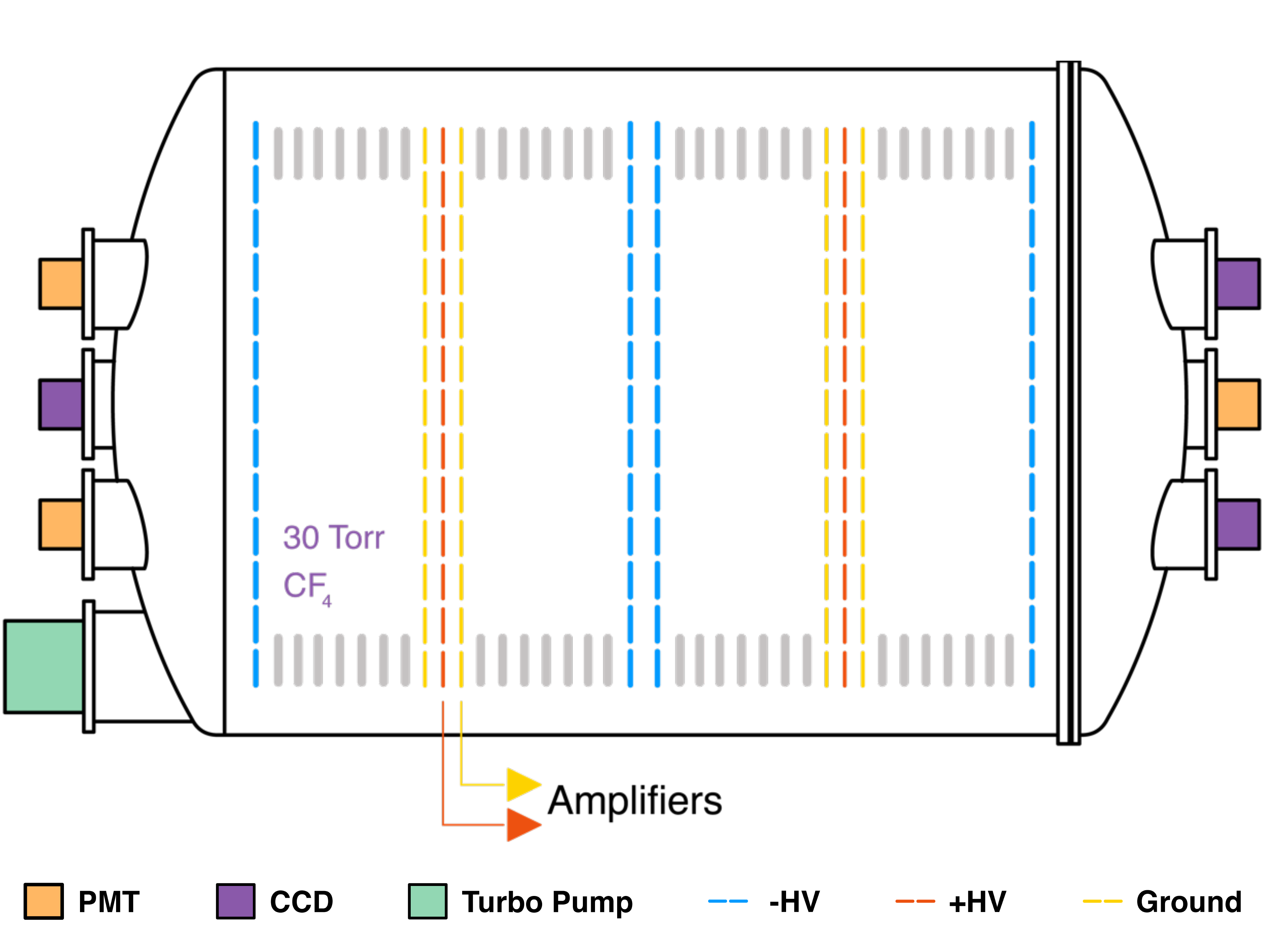}
  \hfill
  \includegraphics[width=0.45\textwidth]{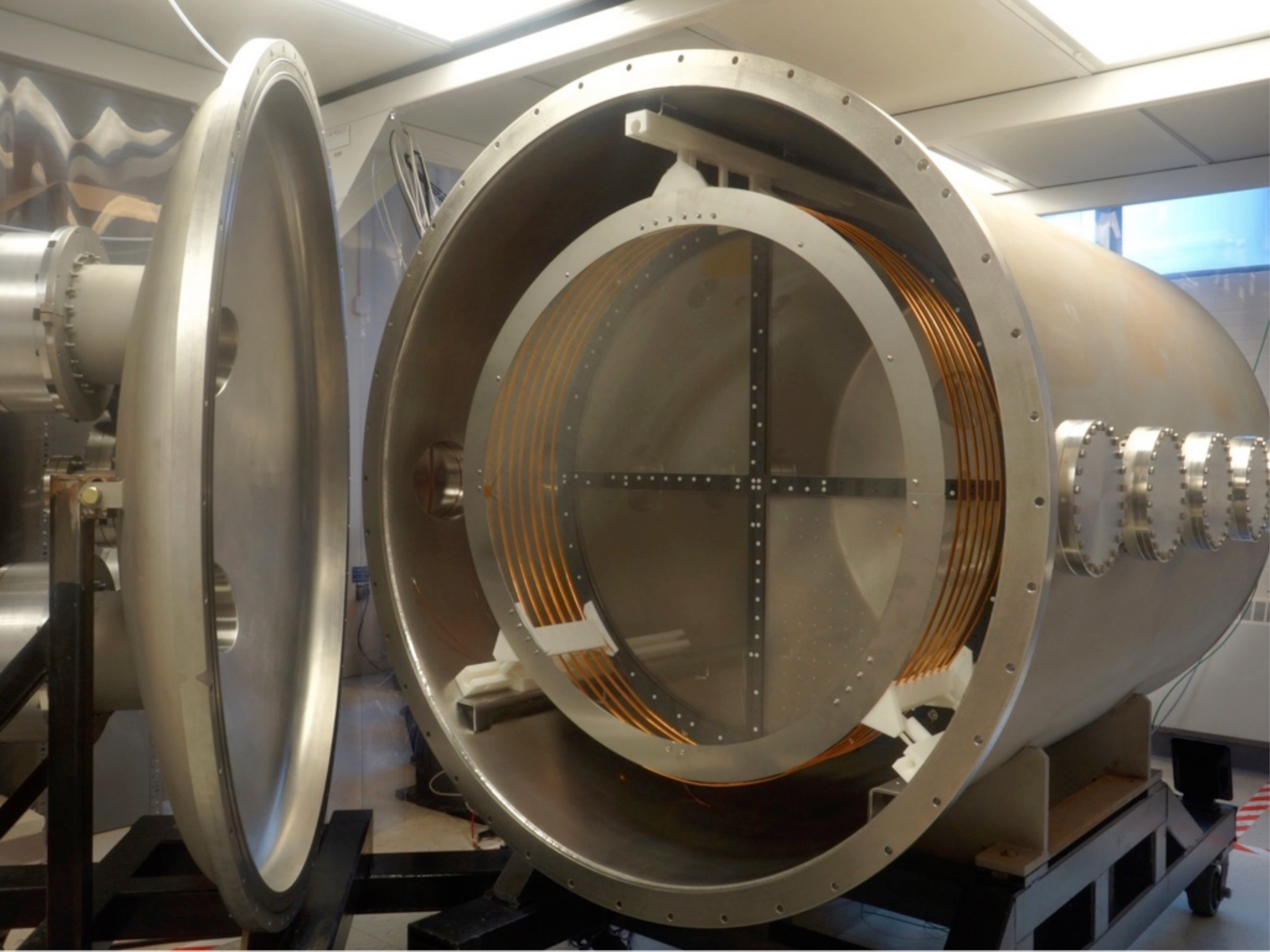}
  \includegraphics[width=0.45\textwidth]{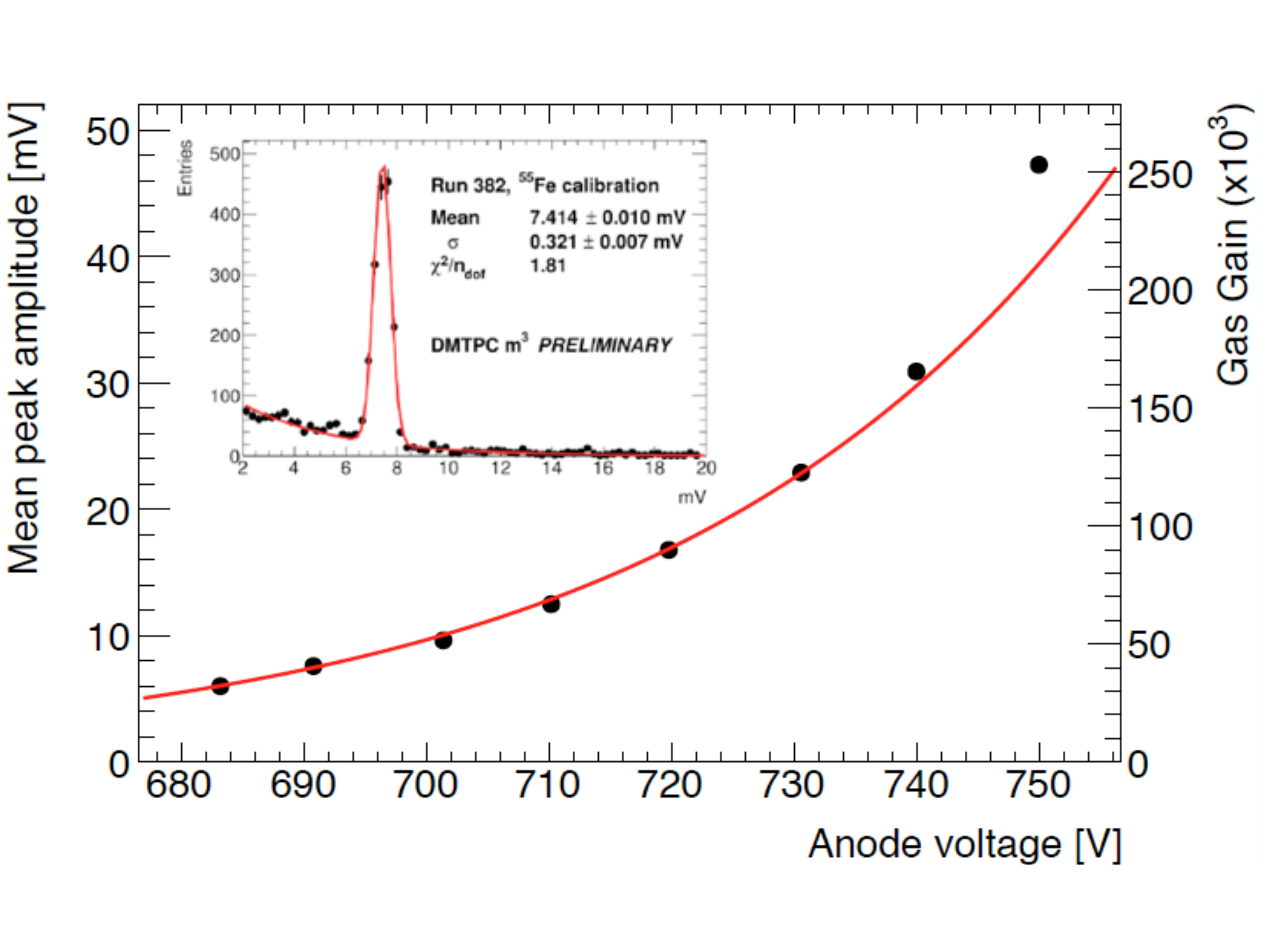}
  \hfill
  \includegraphics[width=0.45\textwidth]{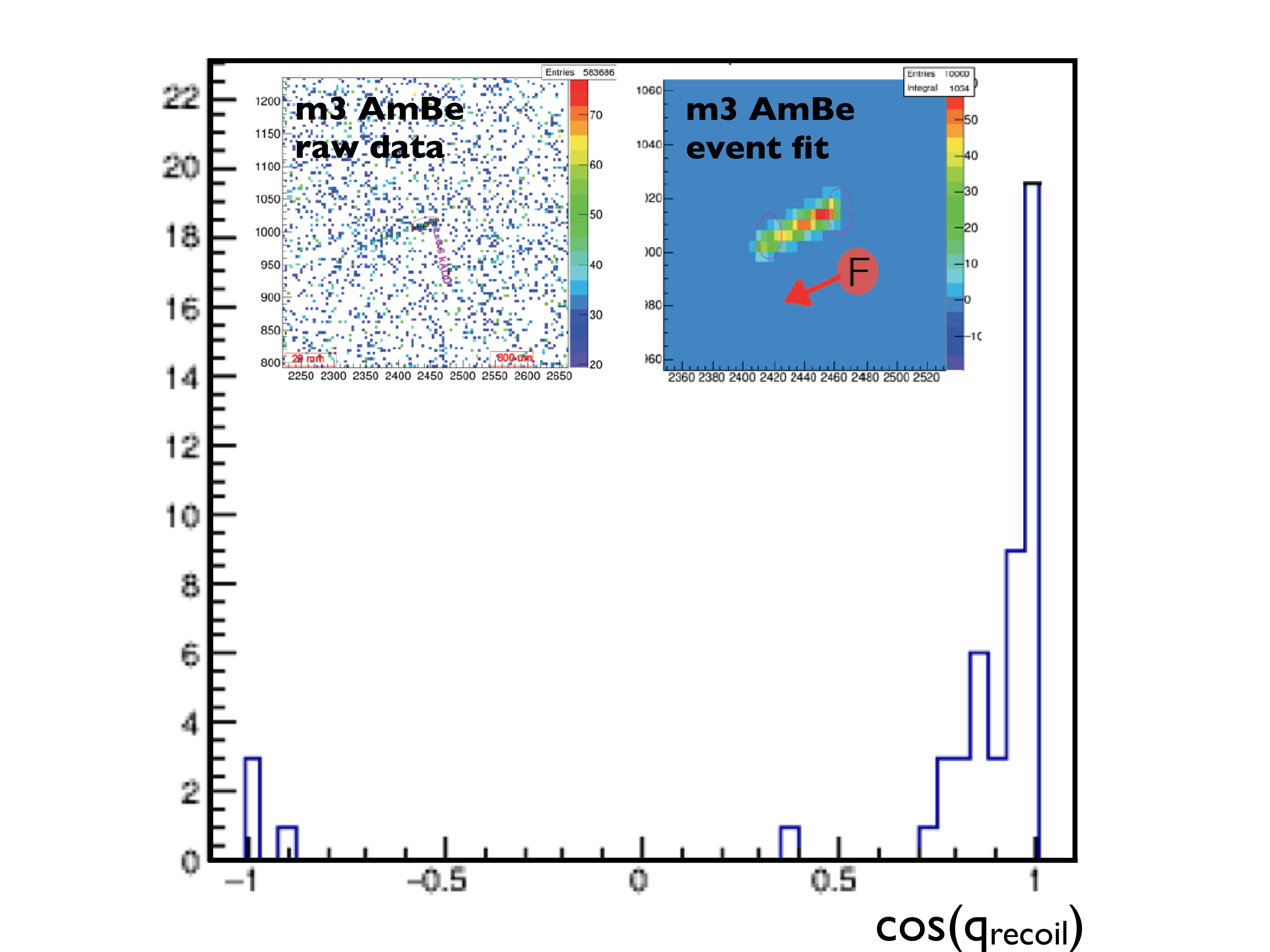}
  \caption{\label{fig:DMTPC_schematic}
Top left: schematic of the m$^3$ DMTPC prototype detector; each of the four TPCs is 1.2\,m in diameter and 27\,cm in length.  One pair of TPCs is imaged by one CCD and four PMTs, while the other pair is imaged by four CCDs and one PMT.  Each camera views the shared anode of each pair of TPCs through the lens, the cathode mesh, and the ground mesh.  Top right:  photo of the detector, showing one TPC.  Bottom left: measured gas gain in the m$^3$ detector as a function of anode voltage.  Inset shows the \isotope{Fe}{55} source spectrum used to measure the peak voltage for the gain calculation.  Bottom right: measured nuclear recoil candidate angle $q_{\mbox{\small recoil}}$ with respect to the neutron source for AmBe neutron calibration data in the m$^3$ detector.  
Inset (left) shows an example nuclear recoil event of candidate energy 60\,\kevr{} in the raw data (with no rebinning or filtering), while inset (right) shows the reconstruction from fitting this event.  Intensity in units of digital CCD counts is indicated by color.}
\end{figure}

The DMTPC optical system employs CCDs and PMTs to image the micromegas-like TPC
amplification region through a vacuum viewport (a window).  DMTPC also
implements charge readout, again external to the vacuum vessel.  The
charge channel comprises a fast amplifier connected via an electrical
feedthrough to the TPC ground electrode, a pre-amplifier to the TPC
anode electrode, and (optionally) a pre-amplifier to an
electrically-isolated outer veto ring of the anode. The primary
tracking information comes from the optical readout, while a
complementary energy measurement comes from the single charge channel
measurement on the TPC anode.  The rise time of the charge mesh
channel is used to distinguish particles with low $dE/dx$, e.g. from
gamma scattering~\cite{Lopez:2013ah}.  The PMT signal is used to
trigger the CCD data collection.  A schematic of the detector readout
is shown in Figure~\ref{fig:DMTPC_schematic}.

DMTPC has built a series of prototypes to develop various aspects of
the optical readout.  The first detector, $(i)$, built in 2007, was a small
MWPC chamber with 2\,mm wire pitch.  This prototype was used to
demonstrate head-tail identification capability using optical
readout~\cite{Dujmic:2008iq}.  In subsequent prototypes, the drift
field is created by a cathode mesh, field-shaping rings attached to a
resistor chain and a ground mesh~\cite{Dujmic:2008ut}.  The cathode
and ground planes of the TPC are formed of meshes with 256\,$\mu$m
pitch in both \x{} and \y{} directions (as is the anode plane in later
prototypes).  This higher pitch amplification scheme was developed in
the second prototype, $(ii)$, which was formed of two optically separate back-to-back TPCs, each imaged by a single CCD camera, for a fiducial volume
of 10L~\cite{Ahlen:2010ub}.  Detector $(ii)$ was also used to
prototype the charge readout scheme.  To scale up to large volumes
while preserving the direction measurement capability, multi-camera
readout of a large amplification region is needed.  The third
prototype $(iii)$ was instrumented with four cameras (and three PMTs) viewing
a common amplification region, for a fiducial volume of 20L~\cite{DMTPC_4shins}.  This detector also explored PMT readout,
and was used to further develop the ground mesh charge readout and
anode veto electrode.  In order to increase the readout volume per
unit cost, an all-mesh amplification region was developed in a small
prototype $(iv)$~\cite{Deaconu:2015vbk}, such that two back-to-back drift regions can be
imaged by one CCD.  These developments were incorporated into
the design of a 1\,m$^3$ fiducial volume detector $(v)$, which DMTPC is
currently commissioning.  The m$^3$ prototype is shown in
Figure~\ref{fig:DMTPC_schematic}.

The typical drift electric field employed is 150--250\,V/cm, chosen to
minimize the transverse diffusion of the drifting
electrons \cite{Christophorou1996} at the chamber operating pressure.
The ground mesh to anode plane separation is 0.3--1\,mm (in various
prototypes), with a typical amplification region field of 15\,kV/cm.
In this arrangement, straggling of the primary ion and diffusion in
the drift region dominate the track width.  The operating anode
voltage is chosen to maximize the gain while limiting the rate of
electronic discharge between the anode and ground plane to a few mHz.
The gas gain in the amplification region is 10$^4$--10$^6$ $(ii-v)$,
measured with an \isotope{Fe}{55} calibration
source~\cite{Ahlen:2010ub,DMTPC_ucladm14}.  The gain dependence on
anode voltage for the m$^3$ prototype operated at 30\,Torr is shown in
Figure~\ref{fig:DMTPC_schematic}.

Scintillation light produced in the amplification region is focused by
a photographic lens onto a CCD.
Prototypes have used various lens/camera combinations, including a
Nikon 55\,mm focal length, $f/1.2$ lens $(ii,iv)$, a Canon 85\,mm,
$f/1.2$ lens $(iii)$, and a Canon 55\,mm, $f/0.95$ lens $(iv)$.
Detector $(v)$ uses four Nikon 55\,mm $f/1.2$ lenses and one Canon
55\,mm $f/0.95$ lens.  Prototypes $(i,ii,iii)$ use Apogee Alta U6
cameras with $1024\,\times\,1024$ pixels ($24\,\times\,24\,\mu$m$^2$
each), and 8\,e$^-$ rms read noise (unbinned, though see
Section~\ref{sec:readnoise}).  The CCD clock rate is 1\,MHz with
16-bit digitization.  Pixels are binned $2\,\times\,2$ or
$4\,\times\,4$ prior to digitization, with a typical binned readout
time of 200\,ms.
Prototype $(iv)$
uses an Andor iKon-L 936 camera with $2048\,\times\,2048$ pixels, each
of size $13.5\,\times\,13.5\,\mu$m$^2$, and 3.9\,e$^-$ rms read noise.
Prototype $(v)$ uses four front-illuminated FLI ProLine 9000 cameras
with $3056\,\times\,3056$ pixels of size $12\,\times\,12\,\mu$m$^2$,
and 10\,e$^-$ rms read noise to image one pair of TPCs (each camera
images a quarter of the amplification region, giving
$\eta=2\times10^{-4}$ for each optical system), and one Spectral
Instruments 1100S camera with a back-illuminated $4096\,\times\,4096$
pixel Fairchild 486 sensor with $15\,\times\,15\,\mu$m$^2$ pixels, and
7\,e$^-$ rms read noise (1\,MHz digitization) to image another pair of
TPCs ($\eta=2\times10^{-4}$ for this optical system as well).  All
four TPCs are situated in the same vacuum vessel, shown schematically
in Figure~\ref{fig:DMTPC_schematic}.

The detector dimensions vary, with drift cage heights from 10\,cm
$(i,iv)$ up to 30\,cm $(iii)$, and diameters of 27\,cm $(ii,iii,iv)$
to 118\,cm $(v)$.  Several of the detectors have back-to-back TPCs
$(ii, iv, v)$, which are optically isolated by a solid anode plane in
$(ii)$, or optically transparent, using a mesh anode plane in
$(iv,v)$.  Using a mesh anode plane allows one CCD camera to image two
drift regions that share a common anode, and therefore doubles the
fiducial volume per readout \cite{Dujmic:2008ut}.  The penalty of this
arrangement is the additional mesh the CCD views the far TPC through,
reducing the photon throughput.  The mesh transparency is 0.8--0.9
\cite{Ahlen:2010ub,DMTPC_4shins}, and therefore optical signals from
the far TPC are attenuated by that corresponding factor.  The area of
the amplification region imaged by a single unbinned CCD pixel is
approximately 50--180\,$\mu$m on a side (the range is for various
prototype detector configurations, where the smallest is $(iv)$ and
the largest is $(ii)$).

The vixel size determines the spatial resolution of the optical
system.  For reference, in 75\,Torr of \cff, a recoiling fluorine
nucleus with 50\,keV kinetic energy travels approximately 1~mm before
stopping.  Therefore with a vixel linear dimension of 150\,$\mu$m,
7--8 points along the track are measured.  Binning $4\,\times\,4$
prior to readout increases the effective vixel linear dimension to
600\,$\mu$m, resulting in 2--3 samples along the track length.

\subsubsection{Optical readout directional sensitivity results}
The directionality achievable with optical readout depends primarily
on the ratio of track length to track width, which in turn depend on
diffusion, straggling and gas pressure.  The ability to correctly measure the track
width and length depend on the signal-to-noise per pixel.

In the track reconstruction, the projected track length on the
amplification plane is calculated by fitting the track with a
two-dimensional track hypothesis based on the Bragg curve convolved
with a Gaussian kernel in width and length to account for
diffusion~\cite{DMTPC_ucladm14}.  The starting guess for the
reconstruction range comes from the maximally separated pixels in the
cluster.  The starting guess for the track angle projected on the
amplification plane, $\phi$, is determined by finding the major axis
angle of an ellipse with the same second moment as the pixels in the
cluster.  The starting guess for the sense of the direction is
estimated from the skewness of the track light yield.  The fit result
for the track projected range, $\phi$, and direction sense, come with
a fit probability, which is used as a cut to select well-reconstructed
tracks.  The angle and sense measurements are made on an
event-by-event basis, rather than statistically.

The directionality results from prototypes $(i)$ and $(ii)$ were
published before the fitting method was developed.  They used the
starting-guess methods above.  In $(i)$, the drift length was 5\,cm,
and the angular resolution was measured with a D-T neutron source to
be $\sim$20$^{\degr}$ above an energy of 100\,\kevee.  The sense of
recoils could be correctly determined 100\% of the time above
200\,\kevee{}~\cite{Dujmic:2008iq}.  In $(ii)$, the drift length is
20\,cm, and the recoil energy and angle reconstruction resolution were
measured to be 15\% and 40$^{\degr}$ at 50\,\kevee{} (80\,\kevr{})
with a \isotope{Cf}{252} neutron calibration
source~\cite{Ahlen:2010ub}.  In $(iii)$, the drift length is
27\,cm~\cite{DMTPC_4shins} and the track reconstruction fitting method
is used to determine the direction and sense.  To measure the axial
angular resolution and sense reconstruction capability, an
\isotope{Am}{241} alpha source was inserted into the vessel at various
orientations such that only the last few hundred keV of ionization was
within the active volume of the TPC~\cite{DMTPC_ucladm14}.  In that
configuration, the axial angular resolution was measured to be
$<$15$^{\degr}$ at 100\,\kevee{}, and the sense of the recoils could
be correctly determined $>$75\% of the time at that
energy~\cite{DMTPC_ucladm14}.  With a high-gain ($3\times 10^5$)
triple mesh amplification region in prototype $(iv)$, axial angular
resolution of 40$^{\degr}$ was measured at 20\,\kevee{} threshold, and
the threshold above which the sense of the recoils could be correctly
determined $>$50\% of the time was measured to be
40\,\kevee{}~\cite{Deaconu:2015vbk}.  The direction measurement is
currently being commissioned in the m$^3$ prototype, using AmBe
neutrons as the calibration source.  The nuclear recoil candidate
event direction distribution for AmBe calibration data in this
detector at 30\,Torr is shown in Figure~\ref{fig:DMTPC_schematic}.

A main result from the studies in prototype $(iv)$ is that in order to
correctly reconstruct the track sense, the ratio of track length to
track width must be $>$3.  This can be achieved at lower energy
thresholds by lowering the pressure.  For example, with the detector
performance of $(iv)$, this can be achieved at 50\,\kevr{} threshold
at approximately 20~Torr~\cite{DMTPC_ucladm14}.  Therefore, to improve
the direction sense measurement, DMTPC is working to lower the gas
pressure.  Stable gas gain of $1.5\times 10^5$ at 30~Torr has been achieved in the m$^3$
prototype (detector $(v)$).  The penalty of lengthened tracks at fixed
recoil energy is a lower ionization density, and therefore lower
signal-to-noise.

Other R\&D results from DMTPC prototypes are beyond the scope of this
review, however for information on backround rejection
see~\cite{Ahlen:2010ub} and~\cite{Lopez:2013ah}.

\subsection{Scalablity of optical readouts}
The great advantage of CCDs for optical readout is their high
granularity, low cost per channel, and ease of data acquisition,
all in a package that is external to the gas volume.  Regarding cost,
the Spectral Instruments 1100S CCD, for example, has 1.7$\times$10$^7$ channels
(pixels) at 0.005\,USD/channel.  The cost is similar for the ProLine 4301E camera.  Many years of commercial R\&D have
led to sensors that are both low-noise and relatively low-cost.  For
directional Dark Matter detection, an ideal imaging sensor would have a
large overall size (to keep $m$ small, and therefore $\eta$ large),
high granularity (to achieve sub-mm vixel sizes), and a low read
noise.  There are a number of drivers for large-format, low-noise CCD
technology beyond Dark Matter applications, such as development for
metrology, machine vision, security, and medicine.  The data
acquisition generally requires nothing more than a USB or Ethernet
connection to a PC.

To instrument very large detector volumes, one might consider
fabricating a set of detector modules, each of $1\,\times\, 1
\,\times\,0.5\,$m$^3$ volume, each read out by a single CCD.  The $1\,
\times \,1$\,m$^2$ dimension is set by the availability of
large-format CCDs, and the practical limit of $f/0.9$ for lens speed.
For example, a $4096\,\times\,4096$ sensor with 24\,$\mu$m square
pixels (this chip exists but is not yet in wide-scale production),
paired with a Canon $f/0.95$ lens (currently in use by DMTPC) would
give $\eta = 6\times10^{-4}$ and a spatial pitch of 0.25\,mm (vixel
size, unbinned).  The ProLine 4301E camera with the same lens would
give $\eta=2\times10^{-4}$, and a spatial pitch of 0.5\,mm.  This can
be compared with the DMTPC 4Shooter detector which had
$\eta=7\times10^{-4}$ and a vixel size of 0.64\,mm (after
$4\,\times\,4$ binning).

The 0.5\,m dimension (two separate drift regions, each 25\,cm long) is determined by the
requirement that the transverse diffusion is less than the track
length for a 50\,\kevr{} recoil \cite{Caldwell:2009si,DMTPC_4shins}.  
This is a pressure- and
electric field-dependent statement, however this assumes that the
reduced drift field (electric field divided by gas number density) is
chosen to minimize the transverse diffusion.
For example, in 30\,Torr \cff{} the transverse diffusion is minimized for a drift field of 120\,V/cm \cite{Christophorou1996}.
All electrodes are made of mesh, with
optical transparency of $\sim$0.9, with the single CCD camera imaging
the central anode (thereby reading out both
TPCs)~\cite{DMTPC_ucladm14}.  The cost for the optical readout (lens \& packaged CCDs) per 0.5\,m$^3$ module would be approximately \$25k--\$50k (depending on
the readout pitch) at current prices.  This does not assume any custom
sensor development or economy of scale, which could potentially reduce
the costs substantially.  A very large detector would consist of many
of these modules.

\subsection{Conclusion}
Offering high-channel-count sensors external to the vessel and a
trival data acquisition system, optical readout is a promising
technology for directional Dark Matter detection.  The main challenge
of optical readouts is the low photon throughput due mostly to
geometric acceptance, though an increased gas amplification factor can
compensate.  The work with GEM and THGEM amplification has shown that
stable gains in excess of $10^5$ are possible, even at low gas
pressures (down to 35\,Torr).  CCDs (or another pixellated, slow
optical readout) provide a 2D projection of the recoil track, though
there are good prospects for recovering the third track dimension
using a companion sensor such as PMTs or charge readout.  Optical
readouts with very high signal to noise and spatial resolution have
achieved electron/nuclear recoil track discrimination down to
25\,\kevr, axial track reconstruction down to 40\,\kevr, and track
sense recognition down to 55\,\kevr, all in a low-pressure gas
(100\,Torr).
Several optical TPCs have been built for Dark Matter searches.  A
detector with a volume of a cubic meter is currently under
commissioning by the DMTPC group.  The upcoming deployment of this
detector underground will provide valuable information about
directional sensitivity and signal-background discrimination at
low-energy in a large-scale detector, as well as the use of a single
camera to image an amplification region with an area of a square
meter.  The prospects for scaling optical readouts to larger volumes
have been helped by the demonstration of triple-mesh amplification
regions with high gains ($>$10$^5$), and would be further improved
with the availability of large-format CCD chips ($>$50\,mm).

\section{Nuclear emulsions}
\label{sec:emulsions}
\subsection{Super-high resolution nuclear emulsion} 
A proposal for a pilot experiment with an emulsion-based detector has been recently proposed by the NEWS Collaboration (Nuclear Emulsions for WIMP Search) \cite{Aleksandrov:2016fyr}.  A nuclear emulsion is a kind of photographic film that can record charged particle tracks with very high spatial resolution. Emulsions consist of silver-halide crystals dispersed in a polymer layer, typically gelatin. Silver-halide crystals are semiconducting with a band gap of 2.7\,eV, and work as a sensor to detect charged particles. An essential step in this detection mechanism is the formation of latent image specks, which are silver clusters several nanometers in size.  These specks are formed by the following reaction between ionized electrons and interstitial silver ions \cite{Powell,Broadhead}, with $n\,\geq\,4$ for typical latent image specs:

\begin{displaymath}
Ag^{+}+e^{-} \rightarrow Ag_{1}  \\
\end{displaymath}
\begin{equation}
Ag_{1}+ Ag^{+}+e^{-} \rightarrow Ag_{2} \cdots Ag_{n}.
\end{equation}
Through a chemical treatment called development, the latent image specks become silver grains.  The size of a silver grain after development is typically around 50\,nm.  A nuclear recoil can be reconstructed by imaging sets of these silver grains.

\begin{figure}
 \centering
 \includegraphics[width=0.9\textwidth,clip]{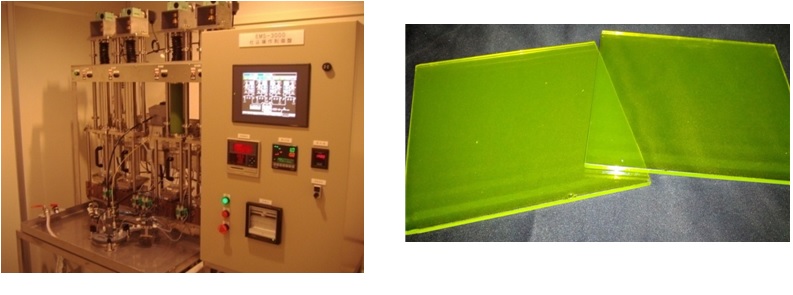}
 \caption{Left: Self-production machine for nuclear emulsion gel.  Right: NIT device coated onto a glass base.}
 \label{fig:NIT}
\end{figure}

The spatial and angular resolutions achievable with an emulsion are determined by the silver halide crystal size and density.  A super-high-resolution nuclear emulsion called the ``Nano-Imaging Tracker'' (NIT) was developed at Nagoya University (Japan) in 2010.  The mean crystal size in the NIT is 40\,nm (6\,nm standard deviation) \cite{FineGrained}, and the overall density of the device is 3.3\,g/cm$^{3}$. Figure\,\ref{fig:NIT} shows the production machines for nuclear emulsions and the NIT. The expected intrinsic tracking performance, taking into account the crystal size distribution and assuming 100\% quantum efficiency (QE), has been simulated with SRIM \cite{SRIM}. The ideal intrinsic direction-sensitive efficiency and angular resolution defined from only the crystal size and density are about 50\% and $30^{\degr}$
for C recoils of 20\,\kevr.

Typically, optical microscopes are used to read out nuclear emulsions. The efficiency in the signal detection, and the discrimination power between signal and background, depends critically on the threshold for detectable track length, and the detection of very short tracks with an optical microscope is very challenging.  Nevertheless, ongoing R\&D has improved the scanning speed and efficiency for all track orientations \cite{fast_scanning,Alexandrov:2016tyi}, making optical microscopy very attractive.

\subsection{Readout strategy for NIT }
\begin{figure}
 \centering
 \includegraphics[width=0.9\textwidth,clip]{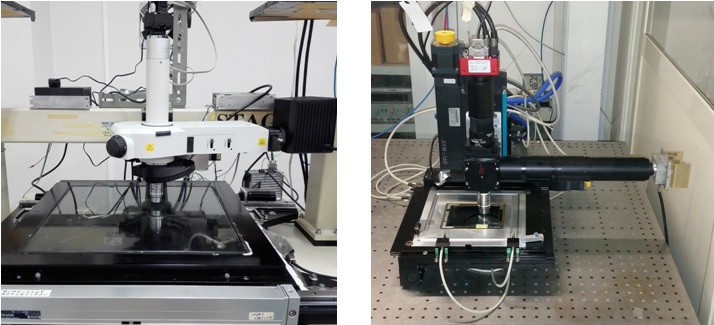}
 \caption{Operational prototype optical microscope systems developed for directional dark matter detection.  Left: A system at the Nagoya University.  Right: The system used at both Naples and LNGS.}
 \label{fig:OP_MS}
\end{figure}

The readout for very short tracks in the NIT uses a multi-level trigger or filter system based on microscopy.
The first step uses fully automated, high-speed optical microscopy, based on techniques developed for the OPERA experiment \cite{EU_scanning,JP_scanning}. The resolution of optical microscopy is limited by the Rayleigh criterion; however, shorter tracks can be distinguished from noise using shape recognition after expanding the emulsion \cite{shape_analysis}.
An optical system capable of shape recognition was constructed at the Nagoya University, as well as at the  University of Napoli and the Gran Sasso National Laboratory (LNGS). Pictures of these prototype systems are shown in Figure\,\ref{fig:OP_MS}.
An ellipse is fit to the optically reconstructed track, and track parameters such as the lengths of the minor and major axes, brightness, and number of pixels making up the cluster are reconstructed.
Candidate events are selected by applying a cut on the ratio of the lengths of the major and minor axes.  The optically measured ellipticity is correlated with the true track length.  The relationship between the two is measured using \xray{} microscopy, which has a resolution of a few tens of nanometers \cite{XMS}. After elliptical shape recognition with optical microscopy has been used to select candidate events, further discrimination is achieved by a higher precision shape analysis of the optical images.  Finally candidate tracks are confirmed by super-high resolution microscopy techniques.

\subsection {Submicron track selection performance } 
A prototype readout system that achieves sub-micron spatial resolution has been constructed for directional dark matter detection. A standard mechanical stage controls motion in the \xy{} plane, while a piezo actuator controls motion in the \z{} direction. Epi-illumination optics were adopted to obtain sufficient contrast to distinguish several 10\,nm silver grains, which is essential for NIT readout. Optical images were taken with a CMOS camera (4\,MPix, 180\,fps). The field of view (FOV) was $110 \times 110$\,$\mu$m$^2$, and the effective pixel size was 55\,nm for a 100$\times$ objective lens. The prototype system was designed to analyze 50\,g of emulsions per year.
 
The spatial resolution of the optical imager is an important factor in the efficiency of candidate selection. In the current prototype, an objective lens of high numerical aperture (NA=1.45) and short wavelength (450\,nm) has been installed. The resulting effective spatial resolution determined from the point-spread-function (PSF) is 230\,nm.   

\begin{figure}
 \centering
 \includegraphics[width=0.9\textwidth,clip]{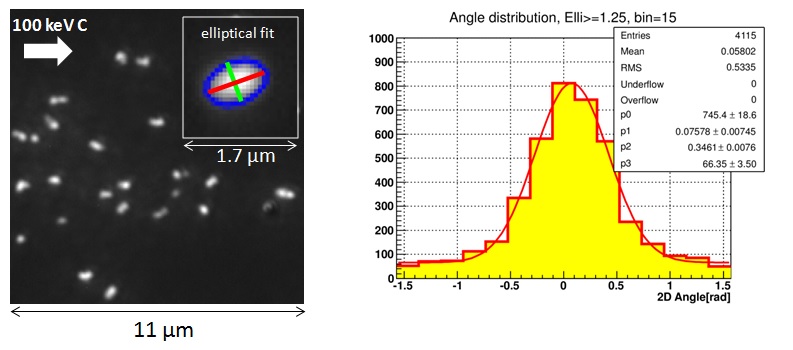}
 \caption{Left: Optical microscope image of 100\,keV C ion tracks.  Right: The corresponding distribution of major axis orientation determined from elliptical shape fitting for events whose ratio of lengths of major and minor axes exceeds 1.25.}
 \label{fig:C100}
\end{figure}

The detection performance has been evaluated experimentally using an ion-implant system to implant monochromatic ions (\eg{} C, O, Kr, F, B, \etc) with energies in the range 10 -- 200\,\kevr, and with uniform direction (angular spread $< 0.6^{\degr}$).
The detection efficiency has been evaluated for C ions introduced from a CO$_{2}$-Ar gas mixture. Absolute tracking efficiencies for carbon ions at 60\,keV, 80\,keV, and 100\,keV is 30\%, 61\% and 73\%, respectively, with a systematic error of about 5--10\%.  These values are consistent with the simulations after taking into account the readout efficiency with the SRIM and silver halide crystal size and density effects, assuming 100\% QE. This consistency indicates that the current device has 100\% QE for carbon ion energies larger than 60\,\kevr, without tuning the background rejection power.  A comparison between data and simulation reveals that the angular resolution is primarily limited by the emulsion crystal size and by straggling, not microscope spatial resolution. The track detection efficiency is therefore limited by pixel resolution, and so finer pixel processing (\eg{} 22\,nm/pix) should improve the track detection efficiency from 30\% to more than 50\% for carbon ion energies of 60\,\kevr.  Sample track images, and the distribution of reconstructed track angles for 100\,\kevr{} carbon ions are shown in Figure\,\ref{fig:C100}.  The measured angular resolution of $20^{\degr}$
includes not only the effects of the intrinsic angular resolution of the imaging system, but also the scattering experienced by low-energy recoils (straggling) \cite{Aleksandrov:2016fyr}.

The elliptical fitting analysis of optical images works for event selection, but a more effective shape analysis after elliptical selection is under development.  Multivariate and neural network analyses are also promising as second triggers for events selection. Such studies are also underway.

\subsection{\xray{} microscope system}
\begin{figure}
 \centering
 \includegraphics[width=0.9\textwidth,clip]{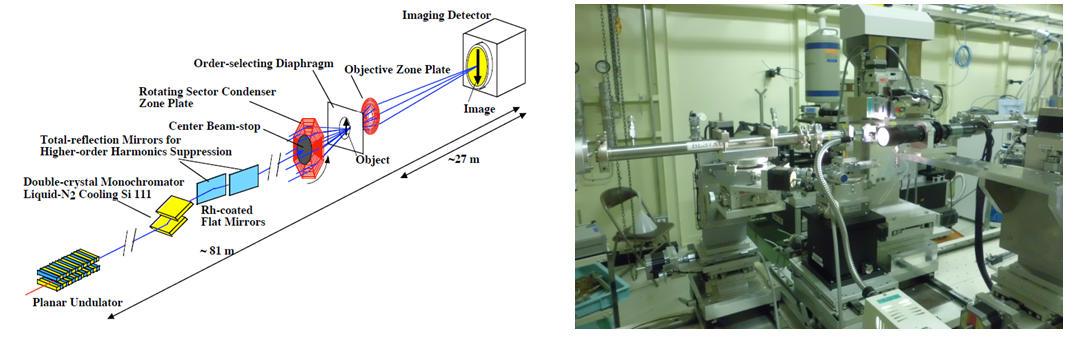}
 \caption{Schematics and picture of a hard \xray{} microscope system at BL37XU, SPring-8.}
 \label{fig:Xray_MS}
\end{figure}

A  hard \xray{} microscope system is one tool used to confirm that candidate events selected via optical microscopy are nuclear recoils. The hard \xray{} microscope provides non-destructive sensing, with a higher spatial resolution than optical microscopy (optical microscopy cannot yet resolve individual grains). The hard \xray{} microscope system was constructed in the BL37XU line of SPring-8 \cite{SP8}.  It uses a Fresnel zone-plate as a condenser, and a set of objective lenses and a CMOS camera to image light converted by a thin phosphor screen made of fine Tb-doped Gd$_{2}$O$_{2}$S powder. In addition, by adopting the phase-contrast method using Zernike phase plates, higher contrast imaging of small silver grains was obtained. The optics were tuned to \xray{} energies of 6 or 8\,keV (6\,keV was used more often because it provides higher contrast).  Measurement of a line-and-space test pattern  made from 100\,nm-thick Ta revealed an effective spatial resolution better than 70\,nm. Figure\,\ref{fig:Xray_MS} shows the scheme and picture of the experimental site.  
  The system includes a mechanical drive stage for semi-automatic scanning and acquisition of coordinate information, moving to the coordinates of candidate events selected by the optical microscope.  The cross-calibration of the two reference systems was achieved thanks to mask patterns printed on the film.  A matching accuracy better than 5\,$\mu$m in a FOV of 20 $\times$ 20\,$\mu$m$^2$ was sufficient for our scientific goals.

Combining the optical and \xray{} microscope system, a readout efficiency of 50\% was achieved for 120\,nm tracks identified by elliptical shape selection with an elliptical cut-off of 1.25, optical resolution of 230\,nm and effective pixel size of 55\,nm \cite{XMS}. Tests with a smaller pixel size have shown even better results.

\subsection {Very-high-precision analysis method using plasmon resonance}
Plasmon resonance is the collective oscillation of free electrons in response to an external electric field.  A localized surface plasmon resonance (LSPR) is an important effect in this system.  Free electrons in nano-metallic particles have a natural frequency due to the binding force responding to displacement generated by electrostatic attraction in an external electric field. For the tens of nanometer-scale silver nano-particles that constitute tracks in a NIT, this creates a resonance condition at visible wavelengths \cite{plasmon}.  In addition, if assuming a non-spherical structure, like an ellipsoid body, the dipole moment depends on the direction, which can be seen by the angle of linear polarization. For example, when the linear polarization aligns with the ellipsoid's major axis, the resonance wavelength tends to be longer than when the polarization aligns with the minor axis.  Thus, plasmon resonance provides information at smaller spatial scales than does standard optical imaging.

\begin{figure}
 \centering
 \includegraphics[width=0.9\textwidth,clip]{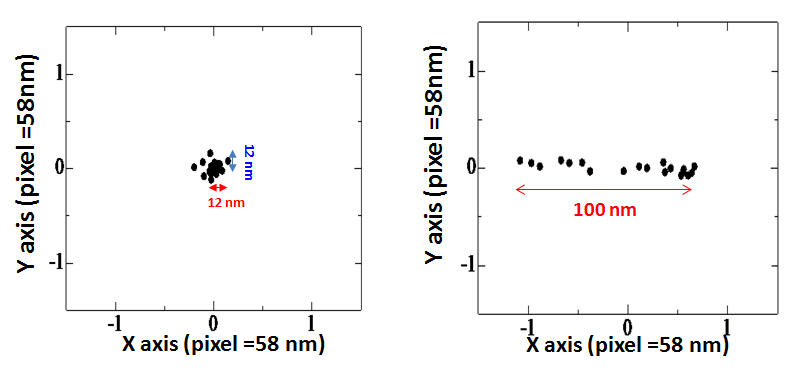}
 \caption{Analysis of polarization dependence due to the plasmon effect. Left: measured positions of simple silver nano-particles.  Right: measured carbon recoil track. }
 \label{fig:plasmon}
\end{figure}

Analysis of the plasmon response for different polarizations should allow for signal-background discrimination and the reconstruction of shorter tracks because the silver grains that form the track have a filamentary structure and encode information about the energy deposited by the particle that passed through the crystal. This is attributed to differences in the number and size of latent image specks on silver halide crystal. For example, developed silver grains due to accidental noise or low $dE/dx$ particles like $\gamma$/$\beta$-ray backgrounds, tend to be small and show spherical structures. In contrast, high $dE/dx$ particles, like nuclear recoils, tend to have very complicated filamentary structure, with several large latent image specks expected to form on the crystal. This difference will be distinguishable by the polarization effect in the plasmon resonance. A super-high-resolution analysis method was proposed using position displacement due to plasmon resonance peak differences, because resonance peaks are not simultaneous, and the position displacement can be obtained for specific wavelengths. The spatial resolution is defined by the position accuracy of 10\,nm, which is possible because this system is not limited by the Rayleigh criterion. Shorter tracks can be distinguished by this effect easily from simple silver clusters, as demonstrated in Figure\,\ref{fig:plasmon}.

\subsection {Future plan for high-speed optical microscope and fine-analysis microscopy  } 
A high-speed microscope system capable of 10\,kg/year scanning throughput has been designed. Such a system would include an imaging sensor with high definition and frame rate, and a $\mu$s pulsed laser or a high intensity pulsed LED synchronized with a camera shutter to acquire images free from the vibrations created by high-speed driving. In addition, since the scanning speed scales with the number of images required, a new objective lens with a large depth of focus (DOF) would be used. A lens with large NA and central obscuring mask is chosen because it provides a larger DOF and higher resolution than a full-aperture lens. For the optics described above, with a FOV of $240 \times 180$\,$\mu$m$^2$ (effective pixel size of 60\,nm), a DOF of 1\,$\mu$m -- 2\,$\mu$m, a 12\,MPix imager operating at 300\,fps, and an image acquisition speed of 5\,Hz (100\,ms for sample translation and 100\,ms for image exposure), the achievable scanning throughput becomes $\sim$2\,kg/year.  The achievable readout speed for an optical microscope readout system using custom-made lenses with a large enough FOV scales linearly with the number of cameras and computers.  The scanning throughput can then be increased from $\sim$2\,kg/year to $\sim$10\,kg/year by using four cameras in parallel (total FOV of 484\,$\times$\,360\,$\mu$m$^2$), and by increasing the DOF with a custom-made objective lens.
 
Plasmon analysis is proposed as the next high-precision analysis of candidate events selected by a high-speed optical readout. The plasmon analysis does not require high-speed scanning power. However, an efficient analysis system using the polarization effect should be constructed. The proposed system is based on a standard epi-illuminated optical microscope, and installing a beam splitter enables simultaneous imaging of different polarization angles by several cameras. These images would be analyzed to exploit the plasmon effect and achieve more efficiently the 10\,nm resolution.

\section{Readouts as used in directional experiments}
\label{sec:expts}
As explained in Sections\,\ref{sec:mwpc} through \ref{sec:emulsions}, each of these readout technologies has been used in the context of a directional Dark Matter search.  In Table~\ref{tab:experiments}, we compile a list of directional Dark Matter detectors that are using these readouts, and present some detector performance metrics.  However, we strongly caution that the table is not a comparison of the readout technologies themselves, because many factors other than readout technology impact the overal detector performance.  Some of the table entries are not a consequence of the readout used, but rather of other particular design choices followed by each experiment.  As an example, the table may seem to imply a link between MWPCs and negative-ion gas operation (since DRIFT uses \cstwo).  But this is just a design choice of the DRIFT experiment.  Conversely, the MIMAC collaboration chooses to operate with an electron-drift TPC (a mixture of \cff, CHF$_3$, and C$_4$H$_{10}$), but micromegas have been successfully operated in electronegative gases \cite{Lightfoot:2007zz}.  Additionally, quantities having to do with track reconstruction (\eg\,angular resolution and directional energy threshold) depend critically on diffusion during drift (and therefore on the total drift length of a given detector, and the particular gas mixture used).  Suffice it to say that there are many factors, separate from the capability of the readout technology, that influence the overall detector performance.

Additionally, members of the directional Dark Matter community are working to define a figure of merit to quantify the sensitivity of a full-scale directional detector to WIMP cross-section as a function of readout technology and detector cost.  The prototypes and technologies summarized in this paper aim at demonstrating the feasibility of track reconstruction at the low energies relevant for dark matter direct detection; we focus on this performance comparison in Table 2.  Most prototype detectors are relatively small and have not yet been through the process of cost-optimization for a full-scale detector design, so a comparison of readout cost goes beyond the scope of this paper.

\newcolumntype{R}{>{\raggedleft\arraybackslash}X}
\newcolumntype{L}{>{\raggedright\arraybackslash}X}

\begin{landscape}
  \noindent
  \captionof{table}{Readout technologies and the collaborations that use them for directional dark matter detection.  Because many factors influence the overall detector performance, this table should not be interpreted as a comparison of the capability of the readout technologies.  In addition to the experiments listed here, the Negative Ion Time Expansion Chamber (NITEC) group has recently begun exploring the use of the TimePix chip with a negative-ion-drift gas for directional DM detection \cite{baracchiniIDM2016}.}
  \begin{tabularx}{\linewidth}{XXXXXXXXXXX}
  \hline \hline
Readout & Experiment & Target\footnote{NI=Negative-ion-drift gas; EG=Electron-drift gas.  It may be possible to use either NI or EG gases with all of the gas-based readouts.} & Granularity\newline $x$, $y$, $z$\footnote{Given as either the pitch of sensing elements, or the time interval between samples.  For DMTPC, the granularity is the size of the imaged area, not the pitch of the CCD pixels.  For NEWS, it is the spatial resolution of the optical microscopy used to scan the emulsions after a dark matter exposure.  NA means no sensitivity in that dimension.} & Area\newline m$^2$ \footnote{Largest practical area for a continuous plane of readout, even if not yet achieved by a directional experiment.  Larger readout areas can be achieved by tiling, potentially with dead space in between.} & Fiducial\newline Volume\footnote{Largest fiducial volume deployed for directional Dark Matter detection.  The DMTPC experiment is currently commissioning a 1\,m$^3$ detector.} & Gas\newline Gain\footnote{Typical operating gas gain.  For pixel chips and optical readout, the gas amplification is distinct from the readout (\eg\,either could be paired with GEMs or micromegas or other devices). Gas gain does not apply to emulsions.} & Energy resolution\footnote{$FWHM/E$, measured with a Fe-55 X-ray source (5.9\,keV), unless otherwise noted.  Energy resolution for energies of interest for dark matter searches may be dominated by the gain resolution of the gas amplification device, and (for the lowest energies) by primary ionization statistics. See study of gain resolution versus energy in Ref.\,\cite{Vahsen:2014fba}.} & Enery threshold\footnote{Because of variations in the way experiments report their data, caution must be used when interpreting these results.  In some cases, the threshold is the smallest energy for a detection of signal.  In others, the threshold is the lowest energy for which a track direction (axis) can be reconstructed 50\% of the time.  Also, some experiments report results in \kevee\, while others report in \kevr.  Since the quenching factor at low recoil energy is not known for some gas mixtures, we present the data as reported by the individual experiments.} & Angular resolution & Sense recognition threshold \\ \hline \hline
MWPC     & DRIFT  & NI\newline\cstwo\newline\cstwo:\cff\newline\cstwo:\cff:O$_2$ & 2\,mm\newline NA\newline 1\,$\mu$s & $2\times2$ & 0.8\,m$^3$ & $\sim$1000  &  42\%\,\cite{Burgos:2009xm} & 50\,\kevr\,\cite{Burgos:2008jm}\newline 30\,\kevr\,\cite{Battat:2016xaw} & unpublished & $50$\,\kevr\,\cite{Burgos:2008jm}\newline 40\,\kevr\,\cite{Battat:2016xaw} \\ \hline
Micromegas & MIMAC  & EG\newline Mix of \cff,\newline CHF$_3$, and\newline C$_4$H$_{10}$ & 0.42\,mm\newline 0.42\,mm\newline 20\,ns & $2\times1$ & 5\,L & $2\times 10^4$  & 22\% \cite{riffardThesis} & $\sim$1\,\kevee\newline\cite{Riffard:2015rga} & unpublished & unpublished \\ \hline
$\mu$PIC   & NEWAGE  &  EG\newline \cff & 0.4\,mm\newline 0.4\,mm\newline 10\,ns & $0.3\times0.3$ & 36\,L & 1000 & 23\% \cite{ref:uPIC_NagayoshiThesis}\newline 47\% at \newline 50\,\kevee\,\cite{ref:NEWAGE_PTEP2015} & 50\,\kevee\newline (directional)\newline\cite{ref:NEWAGE_PTEP2015} & 40$^\circ$ at\newline 50\,\kevee\newline\cite{ref:NEWAGE_PTEP2015} & 75\,\kevee\,\cite{Miuchi:2009vga}  \\ \hline
ATLAS\newline Pixel\newline chips & D$^3$  & EG\newline He+CO$_2$\newline NI\newline\sfsix\,\cite{vahsenPrivate} & 0.05\,mm\newline 0.25\,mm\newline 25\,ns\newline
& NA\footnote{The current-generation pixel chip is $2.0 \times 1.68$\,cm.  The current ATLAS IBL, utilizing this chip, is 0.15\,m$^2$\,\cite{ref:atlasIBL}, and the ATLAS Phase II pixel detector will be 12 to 14\,m$^2$\,\cite{Bates:2015fpj}.} & 50.4\,cm$^3$
& NA & 20\%, 100\,Torr \cff\,\cite{Vahsen:2014mca} & 1--10 primary $e^-$\,\cite{Kim:2008zzi} & $\frac{\sqrt{12}\sigma}{L\sqrt{N}}$ radians\newline\cite{Vahsen:2015oya}\footnote{$\sigma$ is the point resolution, $L$ is the track length, $N$ is the number of pixels hit.}  & unpublished \\ \hline
  Optical    & DMTPC   &  EG\newline \cff & 0.3-0.6\,mm\newline0.3-0.6\,mm\newline NA\footnote{Depends on chosen field of view and CCD binning. See Section\,\ref{sec:opticalSpatialResolution}.} & NA\footnote{The camera field of view can be increased, but at the expense of photon throughput, and therefore signal to noise ratio.  See Section\,\ref{sec:opticalAcceptance}.  Also, gas gain is provided by a separate device (\eg\, a micromegas), which may have a practical size limitation.} & 20\,L\newline (1\,m$^3$) & NA & 35\% at\newline 80\,\kevr\newline \cite{Ahlen:2010ub} & 20\,\kevee\newline \cite{Deaconu:2015vbk} & 15$^\circ$ at\newline 20\,\kevee\newline \cite{Deaconu:2015vbk} & 40\,\kevee\,\cite{Deaconu:2015vbk}  \\ \hline
  Emulsions  & NEWS    &  Solid emulsion & 10\,nm\newline 10\,nm\newline 0.1\,$\mu$m & 100\,g\footnote{Total target mass.  A 10\,kg system is under development.} & NA & NA & unpublished
  & 35\,\kevr\newline Carbon \cite{Aleksandrov:2016fyr} & 13$^\circ$ for\newline 100\,\kevr\newline Carbon\,\cite{Aleksandrov:2016fyr} & unpublished  \\ \hline
\hline
\end{tabularx}
  \label{tab:experiments}
\end{landscape}

\section{Conclusions}
\label{sec:conclusions}
The quest for a competitive WIMP detector with directional sensitivity is well-motivated, as it would give access to an unambiguous signature of Dark Matter.  The benefits of directional detection have been recognized since the observable signature was described in 1988 \cite{Spergel:1987kx}.  The construction of a detector that is sensitive to this signature, however, is technologically challenging.  Most efforts are focused on detectors able to image the short tracks left by the WIMP-induced nuclear recoils. To image these tracks with sufficient resolution to extract the sought directional information, while at the same time sampling the large detector volumes needed for a competitive WIMP signal, puts very strong requirements on the readout technologies. In this paper we have reviewed the various technological options being currently explored in the community. Apart from the nuclear emulsion technique, which aims at imaging recoils in solids, all the other focus on low-pressure TPCs. The TPC readouts being explored include MWPCs, Micromegas, $\mu$PICs, optical CCDs, and pixel chips. Definitive progress with all these devices is being achieved in one or more of the basic requirements: granularity, radiopurity, homogeneity, stability and scalability. However, when combined together the quest remains a formidable one and no clear leading strategy has yet emerged among them all. Work progresses on many fronts and prototypes of medium size (0.1--1\,m$^3$) have been built.  The feedback obtained from their operation will be precious to assess the feasibility of a large-scale experiment with directional sensitivity. The evolution of non-directional detectors in the near future will also be crucial.  The detection, or strong hint, of a WIMP signal in a non-directional detector would stimulate immediate interest in the construction of a large directional detector, not only to cross-check the signal, but also to characterize the local WIMP galactic velocity distribution via ``WIMP astronomy.''

\section*{Acknowledgements}
J.B.R.B. acknowledges the support of the Alfred P. Sloan Foundation (BR2012-011), the National Science Foundation (PHY-1649966), the Research Corporation for Science Advancement (Award \#23325), the Massachusetts Space Grant Consortium, and the Wellesley College Summer Science program.
I.G.I. and the Zaragoza group acknowledges support from the European Research Council (ERC) through the ERC-2009-StG-240054 grant (T-REX project) as well as from the Spanish Ministry of Economy and Competitiveness under grants FPA2008-03456, FPA2011-24058, and FPA2013-41085-P.
This work was partially supported by the European Union's Horizon 2020 research and innovation programme under the Marie Sklodowska-Curie grant agreement No 657751.
This work was supported in part by the Office of High Energy Physics of the U.S. Department of Energy (DoE) under contract DE-AC02-05CH11231.
F.I. acknowledges the support from the Juan de la Cierva program.
D.L. acknowledges support from the National Science Foundation (NSF) under grants 1407773 and 1506329.
This work was partially supported by KAKENHI Grant-in-Aids for Young Scientist(A) (16684004, 19684005, 23684014, 15H05446)\
KAKENHI Grant-in-Aid for Scientific Research on Innovative Area (26104005);
KAKENHI Grant-in-Aids for Scientific Research(A) (16H02189);
KAKENHI Grant-in-Aids for Scientific Research(B) (21340063);
KAKENHI Grant-in-Aids for Challenging Exploratory Research (23654084,15K13485);
KAKENHI Grant-in-Aids for JSPS Fellows;
from the Ministry of Education, Culture, Sports, Science and Technology (MEXT) of Japan.
This work was partially supported by Program for Advancing Strategic International Networks to Accelerate the Circulation of Talented Researchers, JSPS, Japan (R2607).
The Royal Holloway, University of London group is supported by ERC Starting Grant number ERC StG 279980 and STFC grant ST/N00034X/1.
D.P.S.I. acknowledges support from the NSF under grants 1407754, 1103511, 1521027 and 1506237, and undergraduate student support from Occidental's Undergraduate Research Center (URC) summer program.
S.V. and the Hawaii group acknowledge support from the U.S. Department of Homeland
Security under Award Number 2011-DN-077-ARI050-03 and the DoE under Award Numbers DE-SC0007852 and
DE-SC0010504. 
Finally, we thank the anonymous referee for their helpful comments.

\bibliographystyle{elsarticle-num}
\bibliography{DDR}

\begin{thebibliography}{100}
\expandafter\ifx\csname url\endcsname\relax
  \def\url#1{\texttt{#1}}\fi
\expandafter\ifx\csname urlprefix\endcsname\relax\def\urlprefix{URL }\fi
\expandafter\ifx\csname href\endcsname\relax
  \def\href#1#2{#2} \def\path#1{#1}\fi

\bibitem{Bertone:2004pz}
G.~Bertone, D.~Hooper, J.~Silk, {Particle dark matter: Evidence, candidates and
  constraints}, \PHYSREPT 405 (2005) 279--390.
\newblock \href {http://arxiv.org/abs/hep-ph/0404175}
  {\path{arXiv:hep-ph/0404175}}, \href
  {http://dx.doi.org/10.1016/j.physrep.2004.08.031}
  {\path{doi:10.1016/j.physrep.2004.08.031}}.

\bibitem{Jungman:1995df}
G.~Jungman, M.~Kamionkowski, K.~Griest, {Supersymmetric dark matter}, \PHYSREPT
  267 (1996) 195--373.
\newblock \href {http://arxiv.org/abs/hep-ph/9506380}
  {\path{arXiv:hep-ph/9506380}}, \href
  {http://dx.doi.org/10.1016/0370-1573(95)00058-5}
  {\path{doi:10.1016/0370-1573(95)00058-5}}.

\bibitem{Feng:2010gw}
J.~L. Feng, {Dark Matter Candidates from Particle Physics and Methods of
  Detection}, \ARAA 48 (2010) 495--545.
\newblock \href {http://arxiv.org/abs/1003.0904} {\path{arXiv:1003.0904}},
  \href {http://dx.doi.org/10.1146/annurev-astro-082708-101659}
  {\path{doi:10.1146/annurev-astro-082708-101659}}.

\bibitem{Kane:2008gb}
G.~Kane, S.~Watson, {Dark Matter and LHC: What is the Connection?},
  \MODPHYSLETT A23 (2008) 2103--2123.
\newblock \href {http://arxiv.org/abs/0807.2244} {\path{arXiv:0807.2244}},
  \href {http://dx.doi.org/10.1142/S0217732308028314}
  {\path{doi:10.1142/S0217732308028314}}.

\bibitem{Goodman:1984dc}
M.~W. Goodman, E.~Witten, {Detectability of Certain Dark Matter Candidates},
  \PHYSREV D31 (1985) 3059.
\newblock \href {http://dx.doi.org/10.1103/PhysRevD.31.3059}
  {\path{doi:10.1103/PhysRevD.31.3059}}.

\bibitem{Gaitskell:2004gd}
R.~J. Gaitskell, {Direct detection of dark matter}, \ARNPS 54 (2004) 315--359.
\newblock \href {http://dx.doi.org/10.1146/annurev.nucl.54.070103.181244}
  {\path{doi:10.1146/annurev.nucl.54.070103.181244}}.

\bibitem{Wasserman:1986hh}
I.~Wasserman, {Possibility of Detecting Heavy Neutral Fermions in the Galaxy},
  \PHYSREV D33 (1986) 2071--2078.
\newblock \href {http://dx.doi.org/10.1103/PhysRevD.33.2071}
  {\path{doi:10.1103/PhysRevD.33.2071}}.

\bibitem{TuckerSmith:2001hy}
D.~Tucker-Smith, N.~Weiner, {Inelastic dark matter}, \PHYSREV D64 (2001)
  043502.
\newblock \href {http://arxiv.org/abs/hep-ph/0101138}
  {\path{arXiv:hep-ph/0101138}}, \href
  {http://dx.doi.org/10.1103/PhysRevD.64.043502}
  {\path{doi:10.1103/PhysRevD.64.043502}}.

\bibitem{Spergel:1987kx}
D.~N. Spergel, {The motion of the Earth and the detection of WIMPs}, \PHYSREV
  D37 (1988) 1353.
\newblock \href {http://dx.doi.org/10.1103/PhysRevD.37.1353}
  {\path{doi:10.1103/PhysRevD.37.1353}}.

\bibitem{Mayet2016}
F.~Mayet, A.~Green, J.~Battat, J.~Billard, N.~Bozorgnia, G.~Gelmini,
  P.~Gondolo, B.~Kavanagh, S.~Lee, D.~Loomba, J.~Monroe, B.~Morgan, C.~O'Hare,
  A.~Peter, N.~Phan, S.~Vahsen,
  \href{http://www.sciencedirect.com/science/article/pii/S0370157316001022}{A
  review of the discovery reach of directional dark matter detection},
  \PHYSREPT 627 (2016) 1--49.
\newblock \href {http://arxiv.org/abs/1602.03781} {\path{arXiv:1602.03781}},
  \href {http://dx.doi.org/http://dx.doi.org/10.1016/j.physrep.2016.02.007}
  {\path{doi:http://dx.doi.org/10.1016/j.physrep.2016.02.007}}.
\newline\urlprefix\url{http://www.sciencedirect.com/science/article/pii/S0370157316001022}

\bibitem{Ahlen:2009ev}
S.~Ahlen, et~al., {The case for a directional dark matter detector and the
  status of current experimental efforts}, \IJMPA A25 (2010) 1--51.
\newblock \href {http://arxiv.org/abs/0911.0323} {\path{arXiv:0911.0323}},
  \href {http://dx.doi.org/10.1142/S0217751X10048172}
  {\path{doi:10.1142/S0217751X10048172}}.

\bibitem{Cushman:2013zza}
P.~Cushman, et~al.,
  \href{http://inspirehep.net/record/1262767/files/arXiv:1310.8327.pdf}{{Working
  Group Report: WIMP Dark Matter Direct Detection}}, in: {Community Summer
  Study 2013: Snowmass on the Mississippi (CSS2013) Minneapolis, MN, USA, July
  29-August 6, 2013}, 2013.
\newblock \href {http://arxiv.org/abs/1310.8327} {\path{arXiv:1310.8327}}.
\newline\urlprefix\url{http://inspirehep.net/record/1262767/files/arXiv:1310.8327.pdf}

\bibitem{Billard:2013qya}
J.~Billard, L.~Strigari, E.~Figueroa-Feliciano, {Implication of neutrino
  backgrounds on the reach of next generation dark matter direct detection
  experiments}, \PHYSREV D89~(2) (2014) 023524.
\newblock \href {http://arxiv.org/abs/1307.5458} {\path{arXiv:1307.5458}},
  \href {http://dx.doi.org/10.1103/PhysRevD.89.023524}
  {\path{doi:10.1103/PhysRevD.89.023524}}.

\bibitem{Akerib:2015rjg}
D.~S. Akerib, et~al., {Improved WIMP scattering limits from the LUX
  experiment}, \PRL 116~(16) (2016) 161301.
\newblock \href {http://arxiv.org/abs/1512.03506} {\path{arXiv:1512.03506}},
  \href {http://dx.doi.org/10.1103/PhysRevLett.116.161301}
  {\path{doi:10.1103/PhysRevLett.116.161301}}.

\bibitem{Billard:2009mf}
J.~Billard, F.~Mayet, J.~F. Macias-Perez, D.~Santos, {Directional detection as
  a strategy to discover galactic Dark Matter}, \PHYSLETT B691 (2010) 156--162.
\newblock \href {http://arxiv.org/abs/0911.4086} {\path{arXiv:0911.4086}},
  \href {http://dx.doi.org/10.1016/j.physletb.2010.06.024}
  {\path{doi:10.1016/j.physletb.2010.06.024}}.

\bibitem{Mei:2005gm}
D.~Mei, A.~Hime, {Muon-induced background study for underground laboratories},
  \PHYSREV D73 (2006) 053004.
\newblock \href {http://arxiv.org/abs/astro-ph/0512125}
  {\path{arXiv:astro-ph/0512125}}, \href
  {http://dx.doi.org/10.1103/PhysRevD.73.053004}
  {\path{doi:10.1103/PhysRevD.73.053004}}.

\bibitem{Bozorgnia:2011vc}
N.~Bozorgnia, G.~B. Gelmini, P.~Gondolo, {Ring-like features in directional
  dark matter detection}, \JCAP 1206 (2012) 037.
\newblock \href {http://arxiv.org/abs/1111.6361} {\path{arXiv:1111.6361}},
  \href {http://dx.doi.org/10.1088/1475-7516/2012/06/037}
  {\path{doi:10.1088/1475-7516/2012/06/037}}.

\bibitem{Bozorgnia:2012eg}
N.~Bozorgnia, G.~B. Gelmini, P.~Gondolo, {Aberration features in directional
  dark matter detection}, \JCAP 1208 (2012) 011.
\newblock \href {http://arxiv.org/abs/1205.2333} {\path{arXiv:1205.2333}},
  \href {http://dx.doi.org/10.1088/1475-7516/2012/08/011}
  {\path{doi:10.1088/1475-7516/2012/08/011}}.

\bibitem{Billard:2010gp}
J.~Billard, F.~Mayet, D.~Santos, {Exclusion limits from data of directional
  Dark Matter detectors}, \PHYSREV D82 (2010) 055011.
\newblock \href {http://arxiv.org/abs/1006.3513} {\path{arXiv:1006.3513}},
  \href {http://dx.doi.org/10.1103/PhysRevD.82.055011}
  {\path{doi:10.1103/PhysRevD.82.055011}}.

\bibitem{Henderson:2008bn}
S.~Henderson, J.~Monroe, P.~Fisher, {The Maximum Patch Method for Directional
  Dark Matter Detection}, \PHYSREV D78 (2008) 015020.
\newblock \href {http://arxiv.org/abs/0801.1624} {\path{arXiv:0801.1624}},
  \href {http://dx.doi.org/10.1103/PhysRevD.78.015020}
  {\path{doi:10.1103/PhysRevD.78.015020}}.

\bibitem{Billard:2011zj}
J.~Billard, F.~Mayet, D.~Santos, {Assessing the discovery potential of
  directional detection of Dark Matter}, \PHYSREV D85 (2012) 035006.
\newblock \href {http://arxiv.org/abs/1110.6079} {\path{arXiv:1110.6079}},
  \href {http://dx.doi.org/10.1103/PhysRevD.85.035006}
  {\path{doi:10.1103/PhysRevD.85.035006}}.

\bibitem{Green:2010zm}
A.~M. Green, B.~Morgan, {The median recoil direction as a WIMP directional
  detection signal}, \PHYSREV D81 (2010) 061301.
\newblock \href {http://arxiv.org/abs/1002.2717} {\path{arXiv:1002.2717}},
  \href {http://dx.doi.org/10.1103/PhysRevD.81.061301}
  {\path{doi:10.1103/PhysRevD.81.061301}}.

\bibitem{Billard:2010jh}
J.~Billard, F.~Mayet, D.~Santos, {Markov Chain Monte Carlo analysis to
  constrain Dark Matter properties with directional detection}, \PHYSREV D83
  (2011) 075002.
\newblock \href {http://arxiv.org/abs/1012.3960} {\path{arXiv:1012.3960}},
  \href {http://dx.doi.org/10.1103/PhysRevD.83.075002}
  {\path{doi:10.1103/PhysRevD.83.075002}}.

\bibitem{Lee:2012pf}
S.~K. Lee, A.~H.~G. Peter, {Probing the Local Velocity Distribution of WIMP
  Dark Matter with Directional Detectors}, \JCAP 1204 (2012) 029.
\newblock \href {http://arxiv.org/abs/1202.5035} {\path{arXiv:1202.5035}},
  \href {http://dx.doi.org/10.1088/1475-7516/2012/04/029}
  {\path{doi:10.1088/1475-7516/2012/04/029}}.

\bibitem{O'Hare:2014oxa}
C.~A.~J. O'Hare, A.~M. Green, {Directional detection of dark matter streams},
  \PHYSREV D90~(12) (2014) 123511.
\newblock \href {http://arxiv.org/abs/1410.2749} {\path{arXiv:1410.2749}},
  \href {http://dx.doi.org/10.1103/PhysRevD.90.123511}
  {\path{doi:10.1103/PhysRevD.90.123511}}.

\bibitem{Alves:2012ay}
D.~S.~M. Alves, S.~E. Hedri, J.~G. Wacker, {Dark Matter in 3D}, \JHEP 03 (2016)
  149.
\newblock \href {http://arxiv.org/abs/1204.5487} {\path{arXiv:1204.5487}},
  \href {http://dx.doi.org/10.1007/JHEP03(2016)149}
  {\path{doi:10.1007/JHEP03(2016)149}}.

\bibitem{Lee:2014cpa}
S.~K. Lee, {Harmonics in the Dark-Matter Sky: Directional Detection in the
  Fourier-Bessel Basis}, \JCAP 1403 (2014) 047.
\newblock \href {http://arxiv.org/abs/1401.6179} {\path{arXiv:1401.6179}},
  \href {http://dx.doi.org/10.1088/1475-7516/2014/03/047}
  {\path{doi:10.1088/1475-7516/2014/03/047}}.

\bibitem{Grothaus:2014hja}
P.~Grothaus, M.~Fairbairn, J.~Monroe, {Directional Dark Matter Detection Beyond
  the Neutrino Bound}, \PHYSREV D90~(5) (2014) 055018.
\newblock \href {http://arxiv.org/abs/1406.5047} {\path{arXiv:1406.5047}},
  \href {http://dx.doi.org/10.1103/PhysRevD.90.055018}
  {\path{doi:10.1103/PhysRevD.90.055018}}.

\bibitem{O'Hare:2015mda}
C.~A.~J. O'Hare, A.~M. Green, J.~Billard, E.~Figueroa-Feliciano, L.~E.
  Strigari, {Readout strategies for directional dark matter detection beyond
  the neutrino background}, \PHYSREV D92~(6) (2015) 063518.
\newblock \href {http://arxiv.org/abs/1505.08061} {\path{arXiv:1505.08061}},
  \href {http://dx.doi.org/10.1103/PhysRevD.92.063518}
  {\path{doi:10.1103/PhysRevD.92.063518}}.

\bibitem{Ruppin:2014bra}
F.~Ruppin, J.~Billard, E.~Figueroa-Feliciano, L.~Strigari, {Complementarity of
  dark matter detectors in light of the neutrino background}, \PHYSREV D90~(8)
  (2014) 083510.
\newblock \href {http://arxiv.org/abs/1408.3581} {\path{arXiv:1408.3581}},
  \href {http://dx.doi.org/10.1103/PhysRevD.90.083510}
  {\path{doi:10.1103/PhysRevD.90.083510}}.

\bibitem{Belli:1992zb}
P.~Belli, R.~Bernabei, C.~Bacci, A.~Incicchitti, D.~Prosperi, {Identifying a
  'dark matter' signal by nonisotropic scintillation detector}, \NUOVOCIM C15
  (1992) 475--479.

\bibitem{Spooner:1996gr}
N.~J.~C. Spooner, J.~W. Roberts, D.~R. Tovey, {Measurements of carbon recoil
  scintillation efficiency and anisotropy in stilbene for WIMP searches with
  direction sensitivity}, in: {Proceedings, 1st International Workshop on The
  identification of dark matter (IDM 1996)}, 1996, pp. 481--486.

\bibitem{Shimizu:2002ik}
Y.~Shimizu, M.~Minowa, H.~Sekiya, Y.~Inoue, {Directional scintillation detector
  for the detection of the wind of WIMPs}, \NIM A496 (2003) 347--352.
\newblock \href {http://arxiv.org/abs/astro-ph/0207529}
  {\path{arXiv:astro-ph/0207529}}, \href
  {http://dx.doi.org/10.1016/S0168-9002(02)01661-3}
  {\path{doi:10.1016/S0168-9002(02)01661-3}}.

\bibitem{Guillaudin:2011hu}
O.~Guillaudin, J.~Billard, G.~Bosson, O.~Bourrion, T.~Lamy, F.~Mayet,
  D.~Santos, P.~Sortais, {Quenching factor measurement in low pressure gas
  detector for directional dark matter search}, \EASPUB 53 (2012) 119--127.
\newblock \href {http://arxiv.org/abs/1110.2042} {\path{arXiv:1110.2042}},
  \href {http://dx.doi.org/10.1051/eas/1253015}
  {\path{doi:10.1051/eas/1253015}}.

\bibitem{Reinking1986JAP}
G.~F. {Reinking}, L.~G. {Christophorou}, S.~R. {Hunter}, {Studies of total
  ionization in gases/mixtures of interest to pulsed power applications},
  \JAPPLPHYS 60 (1986) 499--508.
\newblock \href {http://dx.doi.org/10.1063/1.337792}
  {\path{doi:10.1063/1.337792}}.

\bibitem{Billard:2012dy}
J.~Billard, F.~Mayet, D.~Santos, {Low energy electron/recoil discrimination for
  directional Dark Matter detection}, \JCAP 1207 (2012) 020.
\newblock \href {http://arxiv.org/abs/1205.0973} {\path{arXiv:1205.0973}},
  \href {http://dx.doi.org/10.1088/1475-7516/2012/07/020}
  {\path{doi:10.1088/1475-7516/2012/07/020}}.

\bibitem{Green:2006cb}
A.~M. Green, B.~Morgan, {Optimizing WIMP directional detectors}, \ASTROPART 27
  (2007) 142--149.
\newblock \href {http://arxiv.org/abs/astro-ph/0609115}
  {\path{arXiv:astro-ph/0609115}}, \href
  {http://dx.doi.org/10.1016/j.astropartphys.2006.10.006}
  {\path{doi:10.1016/j.astropartphys.2006.10.006}}.

\bibitem{Billard:2014ewa}
J.~Billard, {Comparing readout strategies to directly detect dark matter},
  \PHYSREV D91~(2) (2015) 023513.
\newblock \href {http://arxiv.org/abs/1411.5946} {\path{arXiv:1411.5946}},
  \href {http://dx.doi.org/10.1103/PhysRevD.91.023513}
  {\path{doi:10.1103/PhysRevD.91.023513}}.

\bibitem{Phan:2015pda}
N.~S. Phan, R.~J. Lauer, E.~R. Lee, D.~Loomba, J.~A.~J. Matthews, E.~H. Miller,
  {GEM-based TPC with CCD imaging for directional Dark Matter detection}\href
  {http://arxiv.org/abs/1510.02170} {\path{arXiv:1510.02170}}.

\bibitem{Billard:2012bk}
J.~Billard, F.~Mayet, D.~Santos, {Three-dimensional track reconstruction for
  directional Dark Matter detection}, \JCAP 1204 (2012) 006.
\newblock \href {http://arxiv.org/abs/1202.3372} {\path{arXiv:1202.3372}},
  \href {http://dx.doi.org/10.1088/1475-7516/2012/04/006}
  {\path{doi:10.1088/1475-7516/2012/04/006}}.

\bibitem{Charpak:1968kd}
G.~Charpak, R.~Bouclier, T.~Bressani, J.~Favier, C.~Zupancic, {The use of
  multiwire proportional counters to select and localize charged particles},
  \NIM 62 (1968) 262--268.
\newblock \href {http://dx.doi.org/10.1016/0029-554X(68)90371-6}
  {\path{doi:10.1016/0029-554X(68)90371-6}}.

\bibitem{Alme:2010ke}
J.~Alme, et~al., {The ALICE TPC, a large 3-dimensional tracking device with
  fast readout for ultra-high multiplicity events}, \NIM A622 (2010) 316--367.
\newblock \href {http://arxiv.org/abs/1001.1950} {\path{arXiv:1001.1950}},
  \href {http://dx.doi.org/10.1016/j.nima.2010.04.042}
  {\path{doi:10.1016/j.nima.2010.04.042}}.

\bibitem{Anderson:2003ur}
M.~Anderson, et~al., {The STAR time projection chamber: A Unique tool for
  studying high multiplicity events at RHIC}, \NIM A499 (2003) 659--678.
\newblock \href {http://arxiv.org/abs/nucl-ex/0301015}
  {\path{arXiv:nucl-ex/0301015}}, \href
  {http://dx.doi.org/10.1016/S0168-9002(02)01964-2}
  {\path{doi:10.1016/S0168-9002(02)01964-2}}.

\bibitem{Acciarri:2015uup}
R.~Acciarri, et~al., {Long-Baseline Neutrino Facility (LBNF) and Deep
  Underground Neutrino Experiment (DUNE) Conceptual Design Report Volume 2: The
  Physics Program for DUNE at LBNF}\href {http://arxiv.org/abs/1512.06148}
  {\path{arXiv:1512.06148}}.

\bibitem{Battat:2014van}
J.~B.~R. Battat, et~al., {First background-free limit from a directional dark
  matter experiment: results from a fully fiducialised DRIFT detector},
  \PHYSDARKU 9-10 (2014) 1--7.
\newblock \href {http://arxiv.org/abs/1410.7821} {\path{arXiv:1410.7821}},
  \href {http://dx.doi.org/10.1016/j.dark.2015.06.001}
  {\path{doi:10.1016/j.dark.2015.06.001}}.

\bibitem{Sauli:1977mt}
F.~Sauli, {Principles of Operation of Multiwire Proportional and Drift
  Chambers}.

\bibitem{Blum:2008zza}
W.~Blum, W.~Riegler, L.~Rolandi, {Particle Detection with Drift Chambers},
  Particle Acceleration and Detection, Springer, Berlin, 2008.
\newblock \href {http://dx.doi.org/10.1007/978-3-540-76684-1}
  {\path{doi:10.1007/978-3-540-76684-1}}.

\bibitem{Sauli:2014cyf}
F.~Sauli, {Gaseous Radiation Detectors}, Camb. Monogr. Part. Phys. Nucl. Phys.
  Cosmol. 36 (2014) pp.1--497.

\bibitem{DePalma:1983kj}
M.~De~Palma, et~al., {A System of Large Multiwire Proportional Chambers for a
  High Intensity Experiment}, \NIM 217 (1983) 135, [Nucl. Instrum.
  Meth.216,393(1983)].
\newblock \href {http://dx.doi.org/10.1016/0167-5087(83)90121-7}
  {\path{doi:10.1016/0167-5087(83)90121-7}}.

\bibitem{Schilly:1971gi}
P.~Schilly, P.~Steffen, J.~Steinberger, T.~Trippe, F.~Vannucci, H.~Wahl,
  K.~Kleinknecht, V.~Lueth, {Construction and performance of large multiwire
  proportional chambers}, \NIM 91 (1971) 221--230.
\newblock \href {http://dx.doi.org/10.1016/0029-554X(71)90658-6}
  {\path{doi:10.1016/0029-554X(71)90658-6}}.

\bibitem{Charpak:1971ku}
G.~Charpak, G.~Fischer, A.~Minten, L.~Naumann, F.~Sauli, G.~Fluegge,
  C.~Gottfried, R.~Tirler, {Some features of large multiwire proportional
  chambers}, \NIM 97 (1971) 377--388.
\newblock \href {http://dx.doi.org/10.1016/0029-554X(71)90296-5}
  {\path{doi:10.1016/0029-554X(71)90296-5}}.

\bibitem{Veres:1978nf}
I.~Veres, A.~Montvai, {Survey on Multiwire Proportional Chambers}, \NIM 156
  (1978) 73--80.
\newblock \href {http://dx.doi.org/10.1016/0029-554X(78)90694-8}
  {\path{doi:10.1016/0029-554X(78)90694-8}}.

\bibitem{Karagiorgi:2013cwa}
G.~Karagiorgi, {Current and Future Liquid Argon Neutrino Experiments}, AIP
  Conf. Proc. 1663 (2015) 100001.
\newblock \href {http://arxiv.org/abs/1304.2083} {\path{arXiv:1304.2083}},
  \href {http://dx.doi.org/10.1063/1.4919499} {\path{doi:10.1063/1.4919499}}.

\bibitem{SnowdenIfft:2013iy}
D.~P. Snowden-Ifft, J.~L. Gauvreau, {High Precision Measurements of Carbon
  Disulfide Negative Ion Mobility and Diffusion}, \REVSCIINSTRUM 84 (2013)
  053304.
\newblock \href {http://arxiv.org/abs/1301.7145} {\path{arXiv:1301.7145}},
  \href {http://dx.doi.org/10.1063/1.4803004} {\path{doi:10.1063/1.4803004}}.

\bibitem{Charpak:1977sv}
G.~Charpak, G.~Petersen, A.~Policarpo, F.~Sauli, {Progress in High Accuracy
  Proportional Chambers}, \NIM 148 (1978) 471.
\newblock \href {http://dx.doi.org/10.1016/0029-554X(78)91028-5}
  {\path{doi:10.1016/0029-554X(78)91028-5}}.

\bibitem{Snowden-Ifft:2014taa}
D.~P. Snowden-Ifft, {Discovery of multiple, ionization-created CS2 anions and a
  new mode of operation for drift chambers}, \REVSCIINSTRUM 85 (2014) 013303.
\newblock \href {http://dx.doi.org/10.1063/1.4861908}
  {\path{doi:10.1063/1.4861908}}.

\bibitem{Battat:2015rna}
J.~B.~R. Battat, et~al., {Reducing DRIFT Backgrounds with a Submicron
  Aluminized-Mylar Cathode}, \NIM A794 (2015) 33--46.
\newblock \href {http://arxiv.org/abs/1502.03535} {\path{arXiv:1502.03535}},
  \href {http://dx.doi.org/10.1016/j.nima.2015.04.070}
  {\path{doi:10.1016/j.nima.2015.04.070}}.

\bibitem{Schnee:2014eea}
R.~W. Schnee, M.~A. Bowles, R.~Bunker, K.~McCabe, J.~White, P.~Cushman,
  M.~Pepin, V.~E. Guiseppe, {Removal of long-lived $^{222}$Rn daughters by
  electropolishing thin layers of stainless steel}, AIP Conf. Proc. 1549 (2013)
  128--131.
\newblock \href {http://arxiv.org/abs/1404.5843} {\path{arXiv:1404.5843}},
  \href {http://dx.doi.org/10.1063/1.4818092} {\path{doi:10.1063/1.4818092}}.

\bibitem{Battat:2014oqa}
J.~B.~R. Battat, et~al., {Radon in the DRIFT-II directional dark matter TPC:
  emanation, detection and mitigation}, \JINST 9~(11) (2014) P11004.
\newblock \href {http://arxiv.org/abs/1407.3938} {\path{arXiv:1407.3938}},
  \href {http://dx.doi.org/10.1088/1748-0221/9/11/P11004}
  {\path{doi:10.1088/1748-0221/9/11/P11004}}.

\bibitem{Bunker:2014bea}
R.~Bunker, et~al., {The BetaCage, an ultra-sensitive screener for surface
  contamination}, AIP Conf. Proc. 1549 (2013) 132--135.
\newblock \href {http://arxiv.org/abs/1404.5803} {\path{arXiv:1404.5803}},
  \href {http://dx.doi.org/10.1063/1.4818093} {\path{doi:10.1063/1.4818093}}.

\bibitem{SnowdenIfft:1999hz}
D.~P. Snowden-Ifft, C.~J. Martoff, J.~M. Burwell, {Low pressure negative ion
  drift chamber for dark matter search}, \PHYSREV D61 (2000) 101301.
\newblock \href {http://arxiv.org/abs/astro-ph/9904064}
  {\path{arXiv:astro-ph/9904064}}, \href
  {http://dx.doi.org/10.1103/PhysRevD.61.101301}
  {\path{doi:10.1103/PhysRevD.61.101301}}.

\bibitem{Alner:2005xp}
G.~J. Alner, et~al., {The DRIFT-II dark matter detector: Design and
  commissioning}, \NIM A555 (2005) 173--183.
\newblock \href {http://dx.doi.org/10.1016/j.nima.2005.09.011}
  {\path{doi:10.1016/j.nima.2005.09.011}}.

\bibitem{Burgos:2008mv}
S.~Burgos, et~al., {Measurement of the Range Component Directional Signature in
  a DRIFT-II Detector using Cf-252 Neutrons}, \NIM A600 (2009) 417--423.
\newblock \href {http://arxiv.org/abs/0807.3969} {\path{arXiv:0807.3969}},
  \href {http://dx.doi.org/10.1016/j.nima.2008.11.147}
  {\path{doi:10.1016/j.nima.2008.11.147}}.

\bibitem{Burgos:2008jm}
S.~Burgos, E.~Daw, J.~Forbes, C.~Ghag, M.~Gold, C.~Hagemann, V.~A. Kudryavtsev,
  T.~B. Lawson, D.~Loomba, P.~Majewski,
  \href{http://dx.doi.org/10.1016/j.astropartphys.2009.02.003}{First
  measurement of the head-tail directional nuclear recoil signature at energies
  relevant to wimp dark matter searches}, \ASTROPART 31~(4) (2009) 261--266.
\newblock \href {http://dx.doi.org/10.1016/j.astropartphys.2009.02.003}
  {\path{doi:10.1016/j.astropartphys.2009.02.003}}.
\newline\urlprefix\url{http://dx.doi.org/10.1016/j.astropartphys.2009.02.003}

\bibitem{Daw:2010ud}
E.~Daw, et~al., {Spin-Dependent Limits from the DRIFT-IId Directional Dark
  Matter Detector}, \ASTROPART 35 (2012) 397--401.
\newblock \href {http://arxiv.org/abs/1010.3027} {\path{arXiv:1010.3027}},
  \href {http://dx.doi.org/10.1016/j.astropartphys.2011.11.003}
  {\path{doi:10.1016/j.astropartphys.2011.11.003}}.

\bibitem{Burgos:2007gv}
S.~Burgos, et~al., {First results from the DRIFT-IIa dark matter detector},
  \ASTROPART 28 (2007) 409--421.
\newblock \href {http://arxiv.org/abs/0707.1488} {\path{arXiv:0707.1488}},
  \href {http://dx.doi.org/10.1016/j.astropartphys.2007.08.007}
  {\path{doi:10.1016/j.astropartphys.2007.08.007}}.

\bibitem{microbooneTDR}
{MicroBooNE Collaboration}, {The MicroBooNE Technical Design Report}, Tech.
  rep. (2012).

\bibitem{2014arXiv1406.5560P}
J.~{Paley}, D.~{Gastler}, E.~{Kearns}, R.~{Linehan}, R.~{Patterson},
  W.~{Foremen}, J.~{Ho}, D.~{Schmitz}, R.~{Johnson}, J.~{St.~John},
  R.~{Acciarri}, P.~{Adamson}, M.~{Backfish}, W.~{Badgett}, B.~{Baller},
  A.~{Hahn}, D.~{Jensen}, T.~{Junk}, M.~{Kirby}, T.~{Kobilarcik},
  P.~{Kryczynski}, H.~{Lippincott}, A.~{Marchionni}, K.~{Nishikawa}, J.~{Raaf},
  E.~{Ramberg}, B.~{Rebel}, M.~{Stancari}, G.~{Zeller}, M.~{Wascko},
  T.~{Maruyama}, E.~{Iwai}, S.~{Kunori}, C.~{Mauger}, F.~{Blaszczyk},
  W.~{Metcalf}, A.~{Olivier}, M.~{Tzanov}, J.~{Evans}, P.~{Guzowski},
  C.~{Bromberg}, D.~{Edmunds}, D.~{Shooltz}, R.~{Gran}, A.~{Habig}, K.~{Kaess},
  S.~{Dytman}, J.~{Asaadi}, M.~{Soderberg}, J.~{Esquivel}, A.~{Farbin},
  S.~{Park}, J.~{Yu}, J.~{Huang}, K.~{Lang}, R.~{Nichol}, A.~{Holin},
  J.~{Thomas}, M.~{Kordosky}, M.~{Stephens}, P.~{Vahle}, B.~T. {Fleming},
  F.~{Cavanna}, E.~{Church}, E.~{Gramellini}, O.~{Palamara}, A.~{Szelc},
  {LArIAT: Liquid Argon In A Testbeam}, ArXiv e-prints\href
  {http://arxiv.org/abs/1406.5560} {\path{arXiv:1406.5560}}.

\bibitem{ChenPersonal}
{H. Chen}, Private communication.

\bibitem{ifftPrivate}
{D. P. Snowden-Ifft}, Private communication.

\bibitem{battatPrivate}
{J. B. R. Battat}, Private communication.

\bibitem{spoonerPersonalCommunication}
{N. J. C. Spooner}, Private communication.

\bibitem{Oed:1988jh}
A.~Oed, {Position sensitive detector with microstrip anode for electron
  multiplication with gases}, \NIM A263 (1988) 351--359.
\newblock \href {http://dx.doi.org/10.1016/0168-9002(88)90970-9}
  {\path{doi:10.1016/0168-9002(88)90970-9}}.

\bibitem{Titov:2007fm}
M.~P. Titov, {New Developments and Future Perspectives of Gaseous Detectors},
  \NIM A581 (2007) 25--37.
\newblock \href {http://arxiv.org/abs/0706.3516} {\path{arXiv:0706.3516}},
  \href {http://dx.doi.org/10.1016/j.nima.2007.07.022}
  {\path{doi:10.1016/j.nima.2007.07.022}}.

\bibitem{Giomataris:1995fq}
Y.~Giomataris, P.~Rebourgeard, J.~P. Robert, G.~Charpak, {MICROMEGAS: A
  high-granularity position-sensitive gaseous detector for high particle-flux
  environments}, \NIM A376 (1996) 29--35.
\newblock \href {http://dx.doi.org/10.1016/0168-9002(96)00175-1}
  {\path{doi:10.1016/0168-9002(96)00175-1}}.

\bibitem{Sauli:1997qp}
F.~Sauli, {GEM: A new concept for electron amplification in gas detectors},
  \NIM A386 (1997) 531--534.
\newblock \href {http://dx.doi.org/10.1016/S0168-9002(96)01172-2}
  {\path{doi:10.1016/S0168-9002(96)01172-2}}.

\bibitem{rd51}
M.~Alfonci, A.~Bellerive, A.~Breskin, et~al., {R\&D Proposal: Development of
  micro-pattern gas detector technologies}, Tech. rep., CERN-LHCC-2008-011
  (2008).

\bibitem{Micromegasrareevents}
I.~G. {Irastorza}, E.~{Ferrer-Ribas}, T.~{Dafni}, {Micromegas in the Rare Event
  Searches Field}, \MODPHYSLETT A28 (2013) 40026.
\newblock \href {http://dx.doi.org/10.1142/S0217732313400269}
  {\path{doi:10.1142/S0217732313400269}}.

\bibitem{cast1}
T.~{Dafni}, S.~{Aune}, G.~{Fanourakis}, E.~{Ferrer-Ribas}, J.~{Gal{\'a}n},
  A.~{Gardikiotis}, T.~{Geralis}, I.~{Giomataris}, H.~{G{\'o}mez}, F.~J.
  {Iguaz}, I.~G. {Irastorza}, G.~{Luz{\'o}n}, J.~{Morales}, T.~{Papaevangelou},
  A.~{Rodr{\'{\i}}guez}, J.~{Ruz}, A.~{Tom{\'a}s}, T.~{Vafeiadis}, S.~C.
  {Yildiz}, {New micromegas for axion searches in CAST}, \NIM A628 (2011)
  172--176.
\newblock \href {http://dx.doi.org/10.1016/j.nima.2010.06.310}
  {\path{doi:10.1016/j.nima.2010.06.310}}.

\bibitem{Aune:2013nza}
S.~Aune, et~al., {X-ray detection with Micromegas with background levels below
  10$^{-6}$ keV$^{-1}$cm$^{-2}$s$^{-1}$}, \JINST 8 (2013) C12042.
\newblock \href {http://arxiv.org/abs/1312.4282} {\path{arXiv:1312.4282}},
  \href {http://dx.doi.org/10.1088/1748-0221/8/12/C12042}
  {\path{doi:10.1088/1748-0221/8/12/C12042}}.

\bibitem{tpchp}
T.~{Dafni}, E.~{Ferrer-Ribas}, I.~{Giomataris}, P.~{Gorodetzky}, F.~{Iguaz},
  I.~G. {Irastorza}, P.~{Salin}, A.~{Tom{\'a}s}, {Energy resolution of alpha
  particles in a microbulk Micromegas detector at high pressure argon and xenon
  mixtures}, \NIM A608 (2009) 259--266.
\newblock \href {http://arxiv.org/abs/0906.0534} {\path{arXiv:0906.0534}},
  \href {http://dx.doi.org/10.1016/j.nima.2009.06.099}
  {\path{doi:10.1016/j.nima.2009.06.099}}.

\bibitem{T2Kdetect}
A.~{Delbart}, {T2K/TPC Collaboration}, {Production and calibration of 9 m
  $^{2}$ of bulk-micromegas detectors for the readout of the ND280/TPCs of the
  T2K experiment}, \NIM A623 (2010) 105--107.
\newblock \href {http://dx.doi.org/10.1016/j.nima.2010.02.163}
  {\path{doi:10.1016/j.nima.2010.02.163}}.

\bibitem{mimac}
D.~Santos, J.~Billard, G.~Bosson, J.~Bouly, O.~Bourrion, et~al., {MIMAC : A
  micro-tpc matrix for directional detection of dark matter}, \EASPUB 53 (2012)
  25--31.
\newblock \href {http://arxiv.org/abs/1111.1566} {\path{arXiv:1111.1566}},
  \href {http://dx.doi.org/10.1051/eas/1253004}
  {\path{doi:10.1051/eas/1253004}}.

\bibitem{Irastorza:2015dcb}
I.~G. Irastorza, et~al., {Gaseous time projection chambers for rare event
  detection: Results from the T-REX project. I. Double beta decay}, \JCAP 1601
  (2016) 033.
\newblock \href {http://arxiv.org/abs/1512.07926} {\path{arXiv:1512.07926}},
  \href {http://dx.doi.org/10.1088/1475-7516/2016/01/033}
  {\path{doi:10.1088/1475-7516/2016/01/033}}.

\bibitem{Irastorza:2015geo}
I.~G. Irastorza, et~al., {Gaseous time projection chambers for rare event
  detection: Results from the T-REX project. II. Dark matter}, \JCAP 1601~(01)
  (2016) 034.
\newblock \href {http://arxiv.org/abs/1512.06294} {\path{arXiv:1512.06294}},
  \href {http://dx.doi.org/10.1088/1475-7516/2016/01/034}
  {\path{doi:10.1088/1475-7516/2016/01/034}}.

\bibitem{tpc}
D.~{Autiero}, B.~{Beltr{\'a}n}, J.~M. {Carmona}, S.~{Cebri{\'a}n}, E.~{Chesi},
  M.~{Davenport}, M.~{Delattre}, L.~{Di Lella}, F.~{Formenti}, I.~G.
  {Irastorza}, H.~{G{\'o}mez}, M.~{Hasinoff}, B.~{Lakic}, G.~{Luz{\'o}n},
  J.~{Morales}, L.~{Musa}, A.~{Ortiz}, A.~{Placci}, A.~{Rodrigurez}, J.~{Ruz},
  J.~A. {Villar}, K.~{Zioutas}, {The CAST time projection chamber}, \NEWJPHYS 9
  (2007) 171.
\newblock \href {http://arxiv.org/abs/physics/0702189}
  {\path{arXiv:physics/0702189}}, \href
  {http://dx.doi.org/10.1088/1367-2630/9/6/171}
  {\path{doi:10.1088/1367-2630/9/6/171}}.

\bibitem{microbulkRadiopurity}
S.~{Cebri{\'a}n}, T.~{Dafni}, E.~{Ferrer-Ribas}, J.~{Gal{\'a}n},
  I.~{Giomataris}, H.~{G{\'o}mez}, F.~J. {Iguaz}, I.~G. {Irastorza},
  G.~{Luz{\'o}n}, R.~{de Oliveira}, A.~{Rodr{\'{\i}}guez}, L.~{Segu{\'{\i}}},
  A.~{Tom{\'a}s}, J.~A. {Villar}, {Radiopurity of micromegas readout planes},
  \ASTROPART 34 (2011) 354--359.
\newblock \href {http://arxiv.org/abs/1005.2022} {\path{arXiv:1005.2022}},
  \href {http://dx.doi.org/10.1016/j.astropartphys.2010.09.003}
  {\path{doi:10.1016/j.astropartphys.2010.09.003}}.

\bibitem{Charpak:2001tp}
G.~Charpak, J.~Derre, Y.~Giomataris, P.~Rebourgeard, {MICROMEGAS, a
  multipurpose gaseous detector}, \NIM A478 (2002) 26--36.
\newblock \href {http://dx.doi.org/10.1016/S0168-9002(01)01713-2}
  {\path{doi:10.1016/S0168-9002(01)01713-2}}.

\bibitem{Abbon:2007ug}
P.~Abbon, et~al., {The Micromegas detector of the CAST experiment}, \NEWJPHYS 9
  (2007) 170.
\newblock \href {http://arxiv.org/abs/physics/0702190}
  {\path{arXiv:physics/0702190}}, \href
  {http://dx.doi.org/10.1088/1367-2630/9/6/170}
  {\path{doi:10.1088/1367-2630/9/6/170}}.

\bibitem{Pancin:2004vja}
J.~Pancin,
  \href{http://inspirehep.net/record/1325891/files/Thesis-2004-Pancin.pdf}{{D\'{e}tection
  de neutrons avec un d\'{e}tecteur de type Micromegas: de la Physique
  nucl\'{e}aire \`{a} l'imagerie}}, Ph.D. thesis, Saclay, SPhT (2004).
\newline\urlprefix\url{http://inspirehep.net/record/1325891/files/Thesis-2004-Pancin.pdf}

\bibitem{CASTMM}
P.~{Abbon}, S.~{Andriamonje}, S.~{Aune}, T.~{Dafni}, M.~{Davenport},
  E.~{Delagnes}, R.~{de Oliveira}, G.~{Fanourakis}, E.~{Ferrer Ribas},
  J.~{Franz}, T.~{Geralis}, A.~{Giganon}, M.~{Gros}, Y.~{Giomataris}, I.~G.
  {Irastorza}, K.~{Kousouris}, J.~{Morales}, T.~{Papaevangelou}, J.~{Ruz},
  K.~{Zachariadou}, K.~{Zioutas}, {The Micromegas detector of the CAST
  experiment}, \NEWJPHYS 9 (2007) 170.
\newblock \href {http://arxiv.org/abs/physics/0702190}
  {\path{arXiv:physics/0702190}}, \href
  {http://dx.doi.org/10.1088/1367-2630/9/6/170}
  {\path{doi:10.1088/1367-2630/9/6/170}}.

\bibitem{Kunne:2003qe}
F.~Kunne, et~al., {The gaseous microstrip detector Micromegas for the COMPASS
  experiment at CERN}, \NUCLPHYS A721 (2003) 1087--1090.
\newblock \href {http://dx.doi.org/10.1016/S0375-9474(03)01291-0}
  {\path{doi:10.1016/S0375-9474(03)01291-0}}.

\bibitem{NSWTDR}
T.~Kawamoto, S.~Vlachos, L.~Pontecorvo, J.~Dubbert, G.~Mikenberg, P.~Iengo,
  C.~Dallapiccola, C.~Amelung, L.~Levinson, R.~Richter, D.~Lellouch, New small
  wheel technical design report, Tech. rep., CERN-LHCC-2013-006 (2013).

\bibitem{Wotschack}
J.~Wotschack, {The development of large-area Micromegas detectors for the ATLAS
  upgrade}, \MODPHYSLETT A28 (2013) 1340020.
\newblock \href {http://dx.doi.org/10.1142/S0217732313400208}
  {\path{doi:10.1142/S0217732313400208}}.

\bibitem{bulk}
I.~Giomataris, R.~De~Oliveira, S.~Andriamonje, S.~Aune, G.~Charpak, et~al.,
  {Micromegas in a bulk}, \NIM A560 (2006) 405--408.
\newblock \href {http://arxiv.org/abs/physics/0501003}
  {\path{arXiv:physics/0501003}}, \href
  {http://dx.doi.org/10.1016/j.nima.2005.12.222}
  {\path{doi:10.1016/j.nima.2005.12.222}}.

\bibitem{T2K}
N.~Abgrall, et~al., {Time Projection Chambers for the T2K Near Detectors}, \NIM
  A637 (2011) 25--46.
\newblock \href {http://arxiv.org/abs/1012.0865} {\path{arXiv:1012.0865}},
  \href {http://dx.doi.org/10.1016/j.nima.2011.02.036}
  {\path{doi:10.1016/j.nima.2011.02.036}}.

\bibitem{Minos}
A.~Obertelli, A.~Delbart, S.~Anvar, L.~Audirac, G.~Authelet, et~al., {MINOS: A
  vertex tracker coupled to a thick liquid-hydrogen target for in-beam
  spectroscopy of exotic nuclei}, \EURPHYSJ A50 (2014) 8.
\newblock \href {http://dx.doi.org/10.1140/epja/i2014-14008-y}
  {\path{doi:10.1140/epja/i2014-14008-y}}.

\bibitem{CLAS12Mod}
S.~Procureur, {Micromegas Trackers for Hadronic Physics}, \MODPHYSLETT A28
  (2013) 1340024.
\newblock \href {http://dx.doi.org/10.1142/S0217732313400245}
  {\path{doi:10.1142/S0217732313400245}}.

\bibitem{Charles:2013cvy}
G.~Charles, \href{http://tel.archives-ouvertes.fr/tel-00873381}{{Mise au point
  de d\'{e}tecteurs micromegas pour le spectrom\`{e}tre CLAS12 au laboratoire
  Jefferson}}, Ph.D. thesis, U. Paris-Sud 11, Dept. Phys., Orsay (2013).
\newline\urlprefix\url{http://tel.archives-ouvertes.fr/tel-00873381}

\bibitem{Riffard:2013psa}
Q.~Riffard, et~al.,
  \href{http://inspirehep.net/record/1238915/files/arXiv:1306.4173.pdf}{{Dark
  Matter directional detection with MIMAC}}, in: {Proceedings, 48th Rencontres
  de Moriond on Very High Energy Phenomena in the Universe}, 2013, pp.
  227--230.
\newblock \href {http://arxiv.org/abs/1306.4173} {\path{arXiv:1306.4173}}.
\newline\urlprefix\url{http://inspirehep.net/record/1238915/files/arXiv:1306.4173.pdf}

\bibitem{Andriamonje:2010zz}
S.~Andriamonje, D.~Attie, E.~Berthoumieux, M.~Calviani, P.~Colas, et~al.,
  {Development and performance of Microbulk Micromegas detectors}, \JINST 5
  (2010) P02001.
\newblock \href {http://dx.doi.org/10.1088/1748-0221/5/02/P02001}
  {\path{doi:10.1088/1748-0221/5/02/P02001}}.

\bibitem{Iguaz:2012ur}
F.~J. Iguaz, E.~Ferrer-Ribas, A.~Giganon, I.~Giomataris, {Characterization of
  microbulk detectors in argon- and neon-based mixtures}, \JINST 7 (2012)
  P04007.
\newblock \href {http://arxiv.org/abs/1201.3012} {\path{arXiv:1201.3012}},
  \href {http://dx.doi.org/10.1088/1748-0221/7/04/P04007}
  {\path{doi:10.1088/1748-0221/7/04/P04007}}.

\bibitem{ntof1}
F.~Belloni, F.~Gunsing, T.~Papaevangelou, {Micromegas for neutron detection and
  imaging}, \MODPHYSLETT A28 (2013) 1340023.
\newblock \href {http://dx.doi.org/10.1142/S0217732313400233}
  {\path{doi:10.1142/S0217732313400233}}.

\bibitem{ntof3}
C.~Guerrero, E.~Berthoumieux, D.~Cano-Ott, E.~Mendoza, S.~Andriamonje, et~al.,
  {Simultaneous measurement of neutron-induced capture and fission reactions at
  CERN}, \EURPHYSJ A48 (2012) 29.
\newblock \href {http://dx.doi.org/10.1140/epja/i2012-12029-2}
  {\path{doi:10.1140/epja/i2012-12029-2}}.

\bibitem{Iguaz:2015myh}
F.~J. Iguaz, et~al., {TREX-DM: a low-background Micromegas-based TPC for
  low-mass WIMP detection}\href {http://arxiv.org/abs/1512.01455}
  {\path{arXiv:1512.01455}}.

\bibitem{MicrobulkTMA}
S.~Cebrian, T.~Dafni, E.~Ferrer-Ribas, I.~Giomataris, D.~Gonzalez-Diaz, et~al.,
  {Micromegas-TPC operation at high pressure in xenon-trimethylamine mixtures},
  \JINST 8 (2013) P01012.
\newblock \href {http://arxiv.org/abs/1210.3287} {\path{arXiv:1210.3287}},
  \href {http://dx.doi.org/10.1088/1748-0221/8/01/P01012}
  {\path{doi:10.1088/1748-0221/8/01/P01012}}.

\bibitem{NEXTHector}
V.~\'Alvarez, et~al., {Description and commissioning of NEXT-MM prototype:
  first results from operation in a Xenon-Trimethylamine gas mixture}, \JINST 9
  (2014) P03010.
\newblock \href {http://arxiv.org/abs/1311.3242} {\path{arXiv:1311.3242}},
  \href {http://dx.doi.org/10.1088/1748-0221/9/03/P03010}
  {\path{doi:10.1088/1748-0221/9/03/P03010}}.

\bibitem{NEXTDiego}
V.~\'Alvarez, et~al., {Characterization of a medium size Xe/TMA TPC
  instrumented with microbulk Micromegas, using low-energy $\gamma$-rays},
  \JINST 9 (2014) C04015.
\newblock \href {http://arxiv.org/abs/1311.3535} {\path{arXiv:1311.3535}},
  \href {http://dx.doi.org/10.1088/1748-0221/9/04/C04015}
  {\path{doi:10.1088/1748-0221/9/04/C04015}}.

\bibitem{PhDMax}
M.~A. Chefdeville, {Development of Micromegas-like Gaseous Detectors Using a
  Pixel Readout Chip as Collecting Anode}, Ph.D. thesis (2009).

\bibitem{Krieger:2014pea}
C.~Krieger, K.~Desch, J.~Kaminski, M.~Lupberger, T.~Vafeiadis,
  \href{https://inspirehep.net/record/1319671/files/arXiv:1410.0264.pdf}{{An
  InGrid based Low Energy X-ray Detector}}, in: {10th Patras Workshop on
  Axions, WIMPs and WISPs (AXION-WIMP 2014) Geneva, Switzerland, June 29-July
  4, 2014}, 2014.
\newblock \href {http://arxiv.org/abs/1410.0264} {\path{arXiv:1410.0264}}.
\newline\urlprefix\url{https://inspirehep.net/record/1319671/files/arXiv:1410.0264.pdf}

\bibitem{Lupberger:2013lfa}
M.~Lupberger, {The Pixel-TPC: first results from an 8-InGrid module}, \JINST
  9~(01) (2014) C01033.
\newblock \href {http://arxiv.org/abs/1311.3125} {\path{arXiv:1311.3125}},
  \href {http://dx.doi.org/10.1088/1748-0221/9/01/C01033}
  {\path{doi:10.1088/1748-0221/9/01/C01033}}.

\bibitem{Dixit}
M.~Dixit, A.~Rankin, {Simulating the charge dispersion phenomena in micro
  pattern gas detectors with a resistive anode}, \NIM A566 (2006) 281--285.
\newblock \href {http://arxiv.org/abs/physics/0605121}
  {\path{arXiv:physics/0605121}}, \href
  {http://dx.doi.org/10.1016/j.nima.2006.06.050}
  {\path{doi:10.1016/j.nima.2006.06.050}}.

\bibitem{alexpoulos-kobe}
T.~Alexopoulos, G.~Iakovidis, G.~Tsipolitis, {Study of Resistive Micromegas in
  a Mixed Neutron and Photon Radiation Field}, \JINST 7 (2012) C05001.
\newblock \href {http://arxiv.org/abs/1111.3185} {\path{arXiv:1111.3185}},
  \href {http://dx.doi.org/10.1088/1748-0221/7/05/C05001}
  {\path{doi:10.1088/1748-0221/7/05/C05001}}.

\bibitem{genetic}
S.~Procureur, R.~Dupre, S.~Aune, {Genetic multiplexing and first results with a
  $50 \times 50$ $cm^2 $ Micromegas}, \NIM A729 (2013) 888--894.
\newblock \href {http://dx.doi.org/10.1016/j.nima.2013.08.071}
  {\path{doi:10.1016/j.nima.2013.08.071}}.

\bibitem{Billard:2013cxa}
J.~Billard, F.~Mayet, G.~Bosson, O.~Bourrion, O.~Guillaudin, et~al., {In situ
  measurement of the electron drift velocity for upcoming directional Dark
  Matter detectors}, \JINST 9 (2014) 01013.
\newblock \href {http://arxiv.org/abs/1305.2360} {\path{arXiv:1305.2360}},
  \href {http://dx.doi.org/10.1088/1748-0221/9/01/P01013}
  {\path{doi:10.1088/1748-0221/9/01/P01013}}.

\bibitem{Piggyback-2013}
D.~Atti\'{e}, A.~Chaus, D.~Durand, D.~Desforge, E.~Ferrer-Ribas, et~al.,
  {Piggyback resistive Micromegas}, \JINST 8 (2013) C11007.
\newblock \href {http://arxiv.org/abs/1310.1242} {\path{arXiv:1310.1242}},
  \href {http://dx.doi.org/10.1088/1748-0221/8/11/C11007}
  {\path{doi:10.1088/1748-0221/8/11/C11007}}.

\bibitem{Attie:2015uoa}
D.~Atti\'{e}, et~al., {R\&D on a novel spectro-imaging polarimeter with
  Micromegas detectors and a Caliste readout system}, \NIM A787 (2015)
  312--314.
\newblock \href {http://dx.doi.org/10.1016/j.nima.2015.01.007}
  {\path{doi:10.1016/j.nima.2015.01.007}}.

\bibitem{Serrano:2016pfh}
P.~Serrano, D.~Atti\'{e}, D.~Desforge, E.~F. Ribas, F.~Jeanneau, O.~Limousin,
  {Caliste-MM: a spectro-polarimeter based on the micromegas concept for soft
  X-ray astrophysics}, \JINST 11~(04) (2016) P04016.
\newblock \href {http://dx.doi.org/10.1088/1748-0221/11/04/P04016}
  {\path{doi:10.1088/1748-0221/11/04/P04016}}.

\bibitem{trexwebpage}
{http://gifna.unizar.es/trex/}.

\bibitem{Aznar:2013jwa}
F.~Aznar, J.~Castel, S.~Cebri\'{a}n, T.~Dafni, A.~Diago, et~al., {Assessment of
  material radiopurity for Rare Event experiments using Micromegas}, \JINST 8
  (2013) C11012.
\newblock \href {http://dx.doi.org/10.1088/1748-0221/8/11/C11012}
  {\path{doi:10.1088/1748-0221/8/11/C11012}}.

\bibitem{Aune:2013pna}
S.~Aune, et~al., {Low background x-ray detection with Micromegas for axion
  research}, \JINST 9~(01) (2014) P01001.
\newblock \href {http://arxiv.org/abs/1310.3391} {\path{arXiv:1310.3391}},
  \href {http://dx.doi.org/10.1088/1748-0221/9/01/P01001}
  {\path{doi:10.1088/1748-0221/9/01/P01001}}.

\bibitem{mimacelec}
J.~Richer, G.~Bosson, O.~Bourrion, C.~Grignon, O.~Guillaudin, et~al.,
  {Development of a front end ASIC for Dark Matter directional detection with
  MIMAC}, \NIM A620 (2010) 470--476.
\newblock \href {http://arxiv.org/abs/0912.0186} {\path{arXiv:0912.0186}},
  \href {http://dx.doi.org/10.1016/j.nima.2010.04.041}
  {\path{doi:10.1016/j.nima.2010.04.041}}.

\bibitem{Richer:2011pe}
J.~Richer, O.~Bourrion, G.~Bosson, O.~Guillaudin, F.~Mayet, et~al.,
  {Development and validation of a 64 channel front end ASIC for 3D directional
  detection for MIMAC}, \JINST 6 (2011) C11016.
\newblock \href {http://arxiv.org/abs/1110.4579} {\path{arXiv:1110.4579}},
  \href {http://dx.doi.org/10.1088/1748-0221/6/11/C11016}
  {\path{doi:10.1088/1748-0221/6/11/C11016}}.

\bibitem{Bourrion:2011vk}
O.~Bourrion, G.~Bosson, C.~Grignon, J.~Richer, O.~Guillaudin, et~al.,
  {Dedicated front-end and readout electronics developments for real time 3D
  directional detection of dark matter with MIMAC}, \EASPUB 53 (2012) 129--136.
\newblock \href {http://arxiv.org/abs/1109.2002} {\path{arXiv:1109.2002}},
  \href {http://dx.doi.org/10.1051/eas/1253016}
  {\path{doi:10.1051/eas/1253016}}.

\bibitem{Bourrion:2011it}
O.~Bourrion, G.~Bosson, C.~Grignon, J.~Bouly, J.~Richer, et~al., {Data
  acquisition electronics and reconstruction software for real time 3D track
  reconstruction within the MIMAC project}, \JINST 6 (2011) C11003.
\newblock \href {http://arxiv.org/abs/1110.4348} {\path{arXiv:1110.4348}},
  \href {http://dx.doi.org/10.1088/1748-0221/6/11/C11003}
  {\path{doi:10.1088/1748-0221/6/11/C11003}}.

\bibitem{T2Kelec1}
P.~Baron, D.~Besin, D.~Calvet, C.~Coquelet, X.~De~La~Broise, et~al.,
  {Architecture and implementation of the front-end electronics of the time
  projection chambers in the T2K experiment}, \IEEETRANSNUCLSCI 57 (2010)
  406--411.
\newblock \href {http://dx.doi.org/10.1109/TNS.2009.2035313}
  {\path{doi:10.1109/TNS.2009.2035313}}.

\bibitem{T2Kelec2}
P.~Baron, D.~Calvet, E.~Delagnes, X.~de~la Broise, A.~Delbart, et~al., {AFTER,
  an ASIC for the readout of the large T2K time projection chambers},
  \IEEETRANSNUCLSCI 55 (2008) 1744--1752.
\newblock \href {http://dx.doi.org/10.1109/TNS.2008.924067}
  {\path{doi:10.1109/TNS.2008.924067}}.

\bibitem{MIMAC-detector}
F.~Iguaz, D.~Attie, D.~Calvet, P.~Colas, F.~Druillole, et~al., {Micromegas
  detector developments for Dark Matter directional detection with MIMAC},
  \JINST 6 (2011) P07002.
\newblock \href {http://arxiv.org/abs/1105.2056} {\path{arXiv:1105.2056}},
  \href {http://dx.doi.org/10.1088/1748-0221/6/07/P07002}
  {\path{doi:10.1088/1748-0221/6/07/P07002}}.

\bibitem{magboltz}
S.~Biagi, {Monte Carlo simulation of electron drift and diffusion in counting
  gases under the influence of electric and magnetic fields}, \NIM A421~(1-2)
  (1999) 234--240.
\newblock \href {http://dx.doi.org/10.1016/S0168-9002(98)01233-9}
  {\path{doi:10.1016/S0168-9002(98)01233-9}}.

\bibitem{riffardThesis}
Q.~Riffard, \href{https://tel.archives-ouvertes.fr/tel-01258830}{{Non-baryonic
  dark matter directional detection with MIMAC}}, Ph.D. thesis, {Universit{\'e}
  Grenoble Alpes} (Oct. 2015).
\newline\urlprefix\url{https://tel.archives-ouvertes.fr/tel-01258830}

\bibitem{cygnus2011}
F.~Mayet, D.~Santos, {Proceedings, 3rd Workshop on Directional Detection of
  Dark Matter (CYGNUS 2011)}, \EASPUB 53 (2012) pp.1--181.
\newblock \href {http://dx.doi.org/10.1051/eas/1253000}
  {\path{doi:10.1051/eas/1253000}}.

\bibitem{Caldwell:2009si}
T.~Caldwell, et~al., {Transport properties of electrons in CF$_{4}$}\href
  {http://arxiv.org/abs/0905.2549} {\path{arXiv:0905.2549}}.

\bibitem{AlbornozVasquez:2012px}
D.~Albornoz~Vasquez, G.~Belanger, J.~Billard, F.~Mayet, {Probing neutralino
  dark matter in the MSSM \& the NMSSM with directional detection}, \PHYSREV
  D85 (2012) 055023.
\newblock \href {http://arxiv.org/abs/1201.6150} {\path{arXiv:1201.6150}},
  \href {http://dx.doi.org/10.1103/PhysRevD.85.055023}
  {\path{doi:10.1103/PhysRevD.85.055023}}.

\bibitem{Riffard:2015rga}
Q.~Riffard, et~al., {First detection of tracks of radon progeny recoils by
  MIMAC}\href {http://arxiv.org/abs/1504.05865} {\path{arXiv:1504.05865}}.

\bibitem{ref:3uPIC}
A.~Ochi, T.~Nagayoshi, S.~Koishi, T.~Tanimori, T.~Nagae, M.~Nakamura, {A new
  design of the gaseous imaging detector: Micro Pixel Chamber}, \NIM A471
  (2001) 264--267.
\newblock \href {http://dx.doi.org/10.1016/S0168-9002(01)00996-2}
  {\path{doi:10.1016/S0168-9002(01)00996-2}}.

\bibitem{ref:MSGC_tanimori}
T.~Tanimori, A.~Ochi, S.~Minami, T.~Nagae, {Development of imaging microstrip
  gas chamber with 5-cm x 5-cm area based on multichip module technology}, \NIM
  A381 (1996) 280--288.
\newblock \href {http://dx.doi.org/10.1016/S0168-9002(96)00833-9}
  {\path{doi:10.1016/S0168-9002(96)00833-9}}.

\bibitem{ref:10uPIC}
T.~Nagayoshi, H.~Kubo, K.~Miuchi, A.~Ochi, R.~Orito, A.~Takada, T.~Tanimori,
  M.~Ueno, {Performance of large area Micro Pixel Chamber}, \NIM A513 (2003)
  277--281.
\newblock \href {http://arxiv.org/abs/hep-ex/0301008}
  {\path{arXiv:hep-ex/0301008}}, \href
  {http://dx.doi.org/10.1016/j.nima.2003.08.047}
  {\path{doi:10.1016/j.nima.2003.08.047}}.

\bibitem{ref:uPIC_neutronNIM2014}
K.~{Miuchi}, et~al., {Performance of the micro-TPC for the time-resolved
  neutron PSD}, \NIM A517 (2004) 219--225.

\bibitem{ref:uPIC_imaging2003Nagayoshi}
T.~{Nagayoshi}, et~al., {Development of $\mu$-PIC and its imaging properties},
  \NIM A525 (2004) 20--27.

\bibitem{ref:uPIC_imaging2003Ueno}
M.~{Ueno}, et~al., {Application of Micro-Pixel Chambers for X-Ray Polarimetry},
  \NIM A525 (2004) 28--32.

\bibitem{ref:uPIC_imaging2003Orito}
R.~{Orito}, et~al., {Compton gamma-ray imaging detector with electron
  tracking}, \NIM A513 (2003) 408--412.

\bibitem{ref:uPIC_NagayoshiThesis}
T.~{Nagayoshi}, {Development of Micro Pixel Chamber and Systematic Study on the
  Electrode structure}, Kyoto University Doctoral Thesis, 2004.

\bibitem{ref:30uPIC}
A.~Takada, K.~Hattori, S.~Kabuki, H.~Kubo, K.~Miuchi, T.~Nagayoshi,
  H.~Nishimura, Y.~Okada, R.~Orito, H.~Sekiya, A.~Takeda, T.~Tanimori, K.~Ueno,
  \href{http://www.sciencedirect.com/science/article/pii/S016890020602167X}{{A
  very large area Micro Pixel Chamber}}, \NIM A573 (2007) 195--199.
\newblock \href
  {http://dx.doi.org/http://dx.doi.org/10.1016/j.nima.2006.10.283}
  {\path{doi:http://dx.doi.org/10.1016/j.nima.2006.10.283}}.
\newline\urlprefix\url{http://www.sciencedirect.com/science/article/pii/S016890020602167X}

\bibitem{ref:uPIC_MeVBalloon}
A.~{Takada}, et~al., {Observation of Diffuse Cosmic and Atmospheric Gamma Rays
  at Balloon Altitudes with an Electron-tracking Compton Camera}, \APJ 733
  (2011) 13.

\bibitem{ref:uPIC_MeVmedical}
S.~{Kabuki}, et~al., {Electron-tracking compton gamma-ray camera for small
  animal and phantom imaging.}, \NIM A623 (2010) 606.

\bibitem{ref:uPIC_neutron2013}
J.~{Parker}, et~al., {Neutron imaging detector based on the $\mu$-PIC
  micro-pixel chamber}, \NIM A697 (2013) 23.

\bibitem{ref:uPIC_SAX}
K.~{Hattori}, et~al., {Performance of the micro-PIC gaseous detector in
  small-angle X-ray scattering experiments}, J. Synchrotron Radiation 16 (2009)
  231.

\bibitem{ref:uPIC_PMT}
H.~{Sekiya}, et~al., {Development of a large area VUV sensitive PMT with
  GEM/$\mu$-PIC}, \JINST 4 (2009) P11006.

\bibitem{ref:uPIC_PS-TEPC}
Y.~{Kishimoto}, et~al., {Basic performance of a position-sensitive
  tissue-equiavrent proportional chamber (PS-TEPC)}, \NIM A732 (2013) 591.

\bibitem{ref:NEWAGE_PTEP2015}
K.~{Nakamura}, et~al., {Direction-sensitive dark matter search with gaseous
  tracking detector NEWAGE-0.3b'}, Prog. Theore. Exp. Phys. (2015) 043F01.

\bibitem{ref:uPIC_KuboPSD}
H.~{Kubo}, et~al., {Development of a time projection chamber with micro-pixel
  electrodes}, \NIM A513 (2003) 94.

\bibitem{ref:uPIC_Nagayoshi_sim}
T.~{Nagayoshi}, et~al., {Simulation study of electron drift and gas
  multiplication in Micro Pixel Chamber}, \NIM A546 (2005) 457--465.

\bibitem{ref:uPIC_Oleg_sim}
T.~{Nagayoshi}, et~al., {Performance optimasation of the micro pixel chamber},
  \NIM A540 (2005) 266--273.

\bibitem{ref:Garfield}
R.~Veenhof, {GARFIELD, recent developments}, \NIM A419 (1998) 726--730.
\newblock \href {http://dx.doi.org/10.1016/S0168-9002(98)00851-1}
  {\path{doi:10.1016/S0168-9002(98)00851-1}}.

\bibitem{ref:Maxwell}
{Maxwell 3-D Field Simulator, Ansoft Corporation}.

\bibitem{ref:Gmsh}
{http://geuz.org/gmsh}.

\bibitem{ref:Elmer}
{http://www.csc.fi/web/elmer}.

\bibitem{ref:Garfield++}
{http://garfieldpp.web.cern.ch/garfieldpp}.

\bibitem{ref:uPIC_Takada_sim}
A.~{Takada}, et~al., {Simulation of gas avalanche in a micro pixel chamber
  using Garfield++}, \JINST 8 (2013) C10023.

\bibitem{ref:NEWAGE_PLB2004}
T.~Tanimori, H.~Kubo, K.~Miuchi, T.~Nagayoshi, R.~Orito, A.~Takada, A.~Takeda,
  {Detecting the wimp-wind via spin-dependent interactions}, \PHYSLETT B578
  (2004) 241--246.
\newblock \href {http://arxiv.org/abs/astro-ph/0310638}
  {\path{arXiv:astro-ph/0310638}}, \href
  {http://dx.doi.org/10.1016/j.physletb.2003.10.077}
  {\path{doi:10.1016/j.physletb.2003.10.077}}.

\bibitem{ref:NEWAGE_PLB2007}
K.~Miuchi, et~al., {Direction-sensitive dark matter search results in a surface
  laboratory}, \PHYSLETT B654 (2007) 58--64.
\newblock \href {http://arxiv.org/abs/0708.2579} {\path{arXiv:0708.2579}},
  \href {http://dx.doi.org/10.1016/j.physletb.2007.08.042}
  {\path{doi:10.1016/j.physletb.2007.08.042}}.

\bibitem{ref:NEWAGE_PLB2010}
K.~Miuchi, et~al., {First underground results with NEWAGE-0.3a
  direction-sensitive dark matter detector}, \PHYSLETT B686 (2010) 11--17.
\newblock \href {http://arxiv.org/abs/1002.1794} {\path{arXiv:1002.1794}},
  \href {http://dx.doi.org/10.1016/j.physletb.2010.02.028}
  {\path{doi:10.1016/j.physletb.2010.02.028}}.

\bibitem{PoikelaTimepix3}
T.~Poikela, J.~Plosila, T.~Westerlund, M.~Campbell, M.~D. Gaspari, X.~Llopart,
  V.~Gromov, R.~Kluit, M.~van Beuzekom, F.~Zappon, V.~Zivkovic, C.~Brezina,
  K.~Desch, Y.~Fu, A.~Kruth,
  \href{http://stacks.iop.org/1748-0221/9/i=05/a=C05013}{Timepix3: a 65k
  channel hybrid pixel readout chip with simultaneous toa/tot and sparse
  readout}, \JINST 9~(05) (2014) C05013.
\newline\urlprefix\url{http://stacks.iop.org/1748-0221/9/i=05/a=C05013}

\bibitem{Kim:2008zzi}
T.~Kim, M.~Freytsis, J.~Button-Shafer, J.~Kadyk, S.~Vahsen, W.~Wenzel, {Readout
  of TPC tracking chambers with GEMs and pixel chip}, \NIM A589 (2008)
  173--184.
\newblock \href {http://dx.doi.org/10.1016/j.nima.2008.02.049}
  {\path{doi:10.1016/j.nima.2008.02.049}}.

\bibitem{Vahsen:2014mca}
S.~Vahsen, K.~Oliver-Mallory, M.~Lopez-Thibodeaux, J.~Kadyk,
  M.~Garcia-Sciveres, {Tests of gases in a mini-TPC with pixel chip readout},
  \NIM A738 (2014) 111--118.
\newblock \href {http://dx.doi.org/10.1016/j.nima.2013.10.029}
  {\path{doi:10.1016/j.nima.2013.10.029}}.

\bibitem{Vahsen:2015oya}
S.~Vahsen, M.~Hedges, I.~Jaegle, S.~Ross, I.~Seong, et~al., {3-D tracking in a
  miniature time projection chamber}, \NIM A788 (2015) 95--105.
\newblock \href {http://dx.doi.org/10.1016/j.nima.2015.03.009}
  {\path{doi:10.1016/j.nima.2015.03.009}}.

\bibitem{Jaegle:2015beasttpc}
I.~Jaegle, et~al., {High resolution 3-D tracking in a time projection chamber
  with pixel readout}, in preparation.

\bibitem{Aad:2008zz}
G.~Aad, M.~Ackers, F.~Alberti, M.~Aleppo, G.~Alimonti, et~al., {ATLAS pixel
  detector electronics and sensors}, \JINST 3 (2008) P07007.
\newblock \href {http://dx.doi.org/10.1088/1748-0221/3/07/P07007}
  {\path{doi:10.1088/1748-0221/3/07/P07007}}.

\bibitem{GarciaSciveres:2011zz}
M.~Garcia-Sciveres, D.~Arutinov, M.~Barbero, R.~Beccherle, S.~Dube, et~al.,
  {The FE-I4 pixel readout integrated circuit}, \NIM A636 (2011) S155--S159.
\newblock \href {http://dx.doi.org/10.1016/j.nima.2010.04.101}
  {\path{doi:10.1016/j.nima.2010.04.101}}.

\bibitem{Seong:2013nua}
I.~Seong, K.~Beamer, M.~Hedges, I.~Jaegle, M.~Rosen, et~al., {Time projection
  chambers with integrated pixels and their application to fast neutron
  detection and dark matter searches}, \NIM A732 (2013) 260--263.
\newblock \href {http://dx.doi.org/10.1016/j.nima.2013.07.053}
  {\path{doi:10.1016/j.nima.2013.07.053}}.

\bibitem{Lewis:2014poa}
P.~Lewis, S.~Vahsen, M.~Hedges, I.~Jaegle, I.~Seong, et~al., {Absolute Position
  Measurement in a Gas Time Projection Chamber via Transverse Diffusion of
  Drift Charge}, \NIM A789 (2015) 81--85.
\newblock \href {http://arxiv.org/abs/1410.1131} {\path{arXiv:1410.1131}},
  \href {http://dx.doi.org/10.1016/j.nima.2015.03.024}
  {\path{doi:10.1016/j.nima.2015.03.024}}.

\bibitem{Ross:2013bza}
S.~J. Ross, M.~T. Hedges, I.~Jaegle, M.~D. Rosen, I.~S. Seong, T.~N. Thorpe,
  S.~E. Vahsen, J.~Yamaoka,
  \href{https://inspirehep.net/record/1226239/files/arXiv:1304.0507.pdf}{{Charge-Focusing
  Readout of Time Projection Chambers}}, in: {Proceedings, 2012 IEEE Nuclear
  Science Symposium and Medical Imaging Conference (NSS/MIC 2012): Anaheim,
  California, USA, October 29-November 3, 2012}, 2012, pp. 1760--1765.
\newblock \href {http://arxiv.org/abs/1304.0507} {\path{arXiv:1304.0507}},
  \href {http://dx.doi.org/10.1109/NSSMIC.2012.6551412}
  {\path{doi:10.1109/NSSMIC.2012.6551412}}.
\newline\urlprefix\url{https://inspirehep.net/record/1226239/files/arXiv:1304.0507.pdf}

\bibitem{Vahsen:2011qx}
S.~Vahsen, H.~Feng, M.~Garcia-Sciveres, I.~Jaegle, J.~Kadyk, et~al., {The
  Directional Dark Matter Detector ($D^{3}$)}, \EASPUB 53 (2012) 43--50.
\newblock \href {http://arxiv.org/abs/1110.3401} {\path{arXiv:1110.3401}},
  \href {http://dx.doi.org/10.1051/eas/1253006}
  {\path{doi:10.1051/eas/1253006}}.

\bibitem{Jaegle:2012sma}
I.~Jaegle, H.~Feng, S.~Ross, J.~Yamaoka, S.~Vahsen, {Simulation of the
  Directional Dark Matter Detector ($D^{3}$) and Directional Neutron Observer
  (DiNO)}, \EASPUB 53 (2012) 111--118.
\newblock \href {http://dx.doi.org/10.1051/eas/1253014}
  {\path{doi:10.1051/eas/1253014}}.

\bibitem{Shalem:2005ix}
C.~Shalem, R.~Chechik, A.~Breskin, K.~Michaeli, N.~Ben-Haim, {Advances in thick
  GEM-like gaseous electron multipliers. Part II: Low-pressure operation}, \NIM
  A558 (2006) 468--474.
\newblock \href {http://arxiv.org/abs/physics/0601119}
  {\path{arXiv:physics/0601119}}, \href
  {http://dx.doi.org/10.1016/j.nima.2005.12.219}
  {\path{doi:10.1016/j.nima.2005.12.219}}.

\bibitem{ref:Khoa_Mth}
V.~M. Khoa, {A pixel readout LSI with a built-in ADC for particle detector
  applications}, Master Thesis Tokyo Institue of Technology (Feburary 2010) 1.

\bibitem{ref:NEWAGE_cygnus2011}
K.~Miuchi, et~al., {NEWAGE}, \EASPUB 53 (2012) 33--41.
\newblock \href {http://dx.doi.org/10.1051/eas/1253005}
  {\path{doi:10.1051/eas/1253005}}.

\bibitem{ref:Fei_Dth}
F.~LI, {Quasi-3D pixel readout LSIs for gaseous particle detectors}, Doctral
  Thesis Tokyo Institue of Technology (December 2012) 1.

\bibitem{Plackett:2010zz}
R.~Plackett, R.~Ballabriga, M.~Campbell, X.~Llopart, T.~Tick, L.~Tlustos,
  D.~Turecek, S.~Vahanen, W.~Wong, {Current status of the Medipix2, Timepix,
  Medipix3 and Timepix2 pixel readout chips}, PoS VERTEX2010 (2010) 030.

\bibitem{Gromov:2011zz}
V.~Gromov, et~al., {Development and applications of the Timepix3 readout chip},
  PoS VERTEX2011 (2011) 046.

\bibitem{Krummenacher}
F.~Krummenacher, {Pixel detectors with local intelligence: an IC designer point
  of view }, \NIM A305 (1991) 527--532.
\newblock \href {http://dx.doi.org/10.1016/0168-9002(91)90152-G}
  {\path{doi:10.1016/0168-9002(91)90152-G}}.

\bibitem{DeGasperi}
M.~D. Gaspari, J.~Alozy, R.~Ballabriga, M.~Campbell, E.~Fr\"{o}jdh,
  J.~Idarraga, S.~Kulis, X.~Llopart, T.~Poikela, P.~Valerio, W.~Wong,
  \href{http://stacks.iop.org/1748-0221/9/i=01/a=C01037}{{Design of the analog
  front-end for the Timepix3 and Smallpix hybrid pixel detectors in 130 nm CMOS
  technology}}, \JINST 9~(01) (2014) C01037.
\newline\urlprefix\url{http://stacks.iop.org/1748-0221/9/i=01/a=C01037}

\bibitem{Frojdh:2015fta}
{Fr\"{o}jdh, E. and Campbell, M. and De Gaspari, M. and Kulis, S. and Llopart,
  X. and Poikela, T. and Tlustos, L.}, {Timepix3: first measurements and
  characterization of a hybrid-pixel detector working in event driven mode},
  \JINST 10~(01) (2015) C01039.
\newblock \href {http://dx.doi.org/10.1088/1748-0221/10/01/C01039}
  {\path{doi:10.1088/1748-0221/10/01/C01039}}.

\bibitem{Campbell:2004ib}
M.~Campbell, et~al., {The detection of single electrons by means of a
  micromegas-covered MediPix2 pixel CMOS readout circuit}, \NIM A540 (2005)
  295--304.
\newblock \href {http://arxiv.org/abs/physics/0409048}
  {\path{arXiv:physics/0409048}}, \href
  {http://dx.doi.org/10.1016/j.nima.2004.11.036}
  {\path{doi:10.1016/j.nima.2004.11.036}}.

\bibitem{vanderGraaf:2007zz}
H.~van~der Graaf, {GridPix: An integrated readout system for gaseous detectors
  with a pixel chip as anode}, \NIM A580 (2007) 1023--1026.
\newblock \href {http://dx.doi.org/10.1016/j.nima.2007.06.096}
  {\path{doi:10.1016/j.nima.2007.06.096}}.

\bibitem{Lupberger:2014pba}
M.~Lupberger, J.~Bilevych, K.~Desch, T.~Fischer, T.~Fritzsch, J.~Kaminski,
  K.~Kohl, M.~Rogowski, J.~Tomtschak, H.~van~der Graaf, {InGrid: Pixelated
  Micromegas detectors for a pixel-TPC}, PoS TIPP2014 (2014) 225.

\bibitem{Koppert:2013lua}
W.~J.~C. Koppert, et~al., {GridPix detectors: Production and beam test
  results}, \NIM A732 (2013) 245--249.
\newblock \href {http://dx.doi.org/10.1016/j.nima.2013.08.010}
  {\path{doi:10.1016/j.nima.2013.08.010}}.

\bibitem{Krieger:2014wxa}
C.~Krieger, K.~Desch, J.~Kaminski, M.~Lupberger, T.~Vafeiadis, {An InGrid based
  Low Energy X-ray Detector for the CAST Experiment}, PoS TIPP2014 (2014) 060.

\bibitem{Murtas:2014zxa}
F.~Murtas, {Applications of triple GEM detectors beyond particle and nuclear
  physics}, \JINST 9~(01) (2014) C01058.
\newblock \href {http://dx.doi.org/10.1088/1748-0221/9/01/C01058}
  {\path{doi:10.1088/1748-0221/9/01/C01058}}.

\bibitem{Capparelli:2014lua}
L.~M. Capparelli, G.~Cavoto, D.~Mazzilli, A.~D. Polosa, {Directional Dark
  Matter Searches with Carbon Nanotubes}, \PHYSDARKU 9-10 (2015) 24--30,
  [Erratum: Phys. Dark Univ.11,79(2016)].
\newblock \href {http://arxiv.org/abs/1412.8213} {\path{arXiv:1412.8213}},
  \href {http://dx.doi.org/10.1016/j.dark.2015.12.004,
  10.1016/j.dark.2015.08.002} {\path{doi:10.1016/j.dark.2015.12.004,
  10.1016/j.dark.2015.08.002}}.

\bibitem{Jakubek:2014}
J.~Jakubek, et~al., {Large area pixel detector WIDEPIX with full area
  sensitivity composed of 100 Timepix assemblies with edgeless sensors}, \JINST
  9~(04) (2014) C04018.
\newblock \href {http://dx.doi.org/10.1088/1748-0221/9/04/C04018}
  {\path{doi:10.1088/1748-0221/9/04/C04018}}.

\bibitem{Tezcan:2007}
D.~S. Tezcan, et~al., {Sloped through wafer vias for 3D wafer level packaging},
  in: {Proceedings of Electronic Components and Technology Conference, 2007},
  2007, pp. 643 -- 647.
\newblock \href {http://dx.doi.org/10.1109/ECTC.2007.373865}
  {\path{doi:10.1109/ECTC.2007.373865}}.

\bibitem{Bosma:2011zz}
M.~J. Bosma, E.~Heijne, J.~Kalliopuska, J.~Visser, E.~N. Koffeman, {Edgeless
  planar semiconductor sensors for a Medipix3-based radiography detector},
  \JINST 6 (2011) C11019.
\newblock \href {http://dx.doi.org/10.1088/1748-0221/6/11/C11019}
  {\path{doi:10.1088/1748-0221/6/11/C11019}}.

\bibitem{Vykydal:2008}
Z.~Vykydal, et~al., {The RELAXd project: Development of four-side tilable
  photon-counting imagers}, \NIM A591 (2008) 241--244.

\bibitem{Charpak:1987it}
G.~Charpak, W.~Dominik, J.~P. Fabre, J.~Gaudaen, V.~Peskov, F.~Sauli,
  M.~Suzuki, A.~Breskin, R.~Chechik, D.~Sauvage, {Some Applications of the
  Imaging Proportional Chamber}, \IEEETRANSNUCLSCI 35 (1988) 483--486.
\newblock \href {http://dx.doi.org/10.1109/23.12770}
  {\path{doi:10.1109/23.12770}}.

\bibitem{Buckland:1994gc}
K.~N. Buckland, M.~J. Lehner, G.~E. Masek, M.~Mojaver, {Low pressure gaseous
  detector for particle dark matter}, \PRL 73 (1994) 1067--1070.
\newblock \href {http://dx.doi.org/10.1103/PhysRevLett.73.1067}
  {\path{doi:10.1103/PhysRevLett.73.1067}}.

\bibitem{Lehner:1997fs}
M.~J. Lehner, K.~N. Buckland, G.~E. Masek, {Electron diffusion in a low
  pressure methane detector for particle dark matter}, \ASTROPART 8 (1997)
  43--50.
\newblock \href {http://dx.doi.org/10.1016/S0927-6505(97)00036-4}
  {\path{doi:10.1016/S0927-6505(97)00036-4}}.

\bibitem{Breskin:1988sg}
A.~Breskin, New developments in optical imaging detectors, Nucl. Phys. A498
  (1989) 457.
\newblock \href {http://dx.doi.org/10.1016/0375-9474(89)90625-8}
  {\path{doi:10.1016/0375-9474(89)90625-8}}.

\bibitem{Gai:2011yb}
M.~Gai, et~al., {An Optical Readout TPC (O-TPC) for Studies in Nuclear
  Astrophysics With Gamma-Ray Beams at HIgS}, \JINST 5 (2010) P12004.
\newblock \href {http://arxiv.org/abs/1101.1940} {\path{arXiv:1101.1940}},
  \href {http://dx.doi.org/10.1088/1748-0221/5/12/P12004}
  {\path{doi:10.1088/1748-0221/5/12/P12004}}.

\bibitem{Breskin:1988sd}
A.~Breskin, et~al., {A Highly Efficient Low Pressure {UV} Rich Detector With
  Optical Avalanche Recording}, \NIM A273 (1988) 798--804.
\newblock \href {http://dx.doi.org/10.1016/0168-9002(88)90099-X}
  {\path{doi:10.1016/0168-9002(88)90099-X}}.

\bibitem{Lightfoot:2008ig}
P.~K. Lightfoot, G.~J. Barker, K.~Mavrokoridis, Y.~A. Ramachers, N.~J.~C.
  Spooner, {Optical readout tracking detector concept using secondary
  scintillation from liquid argon generated by a thick gas electron
  multiplier}, \JINST 4 (2009) P04002.
\newblock \href {http://arxiv.org/abs/0812.2123} {\path{arXiv:0812.2123}},
  \href {http://dx.doi.org/10.1088/1748-0221/4/04/P04002}
  {\path{doi:10.1088/1748-0221/4/04/P04002}}.

\bibitem{Fraga:2002uc}
F.~A.~F. Fraga, L.~M.~S. Margato, S.~T. Fetal, M.~M. F.~R. Fraga,
  R.~Ferreira-Marques, A.~J. P.~L. Policarpo, B.~Guerard, A.~Oed, G.~Manzini,
  T.~van Vuure, {CCD readout of GEM-based neutron detectors}, \NIM A478 (2002)
  357--361.
\newblock \href {http://dx.doi.org/10.1016/S0168-9002(01)01829-0}
  {\path{doi:10.1016/S0168-9002(01)01829-0}}.

\bibitem{FragaLuminescence:2003}
F.~A.~F. Fraga, L.~M.~S. Margato, S.~T.~G. Fetal, M.~M. F.~R. Fraga,
  R.~Ferreira~Marques, A.~J. P.~L. Policarpo, {Luminescence and imaging with
  gas electron multipliers}, \NIM A513 (2003) 379--387.

\bibitem{Roccaro:2009tg}
A.~Roccaro, et~al., {A Background-Free Direction-Sensitive Neutron Detector},
  \NIM A608 (2009) 305--309.
\newblock \href {http://arxiv.org/abs/0906.3910} {\path{arXiv:0906.3910}},
  \href {http://dx.doi.org/10.1016/j.nima.2009.06.102}
  {\path{doi:10.1016/j.nima.2009.06.102}}.

\bibitem{Deaconu:2015vbk}
C.~Deaconu,
  \href{http://inspirehep.net/record/1418276/files/Thesis-2015-Deaconu.pdf}{{A
  model of the directional sensitivity of low-pressure CF$_4$ dark matter
  detectors}}, Ph.D. thesis, MIT (2015).
\newblock \href {http://dx.doi.org/1721.1/99314} {\path{doi:1721.1/99314}}.
\newline\urlprefix\url{http://inspirehep.net/record/1418276/files/Thesis-2015-Deaconu.pdf}

\bibitem{Rubin:2013yua}
A.~Rubin, L.~Arazi, S.~Bressler, A.~Dery, L.~Moleri, M.~Pitt, D.~Vartsky,
  A.~Breskin, {Optical readout: A tool for studying gas-avalanche processes},
  \JINST 8 (2013) P08001.
\newblock \href {http://arxiv.org/abs/1305.1196} {\path{arXiv:1305.1196}},
  \href {http://dx.doi.org/10.1088/1748-0221/8/08/P08001}
  {\path{doi:10.1088/1748-0221/8/08/P08001}}.

\bibitem{Fetal:2007zz}
S.~T.~G. Fetal, F.~A.~F. Fraga, L.~M.~S. Margato, M.~M. F.~R. Fraga, S.~R.
  Pereira, R.~Ferreira-Marques, A.~J. P.~L. Policarpo, {Towards a PMT based
  optical readout GEM TPC: First results}, \NIM A581 (2007) 202--205.
\newblock \href {http://dx.doi.org/10.1016/j.nima.2007.07.078}
  {\path{doi:10.1016/j.nima.2007.07.078}}.

\bibitem{Lopez:2013ah}
J.~P. Lopez, et~al., {Background Rejection in the DMTPC Dark Matter Search
  Using Charge Signals}, \NIM A696 (2012) 121--128.
\newblock \href {http://arxiv.org/abs/1301.5685} {\path{arXiv:1301.5685}},
  \href {http://dx.doi.org/10.1016/j.nima.2012.08.073}
  {\path{doi:10.1016/j.nima.2012.08.073}}.

\bibitem{Arnold:1992mv}
R.~Arnold, Y.~Giamataris, J.~L. Guyonnet, A.~Racz, J.~Seguinot, T.~Ypsilantis,
  {A Fast cathode pad photon detector for Cherenkov ring imaging}, \NIM A314
  (1992) 465--494.
\newblock \href {http://dx.doi.org/10.1016/0168-9002(92)90239-Z}
  {\path{doi:10.1016/0168-9002(92)90239-Z}}.

\bibitem{Tovey:2000mm}
D.~R. Tovey, R.~J. Gaitskell, P.~Gondolo, Y.~A. Ramachers, L.~Roszkowski, {A
  New model independent method for extracting spin dependent (cross-section)
  limits from dark matter searches}, \PHYSLETT B488 (2000) 17--26.
\newblock \href {http://arxiv.org/abs/hep-ph/0005041}
  {\path{arXiv:hep-ph/0005041}}, \href
  {http://dx.doi.org/10.1016/S0370-2693(00)00846-7}
  {\path{doi:10.1016/S0370-2693(00)00846-7}}.

\bibitem{Pansky:1994zh}
A.~Pansky, A.~Breskin, A.~Buzulutskov, R.~Chechik, V.~Elkind, J.~Va'vra, {The
  Scintillation of CF$_4$ and its relevance to detection science}, \NIM A354
  (1995) 262--269.
\newblock \href {http://dx.doi.org/10.1016/0168-9002(94)01064-1}
  {\path{doi:10.1016/0168-9002(94)01064-1}}.

\bibitem{Kaboth:2008mi}
A.~Kaboth, et~al., {A Measurement of Photon Production in Electron Avalanches
  in CF$_4$}, \NIM A592 (2008) 63--72.
\newblock \href {http://arxiv.org/abs/physics.ins-det/0803.2195}
  {\path{arXiv:physics.ins-det/0803.2195}}, \href
  {http://dx.doi.org/10.1016/j.nima.2008.03.120}
  {\path{doi:10.1016/j.nima.2008.03.120}}.

\bibitem{2010NIMPB.268.1456M}
A.~{Morozov}, M.~M.~F.~R. {Fraga}, L.~{Pereira}, L.~M.~S. {Margato}, S.~T.~G.
  {Fetal}, B.~{Guerard}, G.~{Manzin}, F.~A.~F. {Fraga}, {Photon yield for
  ultraviolet and visible emission from CF$_{4}$ excited with
  {$\alpha$}-particles}, \NIM B268 (2010) 1456--1459.
\newblock \href {http://dx.doi.org/10.1016/j.nimb.2010.01.012}
  {\path{doi:10.1016/j.nimb.2010.01.012}}.

\bibitem{Margato:2013gqa}
L.~M.~S. Margato, A.~Morozov, M.~M. F.~R. Fraga, L.~Pereira, F.~A.~F. Fraga,
  {Effective decay time of CF$_4$ secondary scintillation}, \JINST 8 (2013)
  P07008.
\newblock \href {http://dx.doi.org/10.1088/1748-0221/8/07/P07008}
  {\path{doi:10.1088/1748-0221/8/07/P07008}}.

\bibitem{Ju:2007ra}
Y.~L. Ju, J.~Dodd, R.~Galea, M.~Leltchouk, W.~Willis, L.~X. Jia, P.~Rehak,
  V.~Chernyatin, {Cryogenic design and operation of liquid helium in an
  electron bubble chamber towards low energy solar neutrino detectors},
  Cryogenics 47 (2007) 81--88.
\newblock \href {http://dx.doi.org/10.1016/j.cryogenics.2006.08.008}
  {\path{doi:10.1016/j.cryogenics.2006.08.008}}.

\bibitem{JanesickBook2001}
J.~R. {Janesick}, {Scientific charge-coupled devices}, SPIE press, New York,
  2001.

\bibitem{HowellBook2006}
S.~B. {Howell}, {Handbook of CCD Astronomy}, Cambridge University Press,
  Cambridge, UK, 2006.

\bibitem{DMTPC_wolfethesis}
I.~Wolfe, {Measurement of Work Function in CF$_4$ Gas}, MIT B.Sc. Thesis, 2010.

\bibitem{DMTPC_4shins}
J.~Battat, et~al., {The Dark Matter Time Projection Chamber 4Shooter
  directional dark matter detector: calibration in a surface laboratory}, \NIM
  A565 (2014) 88.

\bibitem{Ahlen:2010ub}
S.~Ahlen, J.~Battat, T.~Caldwell, C.~Deaconu, D.~Dujmic, et~al., {First Dark
  Matter Search Results from a Surface Run of the 10-L DMTPC Directional Dark
  Matter Detector}, \PHYSLETT B695 (2011) 124--129.
\newblock \href {http://arxiv.org/abs/1006.2928} {\path{arXiv:1006.2928}},
  \href {http://dx.doi.org/10.1016/j.physletb.2010.11.041}
  {\path{doi:10.1016/j.physletb.2010.11.041}}.

\bibitem{Fustin_thesis}
D.~A. Fustin, {First Dark Matter limits from the COUPP 4 kg bubble chamber at a
  deep underground site}, University of Chicago Ph.D. Thesis, 2012.

\bibitem{nphan_ccd2}
N.~S. Phan, et~al., {The first optical spectrum from Fe-55 in a TPC}.

\bibitem{Dujmic:2008ut}
D.~Dujmic, et~al., {Charge amplification concepts for direction-sensitive dark
  matter detectors}, \ASTROPART 30 (2008) 58--64.
\newblock \href {http://arxiv.org/abs/astro-ph/0804.4827}
  {\path{arXiv:astro-ph/0804.4827}}, \href
  {http://dx.doi.org/10.1016/j.astropartphys.2008.06.009}
  {\path{doi:10.1016/j.astropartphys.2008.06.009}}.

\bibitem{Dujmic:2008iq}
D.~Dujmic, et~al., {Improved measurement of the head-tail effect in nuclear
  recoils}, J. Phys. Conf. Ser. 120 (2008) 042030.
\newblock \href {http://arxiv.org/abs/astro-ph/0801.2687}
  {\path{arXiv:astro-ph/0801.2687}}, \href
  {http://dx.doi.org/10.1088/1742-6596/120/4/042030}
  {\path{doi:10.1088/1742-6596/120/4/042030}}.

\bibitem{Christophorou1996}
L.~G. {Christophorou}, J.~K. {Olthoff}, M.~V.~V.~S. {Rao}, {Electron
  Interactions with CF$_{4}$}, Journal of Physical and Chemical Reference Data
  25 (1996) 1341--1388.
\newblock \href {http://dx.doi.org/10.1063/1.555986}
  {\path{doi:10.1063/1.555986}}.

\bibitem{DMTPC_ucladm14}
C.~Deaconu, {Recent progress from the DMTPC directional dark matter
  experiment}, in: UCLA 11th Symposium on Sources and Detection of Dark Matter
  and Dark Energy in the Universe, 2014.

\bibitem{Aleksandrov:2016fyr}
A.~Aleksandrov, et~al., {NEWS: Nuclear Emulsions for WIMP Search}\href
  {http://arxiv.org/abs/1604.04199} {\path{arXiv:1604.04199}}.

\bibitem{Powell}
C.~{Powell}, P.~H. {Fowler}, D.~H. {Perkins}, {The Study of Elementary
  Particles by the Photographic Method}, {Peragamon Press}, London, UK, 1959.

\bibitem{Broadhead}
P.~{Broadhead}, {The Theory of the Photographic Process}, Macmillan, New York,
  1977.

\bibitem{FineGrained}
T.~{Naka}, T.~{Asada}, T.~{Katsuragawa}, K.~{Hakamata}, M.~{Yoshimoto},
  K.~{Kuwabara}, M.~{Nakamura}, O.~{Sato}, T.~{Nakano}, Y.~{Tawara}, G.~{De
  Lellis}, C.~{Sirignano}, N.~{D'Ambrossio}, {Fine grained nuclear emulsion for
  higher resolution tracking detector}, \NIM A718 (2013) 519--521.
\newblock \href {http://dx.doi.org/10.1016/j.nima.2012.11.106}
  {\path{doi:10.1016/j.nima.2012.11.106}}.

\bibitem{SRIM}
J.~F. {Ziegler}, P.~{Biersack}, {The Stopping and Range of Ions in Matter},
  Pergamon Press, New York, 1985.

\bibitem{fast_scanning}
A.~Alexandrov, A.~Buonaura, L.~Consiglio, N.~D'Ambrosio, G.~D. Lellis, A.~D.
  Crescenzo, N.~D. Marco, G.~Galati, A.~Lauria, M.~Montesi, F.~Pupilli,
  T.~Shchedrina, V.~Tioukov, M.~Vladymyrov,
  \href{http://stacks.iop.org/1748-0221/10/i=11/a=P11006}{A new fast scanning
  system for the measurement of large angle tracks in nuclear emulsions},
  \JINST 10~(11) (2015) P11006.
\newline\urlprefix\url{http://stacks.iop.org/1748-0221/10/i=11/a=P11006}

\bibitem{Alexandrov:2016tyi}
A.~Alexandrov, et~al., {A new generation scanning system for the high-speed
  analysis of nuclear emulsions}, \JINST 11~(06) (2016) P06002.
\newblock \href {http://dx.doi.org/10.1088/1748-0221/11/06/P06002}
  {\path{doi:10.1088/1748-0221/11/06/P06002}}.

\bibitem{EU_scanning}
L.~Arrabito, et~al., {Hardware performance of a scanning system for high speed
  analysis of nuclear emulsions}, \NIM A568 (2006) 578--587.
\newblock \href {http://arxiv.org/abs/physics/0604043}
  {\path{arXiv:physics/0604043}}, \href
  {http://dx.doi.org/10.1016/j.nima.2006.06.072}
  {\path{doi:10.1016/j.nima.2006.06.072}}.

\bibitem{JP_scanning}
K.~Morishima, T.~Nakano, {Development of a new automatic nuclear emulsion
  scanning system, S-UTS, with continuous 3D tomographic image read-out},
  \JINST 5 (2010) P04011.
\newblock \href {http://dx.doi.org/10.1088/1748-0221/5/04/P04011}
  {\path{doi:10.1088/1748-0221/5/04/P04011}}.

\bibitem{shape_analysis}
M.~Kimura, T.~Naka, {Submicron track readout in fine-grained nuclear emulsions
  using optical microscopy}, \NIM A680 (2012) 12--17.
\newblock \href {http://dx.doi.org/10.1016/j.nima.2012.04.010}
  {\path{doi:10.1016/j.nima.2012.04.010}}.

\bibitem{XMS}
T.~Naka, et~al., {Analysis system of submicron particle tracks in the
  fine-grained nuclear emulsion by a combination of hard X-ray and optical
  microscopy}, \REVSCIINSTRUM 86~(7) (2015) 073701.
\newblock \href {http://dx.doi.org/10.1063/1.4926350}
  {\path{doi:10.1063/1.4926350}}.

\bibitem{SP8}
Y.~{Suzuki}, A.~{Takeuchi}, Y.~{Terada}, K.~{Uesugi}, S.~{Tamaru}, {Development
  of large-field high-resolution hard X-ray imaging microscopy and
  microtomography with Fresnel zone plate objective}, Proceedings of SPIE.
  8851, X-ray Nanoimaging: Instruments and Methods 885109.
\newblock \href {http://dx.doi.org/10.1117/12.2025792}
  {\path{doi:10.1117/12.2025792}}.

\bibitem{plasmon}
J.~J. {Mock}, M.~{Barbic}, D.~R. {Smith}, D.~A. {Schultz}, S.~{Schultz}, {Shape
  effects in plasmon resonance of individual colloidal silver nanoparticles},
  J. of Chem. Phys. 116.
\newblock \href {http://dx.doi.org/10.1063/1.1462610}
  {\path{doi:10.1063/1.1462610}}.

\bibitem{baracchiniIDM2016}
E.~Baracchini, {NITEC: a Negative Ion Time Expansion Chamber for directional
  Dark Matter searches}, identification of Dark Matter 2016 (2016).

\bibitem{Lightfoot:2007zz}
P.~K. Lightfoot, N.~J.~C. Spooner, T.~B. Lawson, S.~Aune, I.~Giomataris, {First
  operation of bulk micromegas in low pressure negative ion drift gas mixtures
  for dark matter searches}, \ASTROPART 27 (2007) 490--499.
\newblock \href {http://dx.doi.org/10.1016/j.astropartphys.2007.02.003}
  {\path{doi:10.1016/j.astropartphys.2007.02.003}}.

\bibitem{Burgos:2009xm}
S.~Burgos, et~al., {Low Energy Electron and Nuclear Recoil Thresholds in the
  DRIFT-II Negative Ion TPC for Dark Matter Searches}, \JINST 4 (2009) P04014.
\newblock \href {http://arxiv.org/abs/0903.0326} {\path{arXiv:0903.0326}},
  \href {http://dx.doi.org/10.1088/1748-0221/4/04/P04014}
  {\path{doi:10.1088/1748-0221/4/04/P04014}}.

\bibitem{Battat:2016xaw}
J.~B.~R. Battat, et~al., {First measurement of nuclear recoil head-tail sense
  in a fiducialised WIMP dark matter detector}\href
  {http://arxiv.org/abs/1606.05364} {\path{arXiv:1606.05364}}.

\bibitem{Miuchi:2009vga}
K.~Miuchi, et~al., {Direction-sensitive Dark Matter Search –NEWAGE–},
  \EASPUB 36 (2009) 243--248.
\newblock \href {http://dx.doi.org/10.1051/eas/0936034}
  {\path{doi:10.1051/eas/0936034}}.

\bibitem{vahsenPrivate}
{S. E. Vahsen}, Private communication.

\bibitem{Vahsen:2014fba}
S.~Vahsen, M.~Hedges, I.~Jaegle, S.~Ross, I.~Seong, et~al., {3-D Tracking of
  Nuclear Recoils in a Miniature Time Projection Chamber}\href
  {http://arxiv.org/abs/1407.7013} {\path{arXiv:1407.7013}}.

\bibitem{ref:atlasIBL}
J.~Jentzsch,
  \href{http://stacks.iop.org/1748-0221/10/i=04/a=C04036}{Performance tests
  during the atlas ibl stave integration}, Journal of Instrumentation 10~(04)
  (2015) C04036.
\newline\urlprefix\url{http://stacks.iop.org/1748-0221/10/i=04/a=C04036}

\bibitem{Bates:2015fpj}
R.~Bates, {ATLAS pixel upgrade for the HL-LHC}, PoS VERTEX2015 (2015) 006.

\end{thebibliography}

\end{document}